\documentclass[manuscript,screen]{acmart}
\usepackage{natbib}
\usepackage{graphicx}

\usepackage[small,compact]{titlesec}
\usepackage{pdflscape}
\usepackage{afterpage}
\usepackage{stfloats}
\usepackage{float}
\usepackage{placeins}
\usepackage{caption}

\usepackage{subcaption}

\usepackage{rotating,tabularx}
\usepackage{adjustbox}
\usepackage{hyperref} 
\hypersetup{backref=true,       
    pagebackref=true,               
    hyperindex=true,                
    colorlinks=false,                
    breaklinks=true,                             
    bookmarks=true,                 
    bookmarksopen=false,
    hyperfootnotes=false
}

\usepackage{multirow}
\usepackage[most]{tcolorbox}
\newtcolorbox{MySummaryBox}[2][]{
  enhanced,
  coltitle=black,
  fonttitle=\bfseries,
  title=#2,
  #1,
  attach boxed title to top left={xshift=0.7em, yshift=-2.5mm},
  boxed title style={size=small, colback=gray!40}
}

\usepackage{soul}
\usepackage{textcomp}
\usepackage{xcolor}
\usepackage{enumitem}

\definecolor{hzcolor}{RGB}{10, 186, 181}

\AtBeginDocument{%
  }

\begin{document}
%
\title{Understanding Open Source Contributor Profiles in Popular Machine Learning Libraries}

\author{Jiawen Liu}
\affiliation{
  \institution{Department of Electrical and Computer Engineering, Queen's University}
  \city{Kingston}
  \country{Canada}}
\email{jiawen.liu@queensu.ca}

\author{Haoxiang Zhang}
\affiliation{
  \institution{Software Analysis and Intelligence Lab (SAIL), Queen's University}
  \city{Kingston}
  \country{Canada}}
  
\author{Ying Zou}
\affiliation{
  \institution{Department of Electrical and Computer Engineering, Queen's University}
  \city{Kingston}
  \country{Canada}}

\renewcommand{\shortauthors}{Liu et al.}

\begin{abstract}
With the increasing popularity of machine learning (ML), many open-source software (OSS) contributors are attracted to developing and adopting ML approaches. Comprehensive understanding of ML contributors is crucial for successful ML OSS development and maintenance. Without such knowledge, there is a risk of inefficient resource allocation and hindered collaboration in ML OSS projects. Existing research focuses on understanding the difficulties and challenges perceived by ML contributors by user surveys. There is a lack of understanding of ML contributors based on their activities tracked from software repositories. In this paper, we aim to understand ML contributors by identifying contributor profiles in ML libraries. We further study contributors’ OSS engagement from three aspects: workload composition, work preferences, and technical importance. By investigating 7,640 contributors from 6 popular ML libraries (TensorFlow, PyTorch, Keras, MXNet, Theano, and ONNX), we identify four contributor profiles: Core-Afterhour, Core-Workhour, Peripheral-Afterhour, and Peripheral-Workhour. We find that: 1) project experience, authored files, collaborations, and geological location are significant features of all profiles; 2) contributors in Core profiles exhibit significantly different OSS engagement compared to Peripheral profiles; 3) contributors’ work preferences and workload compositions significantly impact project popularity; 4) long-term contributors evolve towards making fewer, constant, balanced and less technical contributions.
\end{abstract}




\maketitle

\section{Introduction}
Open Source Software (OSS) has emerged as a dominant model in software development, gaining widespread recognition among enterprises and developers as the preferred approach for software development. The OSS community comprises globally distributed contributors and users with shared interests, who actively participate in knowledge sharing and collaborate on the development and maintenance of software projects. Anyone with the necessary knowledge and skills can be a contributor to OSS projects. They can contribute in various ways, such as writing source code, updating documentation, reporting issues, conducting code reviews, and participating in discussions. These activities are critical to the success of the OSS development process and evolution. The rising popularity of open-source ML libraries, such as Tensorflow\footnote{https://www.tensorflow.org/} and PyTorch\footnote{https://pytorch.org/}, has facilitated ML implementation, attracting a growing number of software developers to learn ML technologies and contribute to ML projects~\cite{blog}.

Developers are the most important resource in software development and maintenance. A deep understanding of developer team composition could provide valuable insights for software development management. This is particularly important in OSS projects, as contributors are distributed around the globe and have diverse backgrounds. Therefore, it is challenging for project managers and maintainers to have comprehensive knowledge to facilitate contributor productivity. In traditional software engineering, extensive research has been conducted to portray or categorize OSS contributors from many specific aspects, 
such as the volume of contribution~\cite{da2014unveiling,mockus2000case, mockus2002two}, technical expertise~\cite{dey2021representation, montandon2019identifying}, social networks~\cite{cohen2018large, el2019empirical}, duration~\cite{fe8a3ca6d43d4253855607d9a2cad816, 6880395}, and contribution dynamics~\cite{yue2022off} . However, few have endeavored to provide a comprehensive understanding of OSS contributors considering their behaviors, expertise, workload, work preferences, and the importance of their contributions. A lack of comprehensive insight may lead to ineffective resource allocation or collaborative challenges. For example, as reported by Balali et al.~\cite{balali2018newcomers}, contributors face challenges in their collaborations such as communication issues caused by timezone, language, and cultural differences, difficulties in managing time to collaborate, mismatch of knowledge background, and harsh project atmosphere. 

To investigate the challenges faced by ML developers, existing studies primarily conduct user surveys and interviews to collect insights into their experiences, demands, and pain points~\cite{mldeveloper, MLdeveloperexp, 8836142}. Han et al.~\cite{10398589} conduct an empirical study on the onboarding process for newcomers to deep learning projects.
However, there is a lack of studies that comprehensively portray ML contributors and their characteristics. To enhance our understanding of ML contributors, we conduct an empirical study on contributors of 6 popular ML libraries (i.e., Tensorflow, PyTorch, Keras, MXNet, Theano, and ONNX). Particularly, we aim to address the following research questions:

\textbf{RQ1: What are the characteristics of contributor profiles?} We identify four distinct contributor profiles (i.e., \textit{Core-Afterhour}, \textit{Core-Workhour}, \textit{Peripheral-Afterhour}, and \textit{Peripheral-Workhour}) based on their working habit, amount of contribution, contributing styles, and technical expertise. To gain a further understanding of these profiles, we build four binary logistic regression models to analyze the important features associated with each profile. We find that project experience, authored files, and collaborations are significant for all profiles, mainly distinguishing core from peripheral profiles. 
 
\textbf{RQ2: What is the OSS engagement of each contributor profile?} We observe in RQ1 that contributors within the same profile still have different OSS engagements. To further study the OSS engagement of contributors in different profiles, we analyze contributors from three aspects: \textbf{workload composition} that describes a contributor's focus of work on the five key OSS activities within a period (i.e., reporting issues, issue discussions, commits, pull request discussions, and code reviews); \textbf{work preferences} that capture the dynamics and balance of a contributor's contributions within a period; and \textbf{technical importance} that measures the importance of a contributor's contribution. We identify five workload composition patterns that capture the contributors' common workload compositions, extract nine work preference features, and introduce four technical importance metrics. 
We apply the Chi-square test to compare the distribution of contributors with different workload compositions across profiles. We apply the Mann-Whitney U test to compare contributors' work preferences and technical importance across profiles. We observe significant differences in the workload composition, work preference, and technical importance between \textit{core} and \textit{peripheral} contributors.

\textbf{RQ3: What are the important factors of contributor OSS engagement for increasing the popularity of a project?}
To understand the impact of various contributor OSS engagements on the project, we investigate the association between the distribution of contributor workload composition and work preferences with the growth of project popularity, measured by star ratings and forks. We build four mixed-effect models to study the relationship between the distribution of contributor's work preferences and workload compositions with the increase of stars and forks respectively. We find that work preference towards balanced contributions is significantly associated with the increase in star ratings and forks. Additionally, a higher presence of Issue Reporters in the project and a higher portion of code reviews made by Issue Discussants and Committers are also associated with the increase in stars and forks.

\textbf{RQ4: How does contributor OSS engagement evolve?}
We observe in RQ2 that contributors' OSS engagement changes over time. To understand the evolution of OSS engagement, we apply the Cox-Stuart trend test to examine the significant trends in workload composition, work preference, and technical importance in contributors' active periods. We observe a portion of long-term contributors evolve towards fewer, constant, and balanced contribution dynamics, and shift away from highly technical and intensive contributions.


In summary, we make the following contributions to the software engineering community:

(1) We provide a comprehensive understanding of ML contributors from both static and dynamic points of view by studying their activities in a period and analyzing their engagement over time. 

(2) We identify four contributor profiles in six popular ML libraries. These profiles help establish an initial understanding of ML contributors with quantitative analysis of their OSS activities. 

(3) We identify the association of contributors' work preferences and workload compositions with the growth of project popularity. 

\textbf{Paper organization.} The remainder of our paper is organized as follows. Section~\ref{sec:experiment} describes data collection and experiment setup. Section~\ref{sec:results} presents the motivations, approaches, and results of our research questions. We describe the implication in Section~\ref{sec:implications} and threats to validity of our study in Section~\ref{sec:threats}. Section~\ref{sec:relatedwork} describes relevant studies. Lastly, we conclude our paper in Section~\ref{sec:conclusion} and provide our replication package in Section~\ref{sec:replication}.

\section{Experiment Setup}
\label{sec:experiment}

\begin{figure}[h!]
\includegraphics[width=\columnwidth]{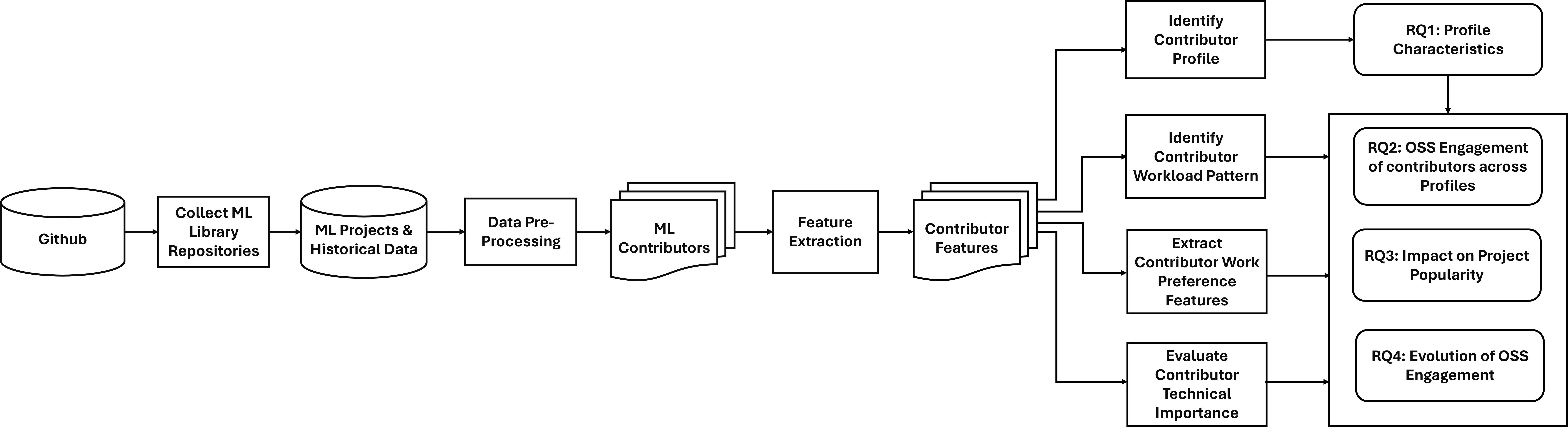}
\caption{The overview of our approach.}
\label{fig:approach_overview}
\end{figure}

In this section, we present data collection and experiment setup. An overview of our approach is shown in Figure~\ref{fig:approach_overview}. We first collect the subject projects and their historical data from Github. Preprocessing is conducted on the collected project data, which involves removing non-human contributors and cleaning noises that may affect contributor feature calculation. Then, we extract ML contributors within the subject projects and extract contributor features from the processed data. Lastly, we identify contributor profiles, workload composition patterns, work preference features, and technical importance.

\subsection{Data Collection}
We select the subject projects and collect the history data from Github using Github REST API~\cite{githubapi} 
and PyDriller~\cite{10.1145/3236024.3264598}. 

\textbf{Project Selection.} To run our experiments on projects that contain rich contributor information and the most up-to-date machine learning technology, we choose to study projects related to ML libraries and frameworks. These projects gather collaborations among ML experts and usually have a larger size than the projects that implement a single machine learning model to solve a problem. We rank the popular ML library projects on Github according to the number of contributors and select six projects with varying sizes of contributor groups. We collect two large-size projects with over 2k contributors (i.e., Tensorflow and Pythoch), two medium-sized projects with 500 to 2k contributors (i.e., Keras and MXNet), and two small-size projects with less than 500 contributors (i.e., Theano and ONNX). All the selected projects contain more than 2k commits and have been active for more than 5 years. 
A detailed overview of the subject projects is presented in Table~\ref{tab:projects}.

\begin{table}[t]
\caption{Descriptive statistics of the selected ML library projects.}
\label{tab:projects}
  \centering
\begin{tabular}{c|c|c|c|c|c}
\hline
\textbf{Project Name} & \textbf{Contributors} & \textbf{Commits} & \textbf{Pull Requests} & \textbf{Issues} & \textbf{Creation Time} \\ \hline
\textbf{Tensorflow}   & 3,222                  & 137,512           & 21,940                  & 33,767           & 2015-11                \\ \hline
\textbf{PyTorch}      & 2,509                  & 53,069            & 59,705                  & 19,499           & 2012-01                \\ \hline
\textbf{Keras}        & 1,072                  & 7,463             & 5,620                   & 11,249           & 2015-03                \\ \hline
\textbf{MXNet}        & 874                   & 11,893            & 10,893                  & 7,751            & 2015-04                \\ \hline
\textbf{Theano}       & 342                   & 28,132            & 4,016                   & 2,086            & 2008-01                \\ \hline
\textbf{ONNX}         & 242                   & 2,085             & 2,360                   & 1,850            & 2012-01                \\ \hline
\end{tabular}
\end{table}

\textbf{Collecting Data.} After the subject projects are selected, we retrieve the project development data from their repositories on Github. We use Github REST API to collect $(i)$ commits; $(ii)$ pull requests; $(iii)$ issue reports of each project, and $(iv)$ the account type of contributors (e.g., bot, organization, or user account) in the projects. PyDriller~\cite{10.1145/3236024.3264598} is used to fetch the commit timestamps in the local time of the commit authors and identify their time zones.

\subsection{Data Preprocessing}

\subsubsection{Removing Nonhuman Contributors}
\hfill

\noindent 
To focus on human contributors, we exclude bot, organization, and enterprise accounts from our analysis. To achieve this, we extract all the usernames of contributors who make at least one commit in the selected ML projects. We then use the Github Users API\footnote{https://docs.github.com/en/rest/users} to check if each username belongs to a user, bot, organization, or enterprise account and only retain those with user accounts. Next, we sort the contributors in each project by their total number of commits and manually investigate the remaining contributors to identify any contributors with abnormal usernames or behaviors (e.g., making a great number of commits without raising any pull requests or making comments). As a result of our investigation, we remove 9 bots (i.e., tensorflower-gardener, tensorflow-jenkins, onnxbot, facebook-github-bot, PyTorchmergebot, docusaurus-bot, theano-bot, deadsnakes-issues-bot, and mxnet-label-bot).

\subsubsection{Cleaning Noises in Pull Request Data}
\hfill

\noindent 
To extract contributor features related to pull request acceptance, such as the number of merged pull requests and the pull request approval ratio for a contributor, it is necessary to determine whether a pull request has been accepted. Typically, we rely on the pull request status on Github, where accepted pull requests are set to 'Merged' status and the rejected pull requests are set to 'Closed'. However, we observe that one of our subject projects, PyTorch, uses different strategies to indicate the approval of pull requests. Prior to July 2018, Pytorch19,499 set accepted pull requests to 'Merged' status, but after July 2018, very few pull requests (i.e., less than 100 each month) were set to 'Merged' status. Starting in May 2019, PyTorch assigns a ‘Merged’ label to the approved and merged pull requests. Therefore, we consider both the 'Merged' status and the ‘Merged’ label as the indication of acceptance for PyTorch pull requests before July 2018 and after May 2019. However, during the period between July 2018 and May 2019, few pull requests were set to 'Merged' status or assigned a ‘Merged’ label, even though the pull request submission rate was similar to other periods. We investigate the comments under the pull requests submitted during this period, and find that many without 'Merged' label or 'Merged' status were actually accepted. This ambiguity makes it challenging to distinguish between accepted and rejected pull requests and is unable to reflect the accurate contributor pull request acceptance. Hence, we exclude the PyTorch pull requests submitted between July 2018 and May 2019 when calculating the features that relate to pull request acceptance (e.g., PR Approval Ratio and PR approval density).

\subsection{Extracting Contributor Features}
We extract 7,640 human contributors from the subject projects, where the cross-project contributors are treated as separate individuals in each project to account for potential variations in their behaviors across projects. For each contributor, we extract contributor features that describe their OSS activities from the project repository and their GitHub profile page. The extracted contributor features encompass four categories: \textit{(i)} Developer, capturing contributors' publicly available personal information defining their identity in the OSS community, such as their timezone and number of Github followers; \textit{(ii)} Commit, including commit-related features, such as the number of code commits and nonfunctional commits made by a contributor; \textit{(iii)} Issue Report, which includes the features related to a contributor's issue report contribution, such as the number of issues raised or solved by a contributor; and lastly \textit{(iv)} Pull Request, which involves features related to a contributor's pull request contribution, such as the number of pull requests submitted or reviewed by a contributor. In total, we extract 31 contributor features. A complete list of the contributor features is presented in Table~\ref{tab:allfeatures} along with their definitions and calculations.

\begin{table}
\caption{List of the 31 extracted contributor features and the descriptions.}
\label{tab:allfeatures}
\begin{adjustbox}{max width=\textwidth}
\begin{tabular}{c|l|l}
\hline
\textbf{Category} & \textbf{Feature} & \textbf{Description} \\ \hline
\multirow{10}{*}{\textbf{Developer}} & Timezone &\begin{tabular}[c]{@{}l@{}} The primary timezone of a developer (i.e., the most frequent timezone where a\\ contributor makes commits). In this study, we define the timezone range of -13 to\\ -2 as Americas, -1 to 3 as Europe/Africa, and 4 to 12 as Asia.\end{tabular}\\ \cline{2-3} 
 & Worktime & \begin{tabular}[c]{@{}l@{}}The primary work time of a developer (i.e., the most frequent commit time of a\\ contributor). Normal work hours are defined as 8h to 18h in this study.\end{tabular}\\ \cline{2-3} 
 & Duration & Number of days a developer stay in a project.\\ \cline{2-3} 
 & Number of followers & Number of Github followers a developer has. \\ \cline{2-3} 
 & \begin{tabular}[c]{@{}l@{}} Number of \\ collaborated developers\end{tabular} & \begin{tabular}[c]{@{}l@{}} Number of developers in the project a developer has collaborated with. A \\collaboration is considered established when one contributor makes comments to \\a pull request or issue report submitted by another, or reviews another contributor's \\pull request.\end{tabular}  \\ \cline{2-3} 
 & Number of authored files &  The number of files in the project repository a developer has made changes to. \\ \cline{2-3} 
 & \begin{tabular}[c]{@{}l@{}} Number of \\ programming languages \end{tabular} & The number of programming languages a developer has used in one's commits. \\ \hline
\multirow{10}{*}{\textbf{Commit}} & Number of commits & The number of commits submitted by a developer. \\ \cline{2-3} 
 & Commit rate & Number of commits / Duration \\ \cline{2-3} 
 & \begin{tabular}[c]{@{}l@{}} Number of \\code commits \end{tabular}& \begin{tabular}[c]{@{}l@{}}The number of commits containing source code changes submitted by a developer\\ in the project.\end{tabular} \\ \cline{2-3} 
 & Code commit rate & The number of code commits / Duration \\ \cline{2-3} 
 & \begin{tabular}[c]{@{}l@{}} Number of \\non-code commits \end{tabular} & \begin{tabular}[c]{@{}l@{}}The number of commits without source code changes submitted by the developer \\ in the project.\end{tabular} \\ \cline{2-3} 
 & Non code commit rate & The number of other commits / Duration \\ \cline{2-3} 
 & Code contribution & Total number of lines of code added or deleted by a developer. \\ \cline{2-3} 
 & Code contribution rate & Code contribution / Duration \\ \cline{2-3}
 & Code contribution density & Average code churn per commit. \\ 
  \hline
\multirow{15}{*}{\textbf{\begin{tabular}[c]{@{}c@{}} Issue\\ Report\end{tabular}}} & Number of issues & Number of issues raised by a developer. \\ \cline{2-3} 
& \begin{tabular}[c]{@{}l@{}} Number of issue \\ participated \end{tabular} & The number of issue reports a developer has participated in. \\ \cline{2-3}
 & Number of issues solved & \begin{tabular}[c]{@{}l@{}}Number of issues the developer has solved in the project in the studied time period.\\ To get this value, we first parse the commit message of all commits made in the\\ studied time period. If the commit message contains keywords such as “fix”, \\ “resolve”, “address”, “close” or “solve” together with a valid issue id in the same\\ line, this commit is considered solving the issue and the author of this commit is \\ considered the one who solves the issue. Note that the issue must be raised \\ before the time of the commit is made. Then, we use the same method to find issue\\ ids from pull request messages.\end{tabular} \\ \cline{2-3} 
 & Issue contribution & Total number of issues raised or issues solved by a developer. \\ \cline{2-3} 
 & Issue contribution rate & Issue Contribution / Duration \\ \cline{2-3} 
 & Issue solving ratio & Issue solved / Number of issues a developer participated in \\ \cline{2-3} 
 & Issue solving density & Issue Solving Ratio / Duration \\ \hline
\multirow{8}{*}{\textbf{\begin{tabular}[c]{@{}c@{}}Pull\\ Request\end{tabular}}} & Number of pull requests & The number of pull requests submitted by a developer \\ \cline{2-3} 
 & Number of PR participated & The number of pull requests a developer has participated \\ \cline{2-3} 
 & Number of PR reviewed & The number of pull requests reviewed by a developer \\ \cline{2-3} 
 & Number of PR merged & The number of pull requests submitted by a developer being merged. \\ \cline{2-3} 
 & PR contribution & Number of pull requests raised or reviewed by a developer in the project \\ \cline{2-3} 
 & PR contribution rate & PR contribution / Duration \\ \cline{2-3} 
 & PR approval ratio & Number of PR merged /Number of pull requests \\ \cline{2-3} 
 & PR approval density & PR approval ratio / Duration \\ \hline
\end{tabular}
\end{adjustbox}
\end{table}

\subsection{Correlation and Redundancy Analysis}
\label{cor-redun}

Considering that correlated features can be expressed by each other and redundant features can be expressed by other features, highly correlated and redundant features blur the importance of the features to our analysis and result in unnecessary computations. Therefore, we conduct a correlation and redundancy analysis to remove the highly-correlated and redundant features. 
\begin{itemize}[leftmargin=*]
\item \textbf{Correlation Analysis:} 
We find that contributor features do not follow a normal distribution, thus we use Spearman Rank Correlation to conduct the correlation analysis~\cite{zar2005spearman}. Two features with a correlation coefficient higher than 0.7 are considered highly correlated~\cite{8613795}. For each pair of highly correlated features, we keep only one in the candidate list and remove the other one. The result of correlation analysis is shown in Figure~\ref{fig:cor}. Eight features (i.e., Code Commits, Code Contribution, Other Commit Rate,  Issue Contribution, Issue Solving Ratio, Issue Solving Density, Number of PR Participated, and PR Contribution) are removed and 23 features remain.

\begin{figure}[t]
\centering
\includegraphics[width = \columnwidth]{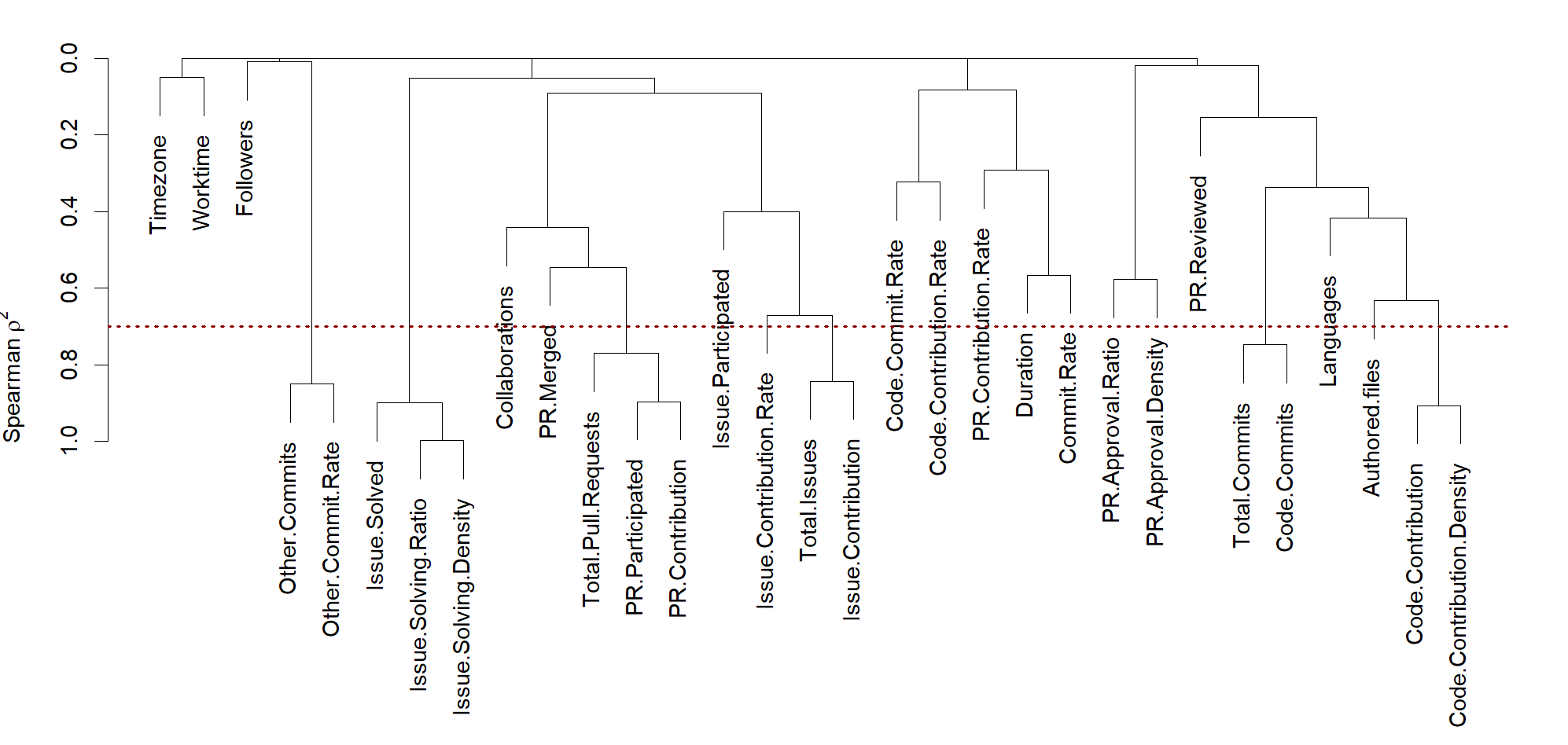}
\caption{Result of the Spearman correlation analysis of contributor features.}
\centering
\label{fig:cor}
\end{figure}

\item \textbf{Redundancy Analysis: } R-squared is a measure that indicates how much variance of a variable can be explained by other variables. We use a R-squared cut-off at 0.9 to identify the redundant features~\cite{miles2005r}. The number of pull requests is found to be redundant. As a result, 22 uncorrelated and non-redundant features remain.
\end{itemize}

\subsection{Identifying Contributor Profiles}
\label{sec:identify_profile}

We aim to study the characteristics of contributors' activities within open-source ML projects. To achieve this, we construct contributor profiles to categorize and capture common characteristics among groups of contributors with similar behaviors. We select the contributor features that capture the key aspects of OSS contributor behaviors and use clustering algorithms on the selected features to identify groups of contributors with similar behaviors and construct their profiles accordingly. The detailed approach involves the following four steps: 

\textbf{Step 1: Select contributor features for clustering.} To group contributors with similar behaviors, we select four contributor features that capture four key aspects of OSS contributors' behaviors to conduct the clustering. We refer to the study by Noei et al.~\cite{noei2023} to select features that effectively capture contributors' characteristics. Specifically, we include worktime, number of commits, and code contribution density, which are relevant to our study. Additionally, we incorporate the number of programming languages to capture the technical expertise of ML contributors. The four selected features and the rationale behind their selection are as follows:
\begin{itemize}
\item \textit{\textbf{Worktime}} to capture the contributor's working habit (i.e., whether one would like to contribute to OSS in a work-like manner), 
\item \textit{\textbf{Number of Commits}} to capture contributor's contribution volume,  
\item \textit{\textbf{Code Contribution Density}} to capture one's contributing style (i.e., whether one would like to make small changes each time or make big changes, this also reflects the complexity of one's contribution), 
\item \textit{\textbf{Number of Programming Languages}} to capture the contributor's technical expertise in ML framework projects. Proficiency in multiple programming languages indicates expertise across various components of ML development. For instance, the implementation of tensors, ML operations, and computational graph execution usually utilizes C++, whereas high-level APIs are commonly written in Python. 
\end{itemize}

\textbf{Step 2: Normalize the selected contributor features.} Clustering algorithms often use distance metrics to group similar data points. When features have different scales, those with larger scales can dominate the clustering process. Therefore, we normalize each selected contributor feature to ensure that all features contribute equally to the clustering. We apply Min-Max normalization to scale each selected feature across all contributors to a range of 0 to 1, while preserving the relative feature relationships between contributors, as shown in Equation~\ref{eq:min_max}.

\begin{equation}
\label{eq:min_max}
\widehat{x_{i}} = \frac{x_{i} - min(x)}{max(x) - min(x)}
\end{equation}
where \( \widehat{x_{i}} \) represents the normalized value of the \(i\)th contributor for a feature, \( x_{i}\) represents the original value of the \(i\)th contributor for the feature, \(min(x)\) and \(max(x)\) represent the minimum and maximum value of the feature across all contributors, \(i \in \{1, 2, 3, \ldots, 7640\}\), and a feature $\in$ \{\textit{Worktime}, \textit{Number of Commits}, \textit{Code Contribution Density}, \textit{Number of Programming Languages}\}.

\textbf{Step 3: Select the clustering algorithm.} We examine 8 clustering algorithms in the Python scikit-learn package~\cite{sklearn}, namely, Kmeans, Affinity Propagation, Mean Shift, Spectral Clustering, Hierarchical Clustering, DBSCAN, OPTICS, and BIRCH, to cluster ML contributors with the four normalized features. The effectiveness of these clustering algorithms is evaluated using the Silhouette score, which is a metric to assess the quality of data point groupings~\cite{sklearn}. Silhouette scores range from -1 to 1, where a score of 1 indicates an optimal clustering and -1 indicates the worst. Scores close to 0 indicate overlapping clusters and negative values indicate that a data point has been incorrectly assigned to a cluster since it is more similar to a different cluster.


For the clustering algorithms requiring a predefined number of clusters (i.e., Kmeans, Spectral Clustering, Hierarchical Clustering, and BIRCH), we experiment with the number of clusters range from 1 to 10 and select the optimal number of clusters with the highest Silhouette score. For the rest of the clustering algorithms, we conduct a gradient search to determine the optimal parameter set with the highest Silhouette score.

\begin{table}[H]
\caption{Eight clustering algorithms and the associated evaluation results.}
\label{tab:cluster_eval}
  \centering
\begin{adjustbox}{max width=\textwidth}
\begin{tabular}{c|cccccccc}
\hline
\textbf{\begin{tabular}[c]{@{}c@{}}Clustering  \\ Algorithms\end{tabular}} & \textbf{Kmeans} & \textbf{\begin{tabular}[c]{@{}c@{}}Affinity \\ Propagation\end{tabular}} & \textbf{\begin{tabular}[c]{@{}c@{}}Mean \\ Shift\end{tabular}} & \textbf{\begin{tabular}[c]{@{}c@{}}Spectral \\ Clustering\end{tabular}} & \textbf{\begin{tabular}[c]{@{}c@{}}Hierarchical \\ Clustering\end{tabular}} & \textbf{DBSCAN} & \textbf{OPTICS} & \textbf{BIRCH} \\ \hline
\textbf{Number of Clusters} & 4 & 3,923 & 4 & 4 & 4 & 11 & 3 & 4 \\
\textbf{Silhouette Score} & 0.508 & 0.01 & 0.507 & \textbf{0.51} & 0.496 & 0.333 & 0.429 & 0.489 \\
\hline
\end{tabular}
\end{adjustbox}
\end{table}

\begin{figure}[H]
\centering
\begin{subfigure}{0.24\textwidth}
\includegraphics[width=\linewidth]{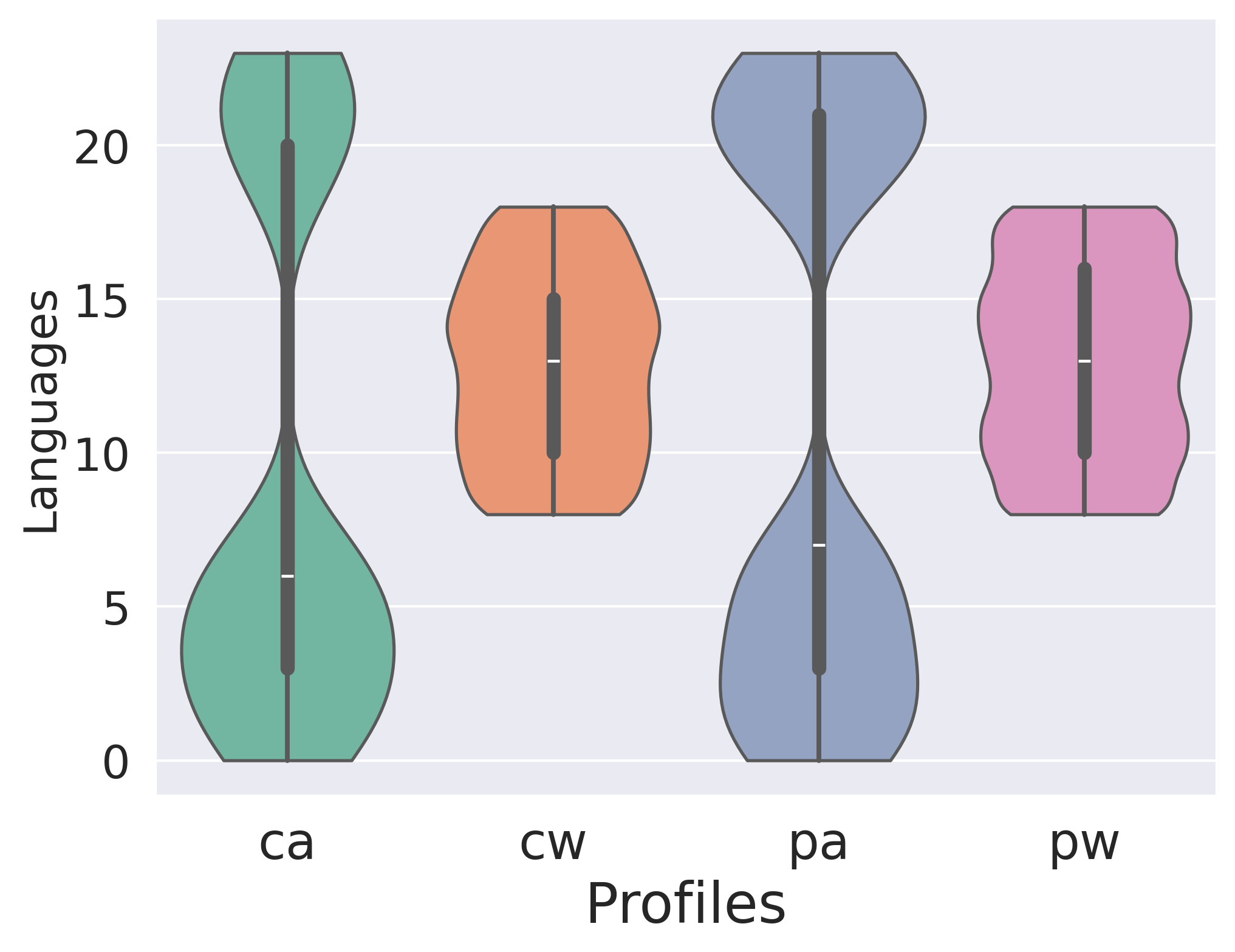}
\caption{Worktime} \label{fig:worktime}
\end{subfigure}
\hspace{\fill}
\begin{subfigure}{0.24\textwidth}
\includegraphics[width=\linewidth]{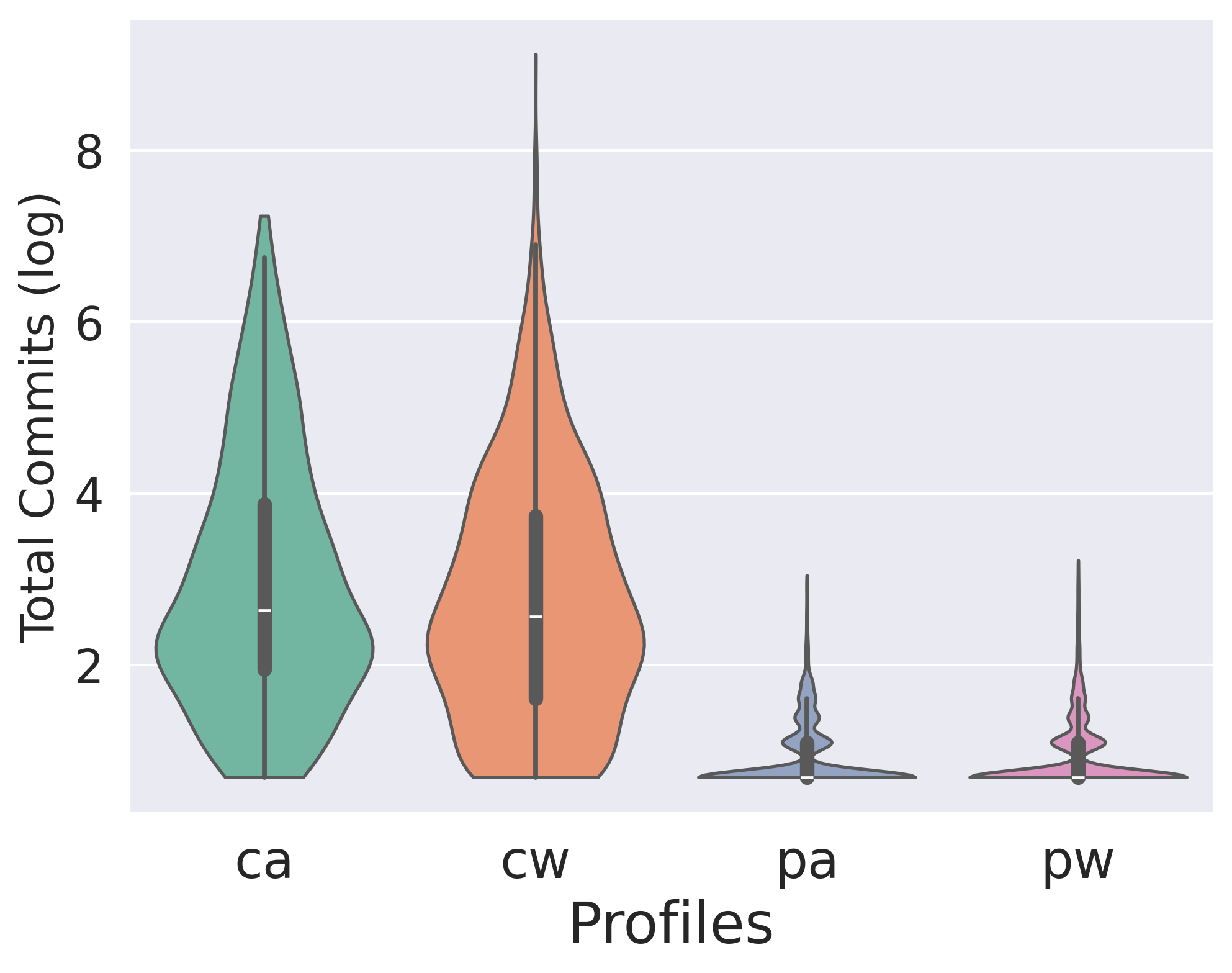}
\caption{Total Commits} \label{fig:commit}
\end{subfigure}
\hspace{\fill}
\begin{subfigure}{0.24\textwidth}
\includegraphics[width=\linewidth]{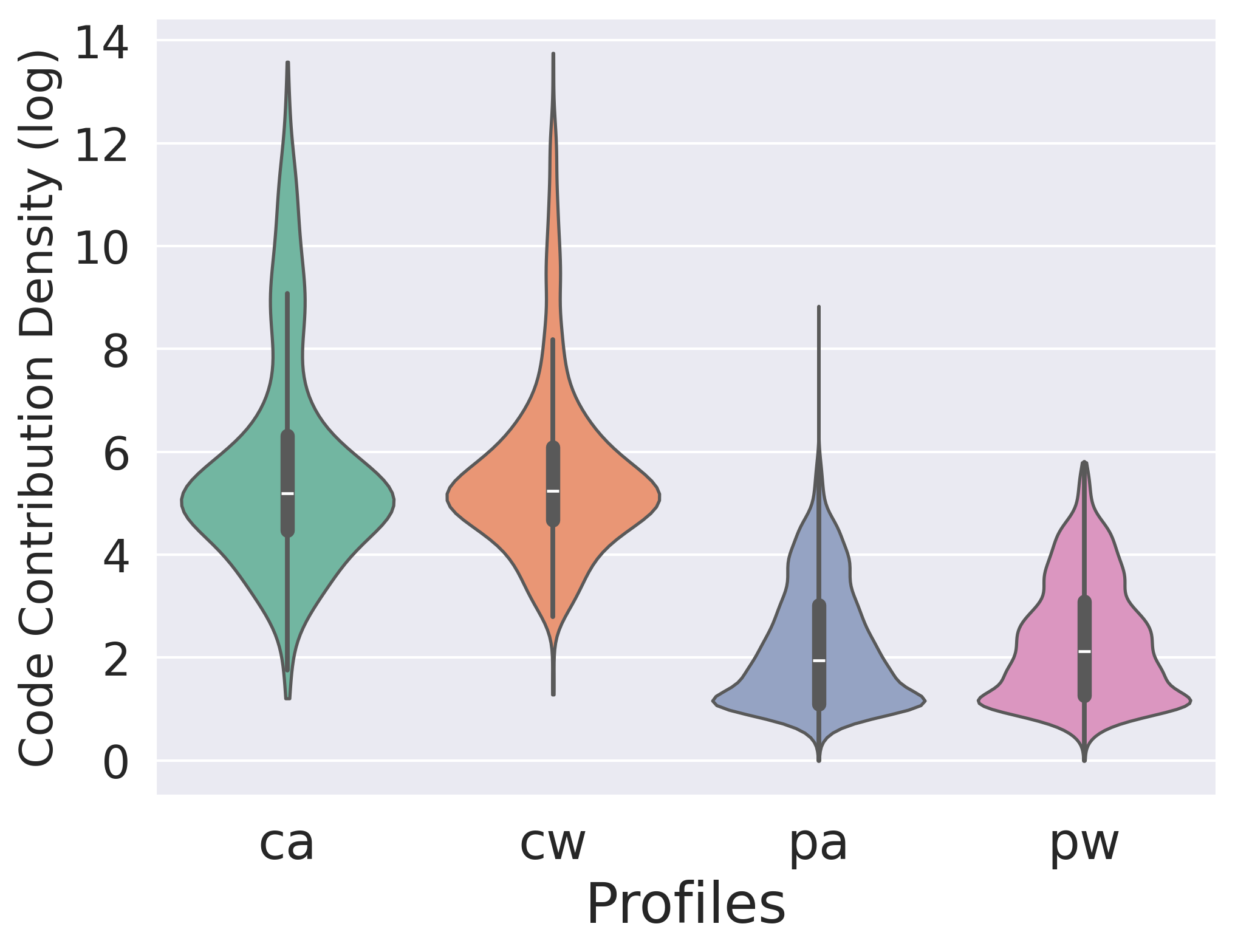}
\caption{Code Contribution Density} \label{fig:code}
\end{subfigure}
\hspace{\fill}
\begin{subfigure}{0.24\textwidth}
\includegraphics[width=\linewidth]{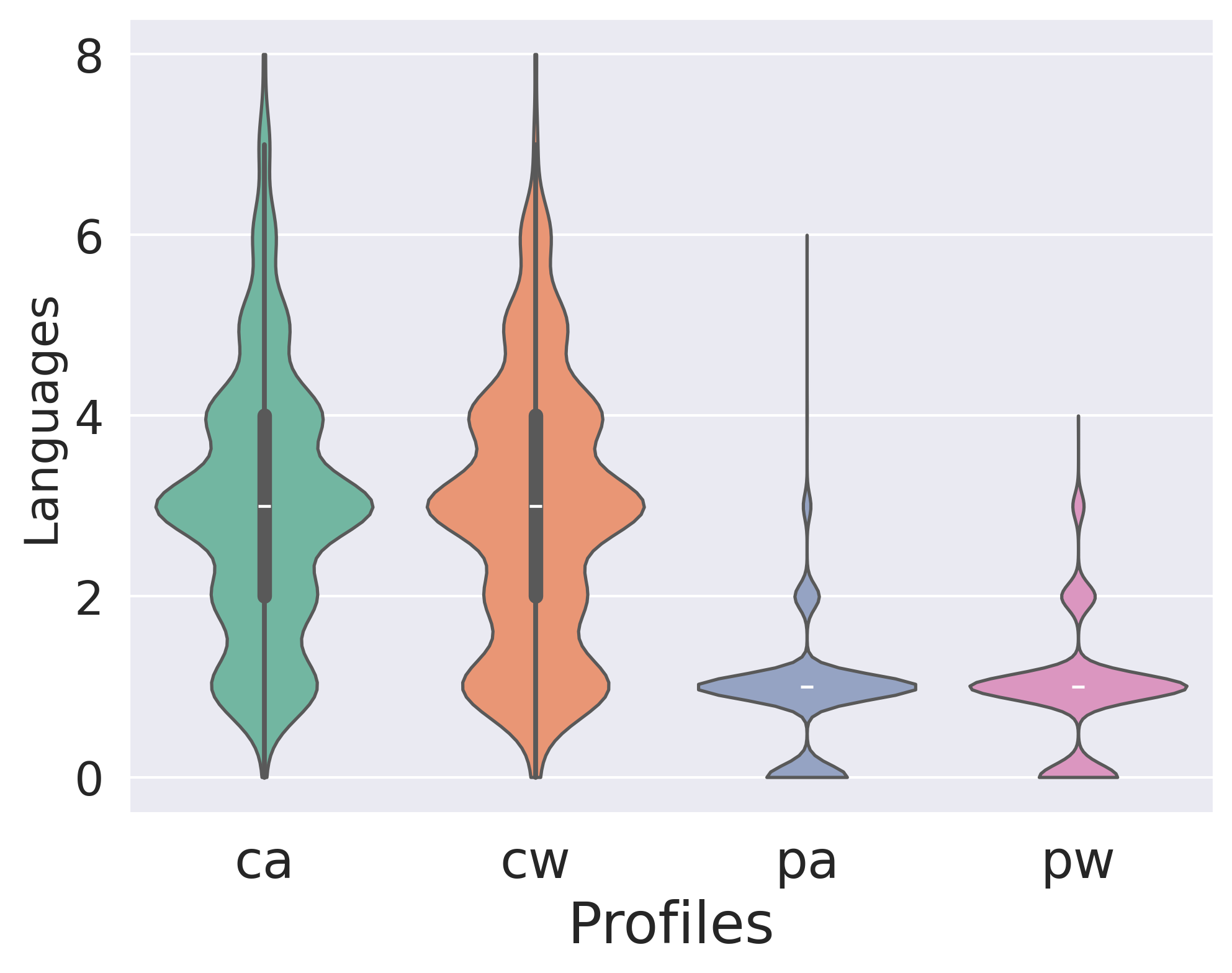}
\caption{Programming Languages} \label{fig:language}
\end{subfigure}
\hspace{\fill}
\caption{Summary of four contributor profiles ('ca' refers to Core-Afterhour, 'cw' refers to Core-Workhour, 'pw' refers to Peripheral-Workhour, and 'pa' refers to Peripheral-Afterhour).} 
\label{fig:profiles}
\end{figure}

\textbf{Step 4: Identify contributor profiles.} Based on the clustering algorithm evaluation results shown in Table~\ref{tab:cluster_eval}, the clusters produced by Spectral Clustering have the highest Silhouette Score of 0.51, indicating the optimal clustering result among all clustering algorithms. Therefore, we identify our contributor profiles from the four clusters generated by Spectral Clustering. Figure \ref{fig:profiles} shows a summary of the profiles in terms of each clustering feature and we name each profile accordingly. As shown in Figure \ref{fig:commit}, \ref{fig:code}, and \ref{fig:language}, the two clusters represented in green and orange have higher total commits, code contribution density, and number of used programming languages compared to the other two clusters, so we name them \textit{Core} and \textit{Peripheral} contributors respectively. \textit{Core} and \textit{Peripheral} contributors can be further divided into \textit{Workhour} and \textit{Afterhour} subgroups based on the time they used to make OSS contributions as shown in \ref{fig:worktime}. We summarize the four profiles in the following:

\begin{itemize}

\item \textbf{Core-Afterhour profile} contains 1,170 (15.3\%) contributors. This profile has a larger number of commits, code contribution density, and programming languages used in comparison to \textit{peripheral} contributors, with a median of 13 commits, 178 LOC per commit, and 3 programming languages. Contributors in this profile tend to make contributions outside normal working hours (i.e., outside 8 AM to 6 PM of their local time).

\item \textbf{Core-Workhour profile} contains 1,249 (16.3\%) contributors. This profile is similar to \textit{Core-Afterhour} in terms of the number of commits, code contribution density, and programming languages, with a median of 12 commits, 188 LOC, and 3 programming languages. Contributors in this profile make contributions within normal working hours.

\item \textbf{Peripheral-Afterhour profile} contains 2,456 (32.1\%) contributors, characterized by few commits, code contribution density, and programming languages, with a median of 1 commit, 6 LOC, and 1 programming language. Contributors within this profile make contributions outside normal working hours.

\item \textbf{Peripheral-Workhour profile} contains 2,765 (36.2\%) contributors. They are similar to \textit{Peripheral-Afterhour} contributors in terms of few commits, code contribution density, and programming languages, with a median of 1 commit, 7 LOC, and 1 programming language. However, contributors in this profile tend to contribute during normal working hours.

\end{itemize} 

\subsection{Contributor OSS Engagement}
\label{sec:oss_engagement}
To gain a further understanding of contributors' engagement in various OSS activities, we study their OSS engagement from three aspects: workload composition, work preferences, and technical importance. In the rest of the section, we describe our methods for identifying contributor workload composition, extracting their work preferences, and evaluating their technical importance.

\subsubsection{Identifying Workload Composition Patterns}
\label{sec:identify_workload_composition_pattern}
\hfill

\noindent  
Workload composition reflects how contributors allocate their effort across key OSS routines, including committing, raising issue reports, participating in issue discussions, participating in pull request discussions, and code reviews within a specific period. We identify workload composition patterns to capture the common workload compositions and study the distribution of contributors with similar workload compositions as well as their impact on the subject project. Our approach to identifying contributor workload composition patterns involves the following four steps:

\textbf{Step 1: Determine the length of the period to analyze contributors' workload composition.} A contributor can be assigned to different tasks at different time periods. Therefore, contributors' workload composition is often dynamic and can evolve over time. We aim to capture their temporary workload compositions that are the focus of a contributor within a time interval, assuming the workload composition remains stable over brief periods. 

We conduct a sensitivity analysis on 30, 60, 90, 120, and 180 days to identify the optimal period length and verify the stability of a contributor's workload composition. We observe that shorter periods, such as 30 and 60 days, lack precision in representing contributors' workload compositions, due to the potential of little activities within short periods. For example, suppose a contributor only makes one commit in 30 days but engages in pull request discussions in the following month (i.e., 30 days), we may observe an exclusive focus on committing for the first period and an exclusive focus on pull request discussions for the subsequent period. Conversely, long time intervals, such as 120 and 180 days, yield too few data points for our analysis. Particularly, in addressing RQ3, where we model the relationship between contributor workload compositions and project popularity (e.g., star rating and number of forks) in each period, the model cannot converge due to insufficient data points (i.e., periods). Therefore, for each subject project, we segment project activities into 90-day intervals from the beginning of the project's lifetime to the date of our data collection and capture the workload composition of active contributors in each period. In the rest of this paper, the term '\textit{period}' refers to these 90-day project intervals.

\textbf{Step 2: Construct contributor workload composition vector space.} We create a 5-dimensional vector space to represent the contributors' workload composition within a period. The five dimensions include the count of commits, raised issue reports, issue comments, pull request comments, and code reviews, which represent the key components of OSS project routines. For each period of a project, we create a workload composition vector for each active contributor. Recognizing that the absolute counts of OSS activities may not accurately reflect the actual focus of contributions (e.g., conducting 2 code reviews might be a considerable contribution while making 2 issue comments might be considered a small contribution), we normalize each dimension of the workload composition vector across contributors. The normalization ensures that each dimension in the vector represents the relative contribution to an OSS activity among contributors and the vector can reflect the actual workload allocations of contributors. Furthermore, the relative contribution 
can vary between projects and even within the same project over different periods. Hence, for each period in each subject project, we apply Min-Max normalization to normalize each dimension of the workload composition vectors across all active contributors. The normalized vector space ensures safe comparisons between vector dimensions and vectors from different projects or periods. As a result, we build 26,335 workload composition vectors from 7,640 contributors across a total of 163 periods from our subject projects. 

\textbf{Step 3: Identify workload composition patterns.} We employ agglomerative hierarchical clustering~\cite{sklearn} to identify the workload composition patterns. Agglomerative hierarchical clustering begins by treating each vector as an individual cluster and successively merges smaller clusters with the highest similarity until all data points belong to a single cluster or a predefined dissimilarity threshold is reached. This iterative process naturally forms cohesive clusters and can produce a dendrogram, which is a hierarchical tree-like structure of datapoints and can be cut at different heights (i.e., dissimilarity threshold) to obtain different numbers of clusters. Compared to other clustering algorithms discussed in Section~\ref{sec:identify_profile}, 
we select hierarchical clustering to identify workload composition patterns because the dendrogram enables us to visualize and explore the similarity of workload compositions at different levels of granularity, thereby determining the
the level of similarity of the resulting patterns.

We use cosine similarity as the similarity measurement for workload composition vectors, as it gauges the similarity of vector dimensions irrespective of the vector scale. This aligns with our goal of identifying contributors with a similar focus across different types of OSS contributions, regardless of their amount of contribution.

We first compute the pairwise cosine similarity of every two workload composition vectors to build a distance matrix. Then, we apply agglomerative hierarchical clustering on the distance matrix to group similar vectors and plot the dendrogram of the clustering. We manually inspect the clusters generated from different cut-off dissimilarity thresholds and calculate the Silhouette scores for the resulting clusters to determine the optimal number of clusters and identify workload composition patterns accordingly. 

\begin{figure}
\centering
\begin{subfigure}{0.19\textwidth}
\includegraphics[width=.95\linewidth]{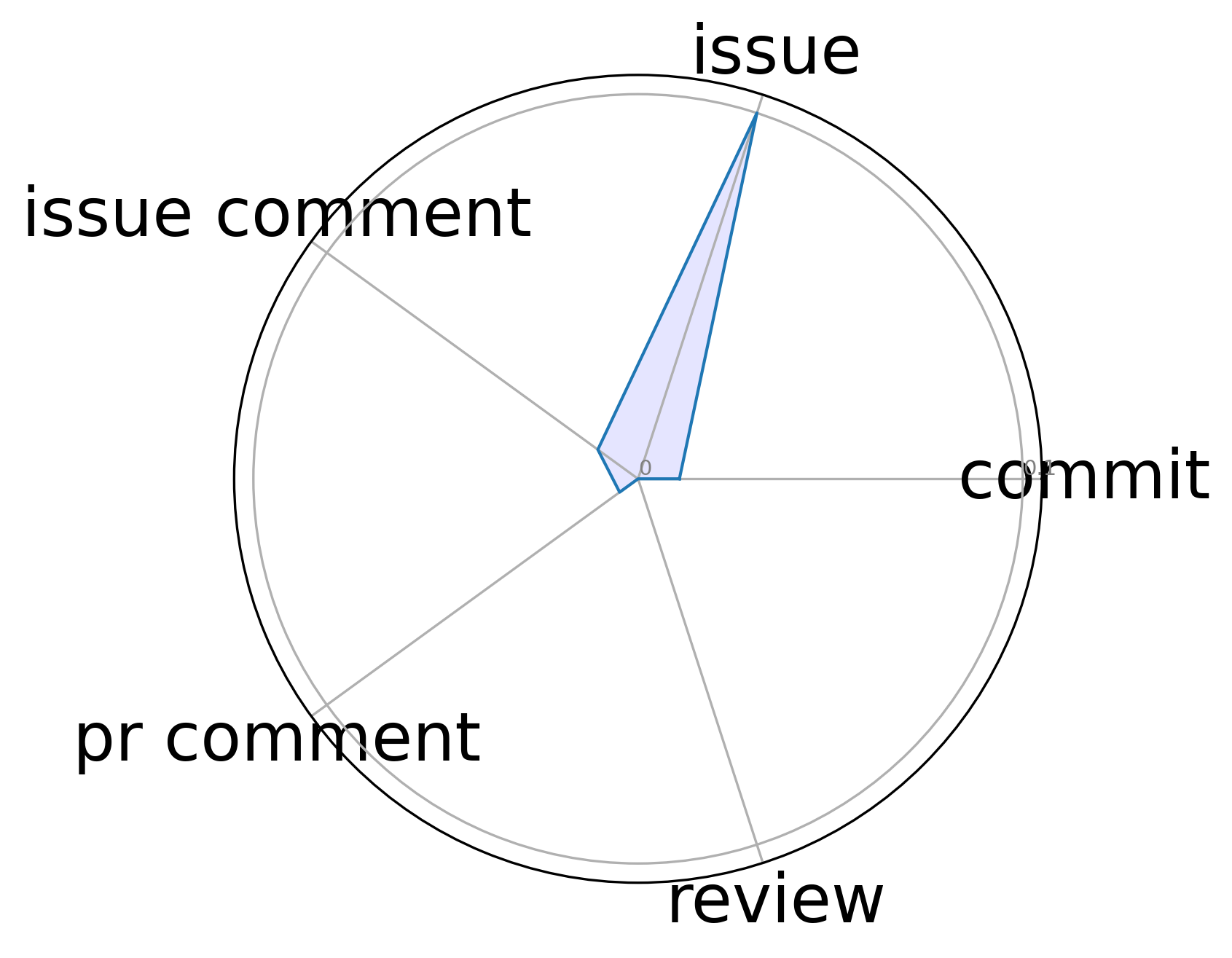}
\caption{Issue Reporter \protect\\ ~} \label{fig:rq2_pattern1}
\end{subfigure}
\hspace{\fill}
\begin{subfigure}{0.19\textwidth}
\includegraphics[width=.95\linewidth]{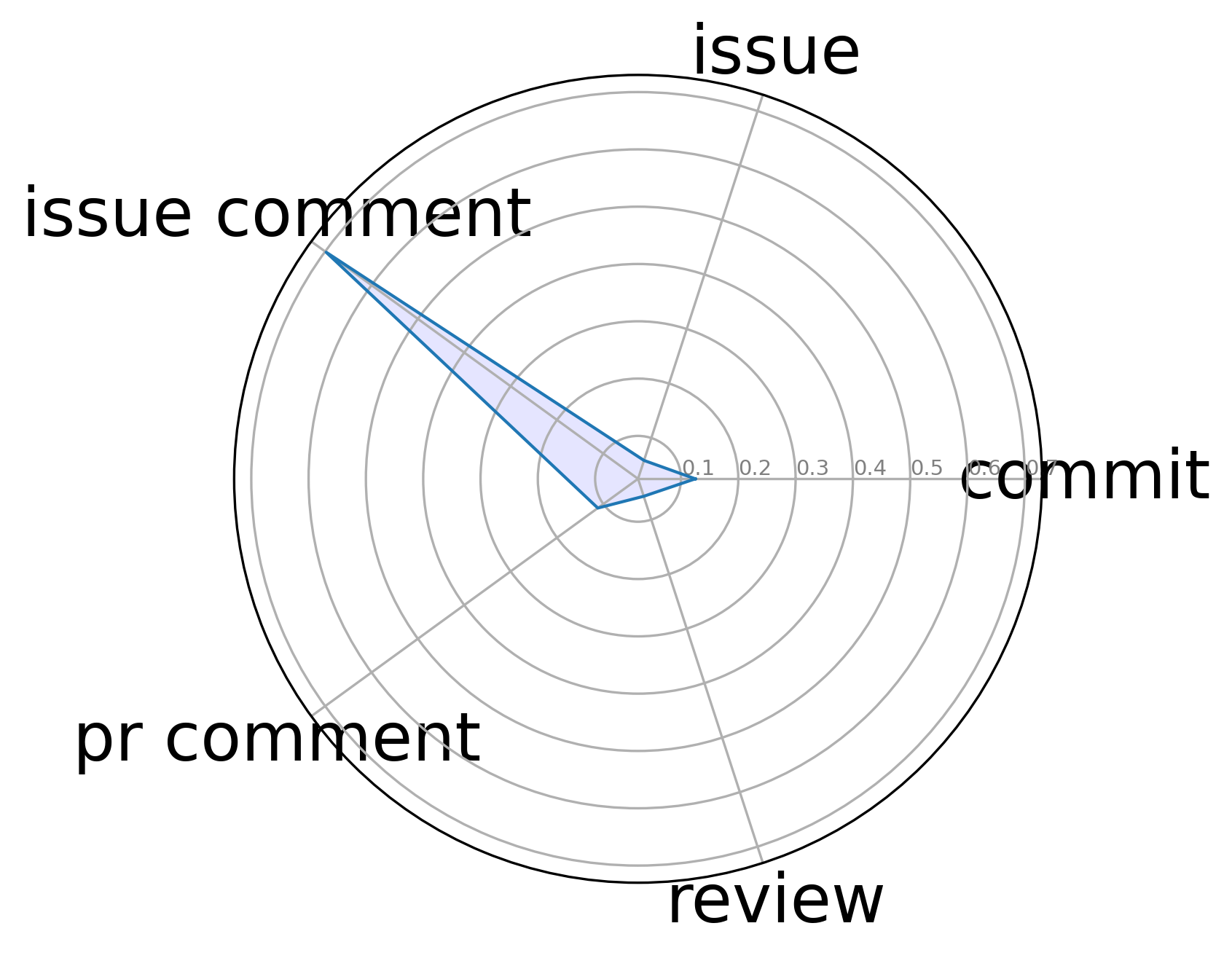}
\caption{Issue Discussant \protect\\ ~} \label{fig:rq2_pattern2}
\end{subfigure}
\hspace{\fill}
\begin{subfigure}{0.19\textwidth}
\includegraphics[width=.95\linewidth]{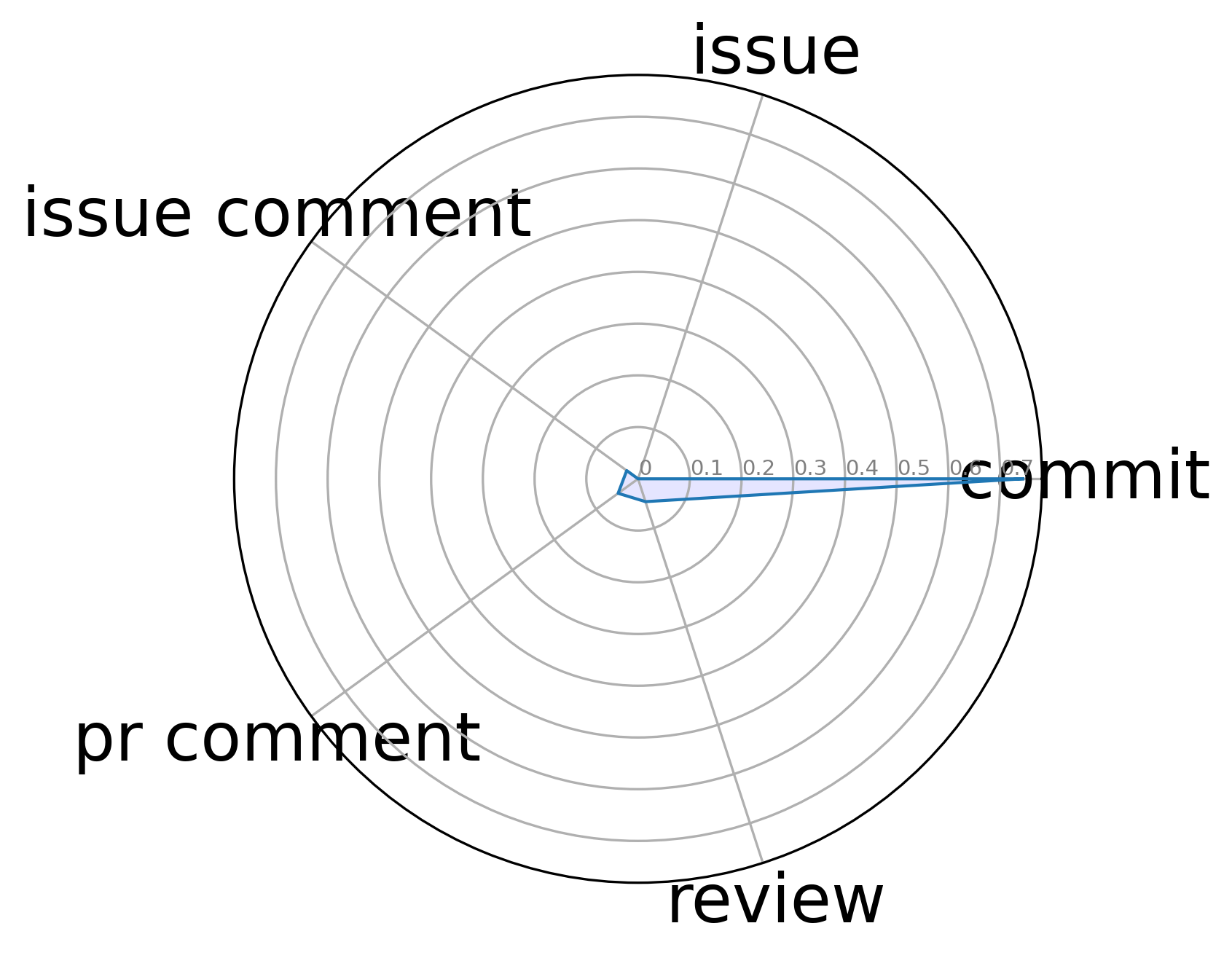}
\caption{Committer \protect\\ ~} \label{fig:rq2_pattern3}
\end{subfigure}
\hspace{\fill}
\begin{subfigure}{0.19\textwidth}
\includegraphics[width=.95\linewidth]{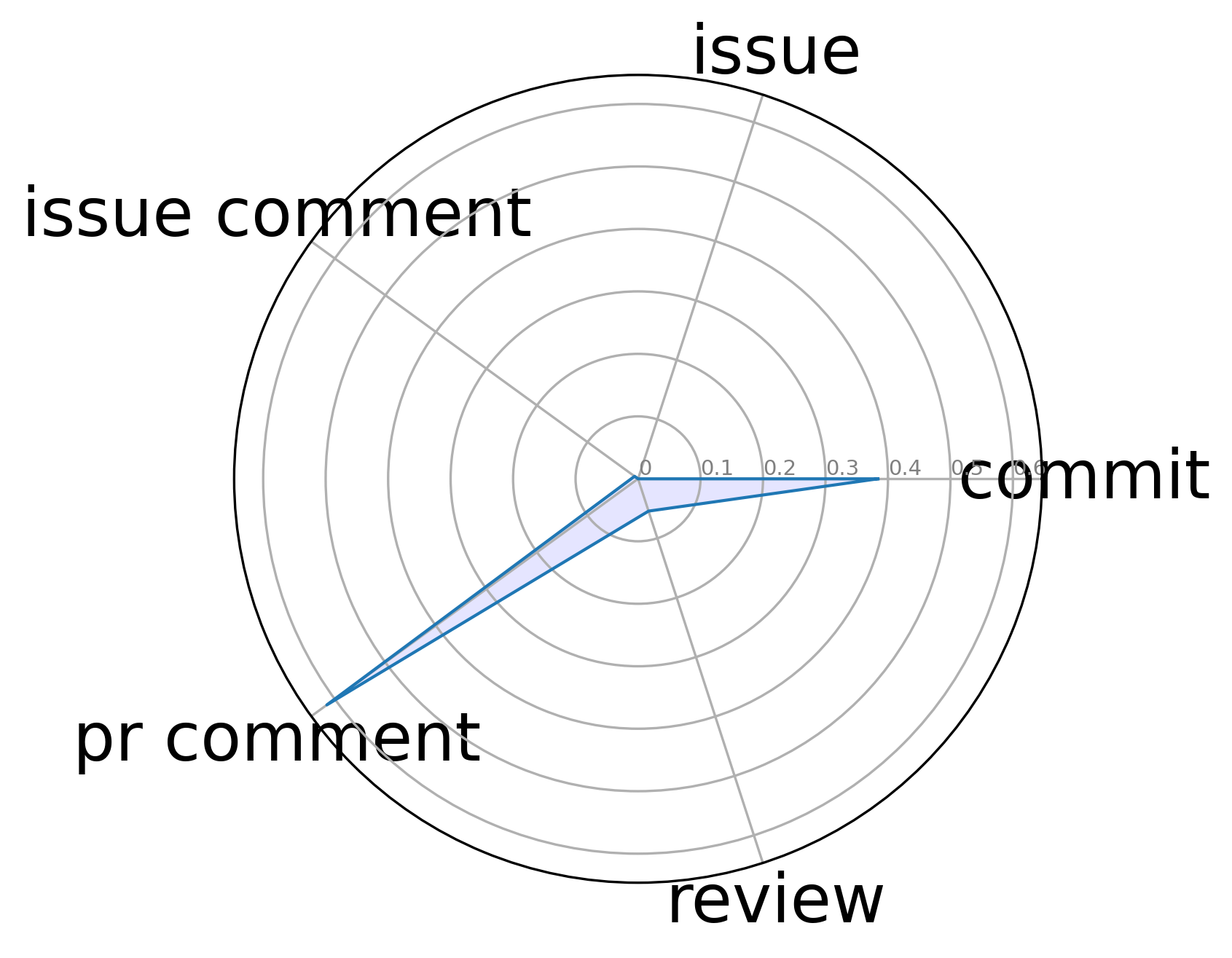}
\caption{Collaborative \protect\\ \hspace*{1.6em}Committer} \label{fig:rq2_pattern4}
\end{subfigure}
\hspace{\fill}
\begin{subfigure}{0.19\textwidth}
\includegraphics[width=.95\linewidth]{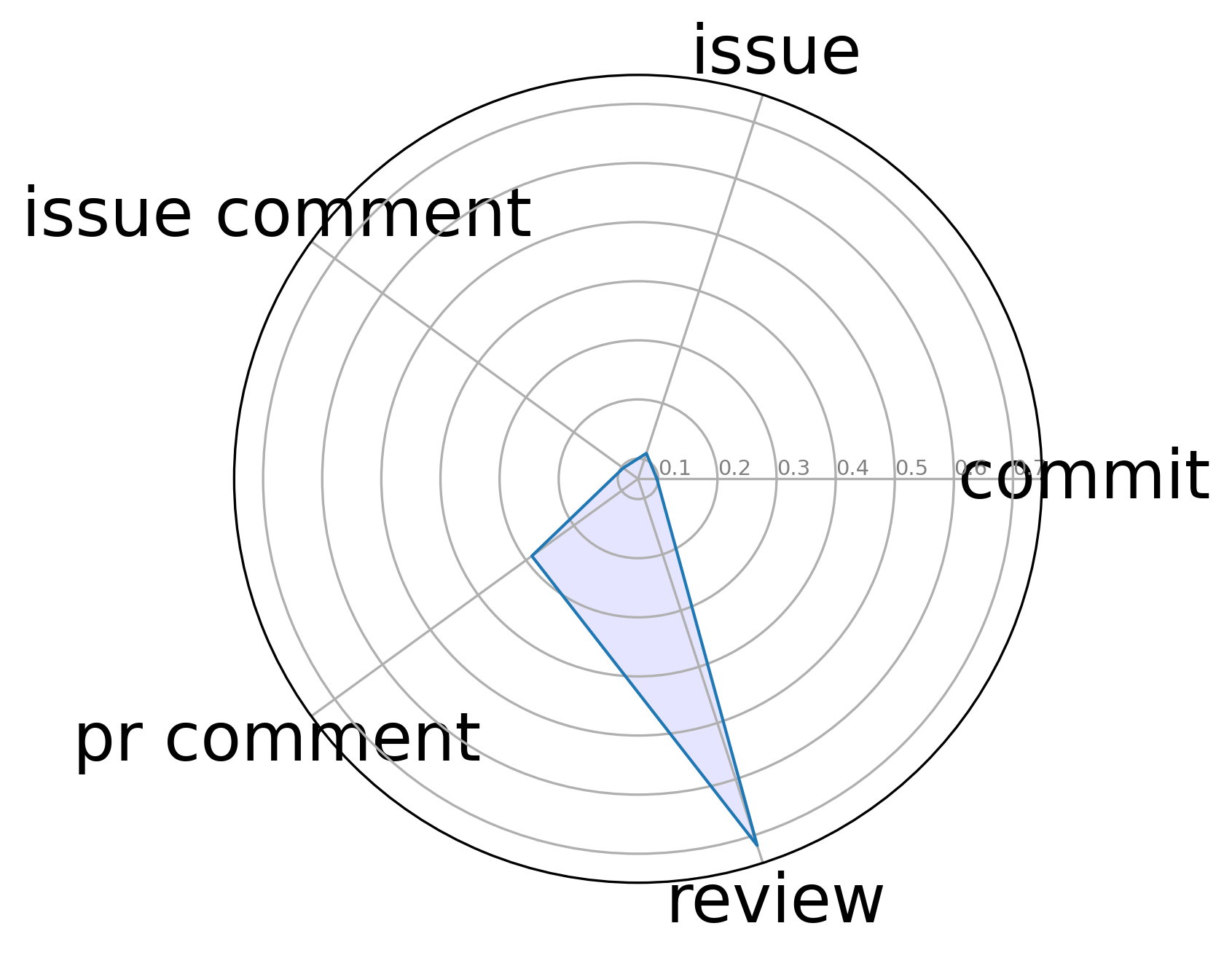}
\caption{Code Reviewer \protect\\ ~} \label{fig:rq2_pattern5}
\end{subfigure}
\hspace{\fill}
\caption{Summary of workload composition patterns.}  
\label{fig:workload_patterns}
\end{figure}

\textbf{Step 4: Summarize workload composition patterns.} We find five clusters as the optimal number of clusters at a cut-off threshold of 0.9, which also achieves the highest Silhouette score of 0.474 compared to other numbers of clusters. From these clusters, we identify five workload composition patterns. Figure~\ref{fig:workload_patterns} shows the centroid workload composition vector in each cluster, which is the point with the highest cosine similarity with other points in the cluster. We summarize the work focus of each cluster according to the centroid behavior in the following: 

\begin{itemize}

\item \textbf{Pattern 1: Issue Reporter} accounts for 19.8\% of occurrences. Contributors in this pattern focus on raising issue reports and participate minimally in other activities.

\item \textbf{Pattern 2: Issue Discussant} is the second most common pattern comprising 22.5\% of occurrences. They focus on discussing issue reports while infrequently engaging in other types of contributions.

\item \textbf{Pattern 3: Committer} is the most common pattern and comprise 33.3\% of occurrences. They only focus on making commits and solely participate in the other four activities.

\item \textbf{Pattern 4: Collaborative Committers} comprise 19.8\% of occurrences. They focus on both making commits and participating in pull request discussions.

\item \textbf{Pattern 5: Code Reviewer} is the least common pattern comprising 4.6\% of occurrences. Contributors following this pattern focus on code reviews and pull request discussions.

\end{itemize}

\subsubsection{Extracting Work Preference Features}
\label{sec:identify_work_preference}
\hfill

\noindent 
While workload composition provides a static view of contributors' focus of OSS activities within a period of time, work preference features capture their contribution dynamics and preferences in that period, such as whether they make consistent contributions (e.g., making one commit daily) versus peak contributions (e.g., making a high number of commits on a single day and remaining inactive afterward), as well as their preference on making specific types of contributions versus an even contribution across various types. 
\begin{figure}
\centering
\begin{subfigure}{0.49\textwidth}
\includegraphics[width=.95\linewidth]{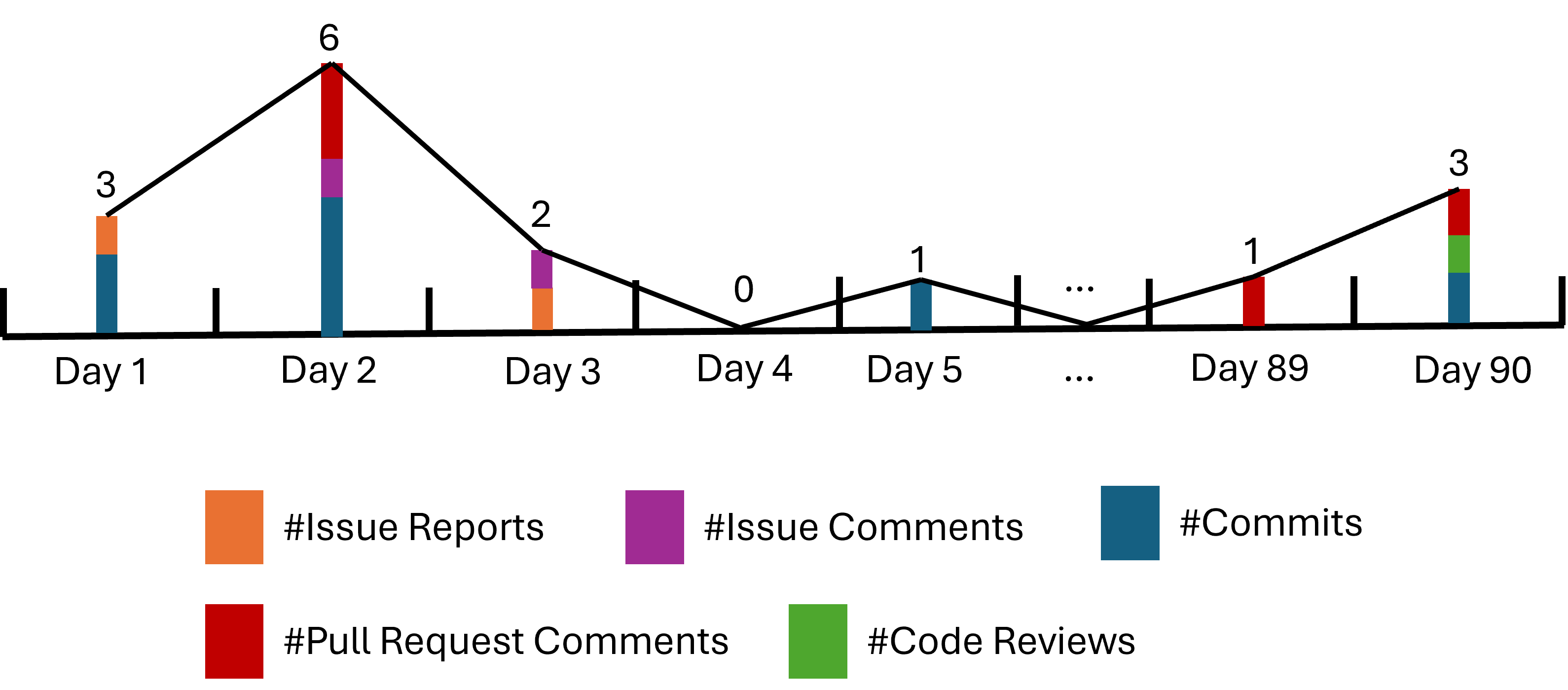}
\caption{Constructing the  OSS activity time ser
ies.}
\label{fig:construct_ts}
\end{subfigure}
\hspace{\fill}
\begin{subfigure}{0.47\textwidth}
\includegraphics[width=.95\linewidth]{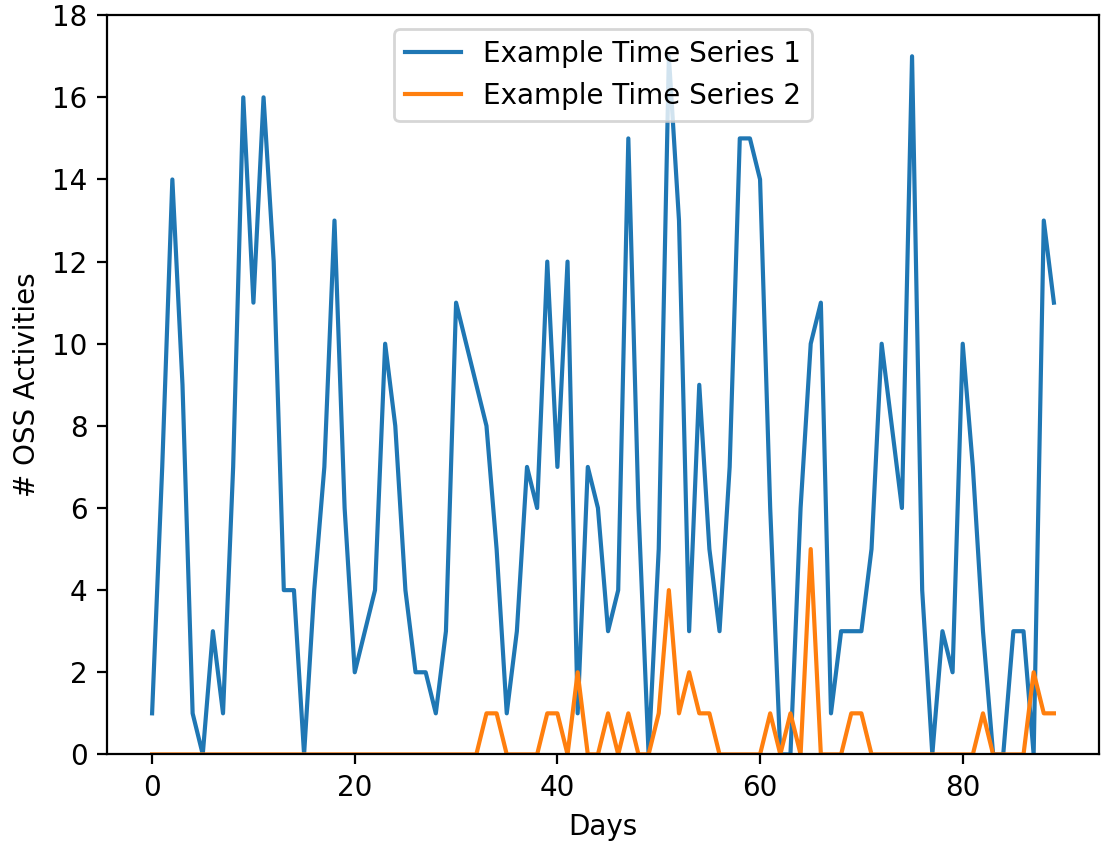}
\caption{Example of a complex and a less fluctuated time series.}
\label{fig:example_ts}
\end{subfigure}

\caption{An example of contributors' OSS activity time series within a 90-day period.}
\label{fig:ts}
\end{figure}

As shown in Figure~\ref{fig:construct_ts}, we count a contributor's daily activity in terms of the five OSS activity dimensions (i.e., the number of issue reports, issue comments, commits, pull request comments, and code reviews) each day and create a time series for each contributor over a 90-day period (as discussed in ~\ref{sec:identify_workload_composition_pattern}). We use the Python \textit{tsfresh} package~\cite{christ2016distributed} to extract time series features including \textit{binned\_entropy}, \textit{c3}, \textit{number\_cwt\_peaks}, \textit{longest\_strike\_above\_mean}, and \textit{longest\_strike\_below\_mean}. These time series features measure a contributor's contribution dynamics. Figure~\ref{fig:example_ts} presents examples of contributors' OSS activities time series with different contribution dynamics, where example time series 1 exhibits complex dynamics and example time series 2 exhibits comparatively fewer fluctuations. We also extract two features, \textit{diverse} and \textit{balance}, to measure the breadth of OSS contribution types made by a contributor and the degree to which they distribute their efforts evenly across various types within each period. We consider these 9 extracted features as contributor work preference features. Detailed descriptions and implications of the work preference features are outlined in Table~\ref{tab:work_preference_features}.

\begin{table}[H]
\caption{Extracted features of contributor work preferences.
}
\label{tab:work_preference_features}
\centering
\begin{adjustbox}{max width=\textwidth}
\begin{tabular}{lll}
\cline{1-2}
\begin{tabular}[c]{@{}l@{}}\textbf{Contributor Work} \\ \textbf{Preference Features}\end{tabular} & \textbf{Description} &  \\ \cline{1-2}
binned\_entropy(10) & \begin{tabular}[c]{@{}l@{}}Binned entropy measures the complexity of the sequence. A higher binned entropy indicates developers \\ have more complex contribution dynamics and make less constant contributions. \end{tabular}  &  \\ \cline{1-2}
c3(1), c3(2), c3(3) & \begin{tabular}[c]{@{}l@{}} C3 statistics measures the degree of nonlinearity or complexity in a time series by calculating the \\ correlation between the original time series and its lag of 1, 2, 3 respectively. Higher C3 statistics \\ indicate developers have more complex contribution dynamics and make less constant contributions. \end{tabular}  &  \\ \cline{1-2}
number\_cwt\_peaks & Number of peaks of contributions. A higher value indicates developers make more peak contributions. &  \\ \cline{1-2}
longest\_strike\_above\_mean & \begin{tabular}[c]{@{}l@{}}The ratio of the longest subsequences above mean. A higher value indicates more consecutive days of \\ high contribution. \end{tabular}&  \\ \cline{1-2}
longest\_strike\_below\_mean & \begin{tabular}[c]{@{}l@{}} The ratio of the longest subsequences below mean. A higher value indicates more consecutive days of \\ low contribution. \end{tabular}&  \\ \cline{1-2}
diverse & The number of different types of contributions a developer has made. &  \\ \cline{1-2}
balance & \begin{tabular}[c]{@{}l@{}}The inverse of the variance of a developer's normalized contribution on the five OSS activities. A higher \\ value indicates developers make a more even amount of different types of contributions.\end{tabular} \\ \cline{1-2} 


\end{tabular}
\end{adjustbox}
\end{table}

\subsubsection{Evaluating Contributor Technical Importance}
\label{sec:evaluate_importance}
\hfill

\noindent 
The technical importance of a contributor can be measured using the importance of the commits made by the contributor. The importance of a commit can be measured with the eigenvector centrality of source code files modified by the commit. Eigenvector centrality measures the importance of a node within a network. In our case, the software repository can be viewed as a network, with the folders and source code files as nodes, and their containment relationships as edges. Therefore, the eigenvector centrality of a source code file measures the importance of the file within the file network of a repository. Contributors who frequently make changes to important files can be considered as possessing higher technical importance to the project.

\begin{figure}
    \centering
    \includegraphics[width=0.8\linewidth]{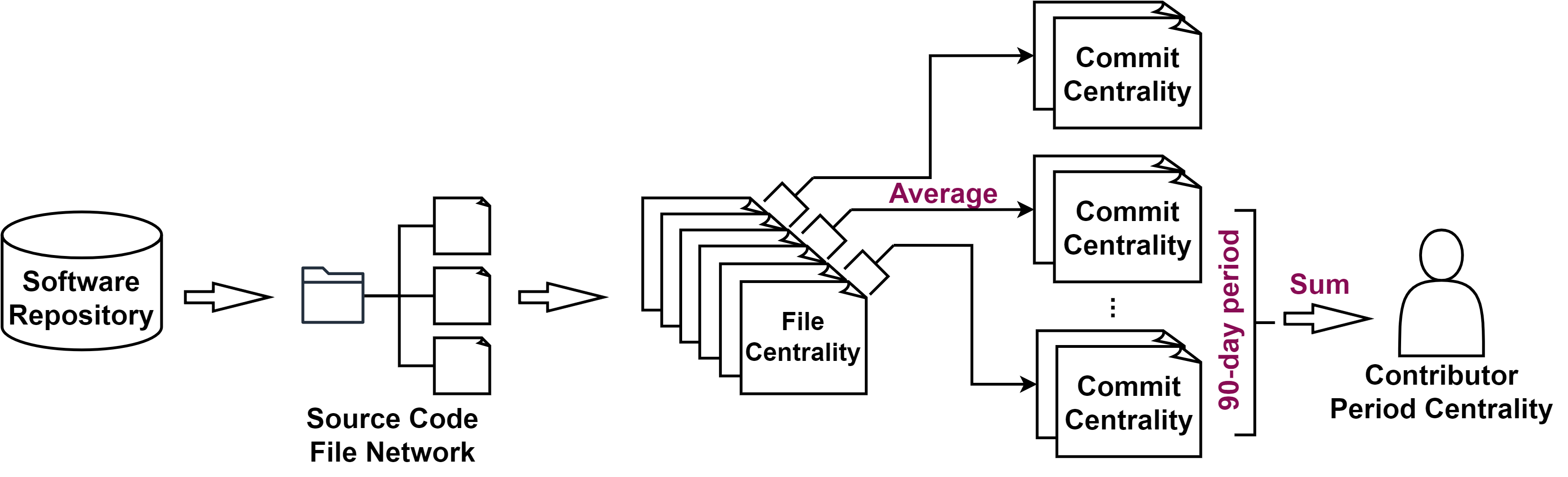}
    \caption{Measure contributor technical importance. 
    }
    \label{fig:centrality_approach}
\end{figure}

Figure~\ref{fig:centrality_approach} shows our approach for measuring contributor technical importance. Following a similar methodology as proposed by Yue et al.~\cite{yue2022off}, we first construct a network with folders and source files as nodes and the relationship of a folder containing a file as edges, then calculate the eigenvector centrality of each source file, as \textit{file centrality}. The \textit{commit centrality} is calculated by averaging the eigenvector centrality of the source files that have been changed in the commit. Higher \textit{commit centrality} means that the commit makes changes to more influential files in the repository and can have a greater impact on the codebase. We sum the \textit{commit centrality} of all commits made by a contributor within a period to calculate the \textit{period centrality}, which measures the technical importance of a contributor within a period, considering both the intensity and importance of their commits. We introduce four metrics to quantify the technical importance of a contributor:
\begin{itemize}
    \item \textbf{\textit{max\_commit\_centrality}}: The highest commit centrality among all commits made by a contributor.
    \item \textbf{\textit{max\_centrality\_day}}: The number of days taken by a contributor to make the commit with the highest centrality.
    \item \textbf{\textit{max\_period\_centrality}}: The highest period centrality among all active periods of a contributor.
    \item \textbf{\textit{max\_centrality\_period}}: The number of periods taken by a contributor to achieve the highest period centrality.

\end{itemize} 

\section{Experiment Results}
\label{sec:results}
\subsection{RQ1: What are the characteristics of each contributor profile?}

\subsubsection{\textbf{Motivation}}
\hfill

\noindent
We identify 4 contributor profiles based on multiple factors including work time, the number of commits, code contribution density, and the number of used programming languages. In this research question, we further delve into the important features of each profile and explore the differences between the profiles as well as their shared attributes. Understanding contributor profiles can equip project managers and maintainers with the knowledge to engage the contributors of different profiles in the project effectively and support tactics to accommodate the varied spectrum of contributors. 

\subsubsection{\textbf{Approach}}
\hfill

\noindent
To gain a further understanding of contributor profiles, we build four different binary logistic regression models to identify the significant contributor features associated with each of the four contributor profiles. Our detailed approach includes the following three aspects:

\textbf{Model Construction:} To build the model for each contributor profile, we exclude the four features used for identifying the profiles (i.e., Worktime, Number of Commits, Code Contribution Density, and Number of Programming Languages) and use the remaining 18 non-correlated and non-redundant contributor features (as described in Section~\ref{cor-redun}) as independent variables. We establish a binary dependent variable by labeling contributors of the target profile (e.g., \textit{Core-Afterhour}) as '1' and contributors of the remaining three profiles (i.e., \textit{Core-Workhour}, \textit{Peripheral-AFterhour}, and \textit{Core-Workhour}) as '0'. Therefore, we create four models. Binary dependent variable allows us to explore the differences between the target profile and the rest, as well as to identify the important features defining the target profile. To avoid bias stemming from the imbalanced labels, we construct a balanced dataset by using an equal number of data points for both '1' and '0' labels. We include all data points labeled '1' and a corresponding number of randomly selected data points labeled '0'. When constructing the logistic regression model, we build the base model using \textit{glm} function in R~\cite{Friedman2010-rt}, and subsequently employ the \textit{step} function to build step models, which iteratively eliminate the features that are not significant to the model. 

\textbf{Model Performance Evaluation:} We evaluate our model's effectiveness using the Area Under the Curve (AUC) metric. The AUC, representing the space beneath the Receiver Operating Characteristic (ROC) curve, quantifies the capability of a binary classifier to differentiate between classes. AUC values range from 0 to 1, with higher values signifying better performance. Specifically, an AUC of 0.5 indicates a random classifier, while an AUC of 1 denotes a perfect classifier~\cite{1388242}.

\textbf{Model Result Analysis:} To identify the significant features of each model and its corresponding profile, we employ Wald statistics ($\tilde{\chi}^2$) to estimate the feature importance and statistical significance (p-values) of each contributor feature, as Wald statistics are often used in regression analysis to assess the significance of individual coefficients or parameters in a regression model~\cite{bewick2005statistics}. A higher $\tilde{\chi}^2$ value indicates a larger impact of the particular feature on the discriminatory capability of the model~\cite{moore1977generalized,toda1995statistical}. We utilize the \textit{ANOVA} function in R to compute $\tilde{\chi}^2$ values and the corresponding p-values. We further calculate the percentage of $\tilde{\chi}^2$ of each feature relative to the sum of $\tilde{\chi}^2$ values to show the proportion to the overall importance for each feature~\cite{hassan2018studying}. The significant features for each profile are marked with asterisks in Table~\ref{tab:rq1_glmresult}, with more asterisks indicating higher statistical significance levels.

We also analyze the coefficients of each feature in the logistic regression models. Positive coefficients imply a direct correlation with the dependent variable, suggesting that as the feature increases, the chances of a contributor being classified into the profile labeled as '1' also tend to increase. Conversely, negative coefficients imply an inverse relationship. The upward ($\nearrow$) and downward ($\searrow$) arrows in Table~\ref{tab:rq1_glmresult} denote the positive and negative correlations.

To validate the findings from the model results, we apply two-sided Mann-Whitney U test~\cite{nachar2008mann} to compare contributor features across contributors in \textit{core} and \textit{peripheral} profiles with the null hypothesis \textit{\(H0_{1}\)}, as well as compare their \textit{workhour} and \textit{afterhour} subgroups with the null hypothesis \textit{\(H0_{2}\)} and \textit{\(H0_{3}\)} respectively. We apply Cliff's delta~\cite{cliff1993dominance} to estimate the effect sizes.

\begin{itemize}
    \item \textit{\(H0_{1}\): The distributions of the given contributor feature are the same across core contributors and peripheral contributors.}
    \item \textit{\(H0_{2}\): The distributions of the given contributor feature are the same across Core-Afterhour and Core-Workhour contributors.}
    \item \textit{\(H0_{3}\): The distributions of the given  contributor feature are the same across Peripheral-Afterhour and Peripheral-Workhour contributors.}
\end{itemize}


\subsubsection{\textbf{Results}}
\hfill

\begin{table}[H]
\caption{AUC of 4 binary logistic regression models.}
\label{tab:glmAUC}
\centering
\begin{adjustbox}{max width=0.5\textwidth}
\begin{tabular}{c|cccc}
\hline
\textbf{Models} & \textbf{\begin{tabular}[c]{@{}c@{}}Model1\\ Core-\\Afterhour\end{tabular}} & \textbf{\begin{tabular}[c]{@{}c@{}}Model2\\ Core-\\Workhour\end{tabular}} & \textbf{\begin{tabular}[c]{@{}c@{}}Model3\\ Peripheral-\\Afterhour\end{tabular}} & \textbf{\begin{tabular}[c]{@{}c@{}}Model4\\ Peripheral-\\Workhour\end{tabular}} \\ \hline
\textbf{AUC} & 0.73 & 0.72 & 0.66 & 0.66 \\ \hline
\end{tabular}
\end{adjustbox}
\end{table}

\begin{table}[H]
\caption{Result of 4 binary logistic regression models.}
\label{tab:rq1_glmresult}
\centering
\begin{adjustbox}{max width=\textwidth}
\begin{tabular}{l|ccc|ccc|ccc|ccc}
\hline
\multirow{2}{*}{\textbf{Contributor Features}} & \multicolumn{3}{c|}{\textbf{Core-Afterhour}} & \multicolumn{3}{c|}{\textbf{Core-Workhour}} & \multicolumn{3}{c|}{\textbf{Peripheral-Afterhour}} & \multicolumn{3}{c}{\textbf{Peripheral-Workhour}} \\ 
 & Chisq\% & Signf. & Rel. & Chisq\% & Signf. & Rel. & Chisq\% & Signf. & Rel. & Chisq\% & Signf. & Rel. \\ \hline
\textbf{Duration} & \textbf{10.33} & *** & $\nearrow$ & 5.15 & *** & $\nearrow$ & \textbf{14.38} & *** & $\searrow$ & 6.75 & *** & $\searrow$ \\
\textbf{Timezone} & \textbf{45.64} & *** & $\nearrow$ & \textbf{24.73} & *** & $\searrow$ & \textbf{34.69} & *** & $\nearrow$ & \textbf{62.2}& *** & $\searrow$ \\
\textbf{Authored files} & 9 & *** & $\nearrow$ & 3.58 & *** & $\nearrow$ & \textbf{13.18} & *** & $\searrow$ & 8.86 & *** & $\searrow$ \\
\textbf{Commit Rate} & 0.61 & . & $\searrow$ & 0.62 &  & $\searrow$ &  &  &  &  &  &  \\
\textbf{Code Commit Rate} & 2.39 & ** & $\nearrow$ & 1.97 & ** & $\nearrow$ & 1.86 & ** & $\searrow$ & 0.42 &  & $\searrow$ \\
\textbf{Other Commits} &  &  &  & 3.84 & *** & $\nearrow$ &  &  &  &  &  &  \\
\textbf{Code Contribution Rate} & 1.69 & ** & $\nearrow$ & 0.89 & . & $\nearrow$ & 1.56 & * & $\searrow$ &  &  &  \\
\textbf{Total Issues} &  &  &  & 2.28 & ** & $\searrow$ & 3.93 & *** & $\nearrow$ & 1.35 & ** & $\nearrow$ \\
\textbf{Issue Contribution Rate} & 0.69 & . & $\searrow$ & 1.29 & * & $\searrow$ &  &  &  &  &  &  \\
\textbf{Issue Solved} & 1.96 & ** & $\searrow$ & 2.21 & ** & $\searrow$ & 0.69 & . & $\nearrow$ & 0.95 & * & $\nearrow$ \\
\textbf{Issue participated} &  &  &  & 1.55 & * & $\searrow$ &  &  &  &  &  &  \\
\textbf{PR Merged} &  &  &  &  &  &  & 0.74 & . & $\searrow$ & 0.66 & . & $\searrow$ \\
\textbf{PR Reviewed} & 1.28 & * & $\searrow$ & 0.77 & . & $\searrow$ & 0.98 & * & $\nearrow$ & 0.9 & * & $\nearrow$ \\
\textbf{PR Contribution Rate} &  &  &  &  &  &  & 6.71 & *** & $\nearrow$ &  &  &  \\
\textbf{PR Approval Ratio} & 4.28 & *** & $\searrow$ & 4.92 & *** & $\searrow$ & \textbf{10.17} & *** & $\nearrow$ & 0.48 &  & $\nearrow$ \\
\textbf{PR Approval Density} & \textbf{15.93} & *** & $\searrow$ & \textbf{11.1} & *** & $\searrow$ &  &  &  & 1.37 & ** & $\nearrow$ \\
\textbf{Followers} & 0.84 & . & $\searrow$ & 0.94 & . & $\searrow$ & 0.74 & . & $\nearrow$ & 0.39 &  & $\nearrow$ \\
\textbf{Collaborations} & 5.37 & *** & $\nearrow$ & \textbf{34.17} & *** & $\nearrow$ & \textbf{10.36} & *** & $\searrow$ & \textbf{15.66} & *** & $\searrow$ \\ \hline
\multicolumn{13}{l}{Signif. codes: \textbf{0 ‘***’ 0.001 ‘**’ 0.01 ‘*’ 0.05} ‘.’ 0.1 ‘ ’ 1} \\
\multicolumn{13}{l}{\begin{tabular}[c]{@{}l@{}}Chisq\%:  The importance of a feature in modeling a profile in percentage.\end{tabular}}\\
\multicolumn{13}{l}{\begin{tabular}[c]{@{}l@{}}Rel.:  $\nearrow$ stands for a positive correlation between a feature and the likelihood of being a profile, and $\searrow$ stands for a negative correlation.\end{tabular}}\\  \hline

\end{tabular}
\end{adjustbox}
\end{table}

\noindent
\textbf{The constructed logistic regression models have the highest discrimination performance for classifying Core-Afterhour contributors.} Table~\ref{tab:glmAUC} presents the AUC values of the models. The models for \textit{Core-Afterhour} and \textit{Core-Workhour} contributors have the top AUC values of 0.73 and 0.72 respectively. Both \textit{Peripheral-Afterhour} and \textit{Peripheral-Workhour} profile models achieve AUC values over 0.6, indicating the effectiveness of the models in classifying the respective profiles beyond random classification. In the rest of the section, we describe our analysis of the logistic regression models and identify the important features and behaviors associated with each profile. Additionally, we validate our analysis with the results of the Mann-Whitney U tests and Cliff's delta values. We find that the null hypothesis \(H0_{1}\) is rejected for 26 (out of 27) contributor features, which shows significant differences between \textit{core} and \textit{peripheral} contributors regarding the majority of contributor features. Similarly, the null hypothesis \(H0_{2}\) is rejected for 20 contributor features and \(H0_{3}\) is rejected for 13 contributor features, highlighting the significant differences between \textit{Core-Afterhour} and \textit{Core-Workhour} contributors and between \textit{Peripheral-Afterhour} and \textit{Peripheral-Workhour} contributors respectively. The complete list of test results is presented in Table~\ref{tab:rq1_mann} in Appendix~\ref{appendix:rq1}.

\textbf{Duration, authored files, and collaboration are common significant features for all four contributor profiles, serving as the main factors to differentiate core and peripheral contributors.} As shown in Table \ref{tab:rq1_glmresult}, both \textit{Core} profiles are associated with increased project duration, number of authored files, and collaborated contributors, while \textit{Peripheral} profiles have negative associations with these three features. The Mann-Whitney U test results show that \textbf{core contributors have a significantly longer duration, larger number of authored files, and more collaborations than peripheral contributors}, with a large effect size for the duration and number of authored files, and medium for collaborations. This suggests that \textit{Core} contributors are characterized by extensive project experience, diverse source file authorship, and a notable social influence within the project, compared to \textit{Peripheral} contributors. Furthermore, the difference in collaborations also indicates that \textit{Core} contributors tend to have active involvement in project coordination, while \textit{Peripheral} contributors tend to focus on their specific contributions to the project without involving others.

\textbf{Timezone is the main factor to differentiate workhour and afterhour contributors, while working hours have negligible 
common association with other contributor features.} Table~\ref{tab:rq1_glmresult} shows that timezone is significant for all profiles with positive associations with two \textit{Afterhour} profiles (i.e., \textit{Core-Afterhour} and \textit{Peripheral-Afterhour}) and negative associations with two \textit{Workhour} profiles (i.e., \textit{Core-Workhour} and \textit{Peripheral-Workhour}). However, the significance of time zones lies in their representation of geographical regions. We categorize contributor timezones into three regions: Americas (timezone from -12 to -2, containing 53.2\% of contributors), Europe/Africa (-1 to 3, containing 16.5\% contributors), and Asia (4 to 12, containing 30.2\% contributors). We find that 67.4\% of contributors from the Americas are \textit{Workhour} contributors, contrasting with 28.3\% from Asia and 49.1\% from Europe/Africa. This suggests a higher inclination among Americas-located contributors towards adopting a regular job-like approach to OSS ML projects compared to other regions. \textit{Afterhour} contributors might have flexible contribute schedules or prioritize their contributions to the project during their free time or outside their regular job commitments. It is also possible that \textit{Afterhour} contributors from Europe/Africa or Asia attempt to align their contributions with work hours in the Americas to ensure quick responses from the majority of the development team. These findings indicate the predominant influence of contributors from the Americas in driving the selected ML projects. 
 
From the Mann-Whitney U test results, we find that \textit{Workhour} contributors have a significantly larger number of issues solved, a larger number of pull requests reviewed, and more collaborations than \textit{Afterhour} contributors with a small or negligible effect. Notably, the effect size of collaboration for \textit{Core-Workhour} and \textit{Core-Afterhour} contributors is small, indicating that \textit{Core-Workhour} contributors are more intensely involved in collaborations to a considerable extent.

\textbf{Core-Afterhour contributors focus on developmental contributions, while Core-Workhour contributors participate in both developmental and collaborative activities.} Logistic regression model results show two \textit{core} profiles tend to have the same directions of correlation on commit-related, issue-related, and pull request-related features. The Mann-Whitney U test results show that \textit{Core-Afterhour} has a statistically significantly higher number of authored files and non-code commits, with negligible effect sizes. \textit{Core-Workhour} contributors are statistically higher in
terms of the number of issues raised, issue contribution
rate, the number of issues solved, issues participated, pull requests merged, pull requests reviewed, pull request contribution rate, and collaborations. The effect sizes are small for the number of pull requests merged, the pull request contribution rate, and the number of collaborations and negligible for other features. There is no statistical difference in the commit rate, code commit rate, and code contribution rate between the two \textit{core} profiles. These findings indicate that both \textit{core} profiles have similar extent of participation in code contributions. However, \textit{Core-Workhour} contributors are more actively participating in collaborative activities such as issue and pull request-related activities. This indicates that \textit{Core-Afterhour} contributors tend to focus on developmental tasks, and \textit{Core-Workhour} contributors engage in a mix of development and project maintenance activities.

\textbf{Despite sharing similar work time, commit volume, code contribution density, and proficiency in multiple programming languages, contributors within the same profile may have different preferences and areas of focus.} Logistic regression model results show that the two \textit{core} profiles tend to be characterized by positive association with commit-related features indicating their high code contributions. Moreover, the two \textit{peripheral} profiles tend to be characterized by positive association with issue-related and pull request-related features, indicating active participation in collaborative activities. However, the Mann-Whitney U test results show that \textit{core} contributors are statistically higher in all the issue-related and pull request-related features than \textit{peripheral} contributors excluding pull request approval ratio and pull request approval density. This contradicted observation indicates that the logistic regression models can effectively leverage features related to the code contributions of \textit{core} contributors but may overlook the characteristics of those \textit{core} contributors who actively participate in other OSS activities.

For instance, we find that 42.9\% \textit{Core-Afterhour} and 52.5\% \textit{Core-Workhour} contributors have raised at least one issue report. We apply the Mann-Whitney U test and calculate Cliff's delta to compare the number of commits made by the \textit{Core-Workhour} contributors who have raised at least one issue report and the \textit{Core-Workhour} contributors who have not raised any issue report. We conduct the same comparison for \textit{Core-Afterhour} contributors. We find that, in the \textit{Core-Workhour} profile, contributors who have raised issue reports make significantly more commits than those who have not, with a small effect size. Conversely, in the \textit{Core-Afterhour} profile, there is no statistical difference in the number of commits made by contributors regardless of raising issue reports. Therefore, contributors of the same profile may have different focuses on participation in various OSS activities. Further investigation is needed to comprehend the actual workload of contributors in each profile and their engagement in different OSS activities.

\begin{MySummaryBox}{RQ1 Summary}
Project experience, diversity of source file authorship, and collaborations are important features common to all four contributor profiles, primarily distinguishing core from peripheral profiles. Contributors in workhour and afterhour profiles mainly differ in their geological locations. Core-Afterhour contributors tend to focus on developmental contributions. Core-Workhour contributors actively participate in both developmental and collaborative activities. Even within the same profile, contributors exhibit varied engagements in OSS activities.
\end{MySummaryBox}

\subsection{RQ2: What is the OSS engagement of each contributor profile?}
\label{sec:rq2}

\subsubsection{\textbf{Motivation}}
\hfill

\noindent 
In RQ1, we observe that contributors within the same profile may differ in their involvement in different OSS activities. Given the broad range of open-source tasks and the limited time OSS contributors can allocate to such activities, they often prioritize certain tasks over others based on their needs and preferences. Understanding contributor workload compositions is essential to enrich our comprehension of contributor profiles, as it provides detailed insights into their individual preferences and priorities. Additionally, while workload composition provides a static view of contributors' activities in their overall engagement in the OSS activities within a period, the dynamics of their activities within that period remain unexplored. With the same amount of workload, contributors may choose to fulfill their workload gradually over a period or complete tasks in a single burst, depending on their preferences. Exploring work preferences is important as reported by Yue et al.~\cite{yue2022off} that contribution dynamics can affect the performance of contributors and their technical importance to a project. Moreover, committing is often considered the most important OSS workload. Although we have analyzed the volume of commits made by contributors across different profiles, the technical importance of their commits to the project remains unclear. Understanding contributor workload composition and work preferences can assist project managers in fostering collaborations among contributors with shared interests and similar work patterns. Moreover, understanding the technical importance of contributors can help project maintainers recognize important contributions. 

\subsubsection{\textbf{Approach}}
\hfill

\noindent 
We explore and compare the behaviors of contributors across different profiles during their engagements in OSS activities from three aspects: their workload compositions, work preferences, and technical importance. Our detailed approach is described as follows:

\noindent \paragraph{Workload Composition Analysis}
\hfill

As described in Section~\ref{sec:identify_workload_composition_pattern}, we identify five workload composition patterns that reflect five types of workload compositions of contributors within a period. 
We investigate the frequency of each type of workload composition for each contributor during one's active periods. The most frequent workload composition pattern of a contributor is identified as one's major pattern. For each profile, we calculate the proportion of contributors with each workload composition pattern as their major pattern and compare the proportions across profiles.


Chi-square test~\cite{Pearson1900} is a statistical test that can determine whether the proportions of categorical variables are the same across different groups or populations. We use the Chi-Square test to compare the distributions of major workload compositions (i.e., most frequent workload composition) of core and peripheral contributors with the null hypothesis \textit{\(H0_{4}\)} (with a significance threshold of p-value \textless{} 0.05), as well as their respective \textit{workhour} and \textit{afterhour} subgroups with the null hypothesis \textit{\(H0_{5}\)} and \textit{\(H0_{6}\)} . 

\begin{itemize}[noitemsep,topsep=0pt]
    \item \textit{\(H0_{4}\): The distributions of the major workload composition patterns are the same across core contributors and peripheral contributors.}
    \item \textit{\(H0_{5}\): The distributions of the major workload composition patterns are the same across Core-Afterhour and Core-Workhour contributors.}
    \item \textit{\(H0_{6}\): The distributions of the major workload composition patterns are the same across Peripheral-Afterhour and Peripheral-Workhour contributors.}
\end{itemize}

Cramér’s V is an effect size measure derived from the chi-squared statistic and is often used to quantify the magnitude of the difference identified by the Chi-square test. We calculate Cramér’s V for the groups with significant differences identified by our Chi-square tests to measure effect sizes. We employ the same interpretation of Cramér’s V as~\cite{cohen2013statistical} at degrees of freedom of 4 (i.e., negligible: 0 \textless{} .05, small: .05 \textless{} .15, medium .15 \textless{} .25, and large: \textgreater{}= .25).

\paragraph{Work Preference Analysis}
\hfill 

To measure the temporal contribution dynamic and preferences of contributors, we extract 9 work preference features from the OSS activity time series of each contributor within each 90-day period as described in Section~\ref{sec:identify_work_preference}. We apply two-sided Mann-Whitney U test~\cite{nachar2008mann} to compare each work preference feature between contributors in \textit{core} and \textit{peripheral} profiles with the null hypothesis \textit{\(H0_{7}\)}. We also compare their respective \textit{workhour} and \textit{afterhour} subgroups with the null hypothesis \textit{\(H0_{8}\)} and \textit{\(H0_{9}\)}, as specified below. We apply Cliff's delta~\cite{cliff1993dominance} to estimate the effect sizes.

\begin{itemize}
    \item \textit{\(H0_{7}\): The distributions of the given work preference feature are the same across core contributors and peripheral contributors.}
    \item \textit{\(H0_{8}\): The distributions of the given work preference feature are the same across Core-Afterhour and Core-Workhour contributors.}
    \item \textit{\(H0_{9}\): The distributions of the given work preference feature are the same across Peripheral-Afterhour and Peripheral-Workhour contributors.}
\end{itemize}


\paragraph{Technical Importance Assessment}
\hfill 

Four technical importance metrics are introduced in Section~\ref{sec:evaluate_importance} to evaluate the importance of a contributor's code contributions. We apply the Mann-Whitney U test to compare each technical importance metric across \textit{core} and \textit{peripheral} contributors with the null hypothesis \textit{\(H0_{10}\)}, as well as their respective \textit{workhour} and \textit{afterhour} subgroups with the null hypothesis \textit{\(H0_{11}\)} and \textit{\(H0_{12}\)} respectively. We also calculate Cliff's delta to estimate the effect sizes.

\begin{itemize}
    \item \textit{\(H0_{10}\): The distributions of the given technical importance metric are the same across core contributors and peripheral contributors.}
    \item \textit{\(H0_{11}\): The distributions of the given technical importance metric are the same across Core-Afterhour and Core-Workhour contributors.}
    \item \textit{\(H0_{12}\): The distributions of the given technical importance metric are the same across Peripheral-Afterhour and Peripheral-Workhour contributors.}
\end{itemize}

\subsubsection{\textbf{Results}}

\paragraph{Workload Composition} 
\hfill 

\begin{table}[H]
\caption{Distribution of major workload composition patterns across contributor profiles.}
\label{tab:workload_distribution}
\centering
\begin{adjustbox}{max width=0.7\textwidth}
\begin{tabular}{l|c|c|c|c}
\hline
\textbf{\begin{tabular}[c]{@{}l@{}}Major   Workload \\ Composition Pattern\end{tabular}} & \textbf{\begin{tabular}[c]{@{}c@{}}Core-\\ Afterhour\end{tabular}} & \textbf{\begin{tabular}[c]{@{}c@{}}Core-\\ Workhour\end{tabular}} & \textbf{\begin{tabular}[c]{@{}c@{}}Peripheral-\\ Afterhour\end{tabular}} & \textbf{\begin{tabular}[c]{@{}c@{}}Peripheral-\\ Workhour\end{tabular}} \\ \hline
\textbf{Issue Reporter} & 17.6\% & 18.8\% & \textbf{29.9\%} & \textbf{27.8\%} \\
\textbf{Issue Discussant} & 15.7\% & 17.5\% & 18.7\% & 17.7\% \\
\textbf{Committer} & \textbf{44.1\%} & \textbf{41.9\%} & \textbf{35.5\%} & \textbf{39.2\%} \\
\textbf{Collaborative Committer} & \textbf{20.5\%} & \textbf{19.1\%} & 15.6\% & 14.8\% \\
\textbf{Core Reviewer} & 2.2\% & 2.8\% & 0.2\% & 0.5\% \\ \hline
\end{tabular}
\end{adjustbox}
\end{table}

\noindent \textbf{Committer is the most frequent workload composition pattern for all contributor profiles, whereas Collaborative Committer is the second most frequent for core contributors and Issue Reporter is the second most for peripheral contributors.} As shown in Table~\ref{tab:workload_distribution}, contributors with Committer as their major pattern constitute the highest proportion in all profiles. Collaborative Committers and Issue Reporters constitute the second highest proportion of \textit{core} and \textit{peripheral} contributors respectively. This finding indicates that although the majority of contributors in all profiles only focus on committing, \textit{core} contributors often engage in a mix of coding and project management activities, whereas \textit{peripheral} contributors may contain a higher proportion of project users who frequently identify issues or request specific features. 

Based on our Chi-Squared test and Cramér's V result, we reject the null hypothesis \textit{\(H0_{4}\)} and \textit{\(H0_{6}\)}, while accepting \textit{\(H0_{5}\)}. The test results show a significant difference in the distribution of major workload composition patterns between \textit{core} and \textit{peripheral} contributors with a medium effect size. The distribution of major workload composition patterns between \textit{Core-Workhour} and \textit{Core-Afterhour} contributors has no significant difference, while the difference between \textit{Peripheral-Workhour} and \textit{Peripheral-Afterhour} contributors is negligible. Additionally, the rejection to \textit{\(H0_{4}\)} supports our findings on the differences between \textit{core} and \textit{peripheral} contributors in their major workload compositions. 


\paragraph{Work Preferences}
\hfill 

\begin{table}
\caption{Result of Mann-Whitney U test and effect size results for work preference features.}
\label{tab:work_preference_mann}
\centering
\begin{adjustbox}{max width=\textwidth}
\begin{tabular}{l|c|c|c}
\hline
\multicolumn{1}{c|}{\multirow{2}{*}{\textbf{Work Preference Features}}} & \textbf{Group1: Core} & \textbf{Group1:   Core-Afterhour} & \textbf{Group1: Peripheral-Afterhour} \\
\multicolumn{1}{c|}{} & \textbf{Group2: Peripheral} & \textbf{Group2: Core-Workhour} & \textbf{Group2: Peripheral-Workhour} \\ \hline
\textbf{binned\_entropy} & large (0.497) & negligible (-0.055) & not significant \\
\textbf{c3(1)} & medium (0.343) & negligible (-0.079) & negligible (0.016) \\
\textbf{c3(2)} & small (0.292) & negligible (-0.066) & negligible (0.01) \\
\textbf{c3(3)} & small (0.292) & negligible (-0.07) & negligible (0.012) \\
\textbf{number\_cwt\_peaks} & large (0.491) & negligible (-0.034) & not significant \\
\textbf{longest\_strike\_above\_mean} & medium (0.377) & negligible (-0.052) & negligible (0.026) \\
\textbf{longest\_strike\_below\_mean} & medium (-0.444) & not significant & not significant \\
\textbf{diverse} & small (0.284) & negligible (-0.144) & not significant \\
\textbf{balance} & medium (-0.339) & negligible (0.042) & negligible (-0.045) \\ \hline
\multicolumn{4}{l}{\begin{tabular}[c]{@{}l@{}}Groups with no significant difference (i.e., null hypothesis is accepted) are indicated with 'not significant'. Groups \\that are significantly different examined by Mann-Whitney U test are indicated with Cliff's delta value and its \\ interpretation (i.e., negligible, small, medium, or large). A positive value means group1 is greater than group2, and \\vice versa. Three columns from left to right correspond to the null hypothesis \textit{\(H0_{7}\)}, \textit{\(H0_{8}\)}, and \textit{\(H0_{9}\)} respectively. \end{tabular}} \\ \hline

\end{tabular}
\end{adjustbox}
\end{table}
As shown in Table~\ref{tab:work_preference_mann}, the null hypothesis \textit{\(H0_{7}\)} is rejected for all work preference features. This indicates that \textit{core} and \textit{peripheral} contributors have significantly different work preferences, in terms of their contribution dynamics and the level of balance of their contributions across various types of OSS activities.

\textbf{Core contributors exhibit more complex contribution dynamics compared to peripheral contributors.} Contribution dynamics of \textit{core} contributors, measured by \textit{binned\_entropy}, \textit{c3(1)}, \textit{c3(2)}, and \textit{c3(3)}, exhibit significantly higher complexity than \textit{peripheral} contributors, indicating greater fluctuations in their period OSS activity time series. Moreover, \textit{core} contributors have significantly more peaks (i.e., \textit{number\_cwt\_peaks}), a higher ratio of continuous time above mean (i.e., \textit{longest\_strike\_above\_mean}), and a lower ratio of the continuous time below mean (i.e., \textit{longest\_strike\_below\_mean}) in their OSS activities time series, compared to \textit{peripheral} contributors. This indicates that \textbf{ \textit{core} contributors tend to make more peak contributions within a period and engage in longer times of continuous intensive contributions, while \textit{peripheral} contributors tend to maintain constant contribution levels and have longer continuous inactive time. }

\textbf{Core contributors make less balanced contributions than peripheral contributors.} As shown in Table~\ref{tab:work_preference_mann}, \textit{Core} contributors are significantly higher than \textit{Peripheral} contributors on \textit{diverse} with a small effect size and significantly lower on \textit{balance} with a medium effect size. \textbf{This indicates that \textit{Core} contributors tend to make more diverse contributions across various types of OSS activities with a stronger focus on one or few types of OSS activities, while peripheral contributors make more even amount of contributions across fewer types of OSS activities. }

\textbf{The work preference differences between Workhour and Afterhour subgroups are negligible.} The null hypothesis \textit{\(H0_{8}\)} is rejected for 8 (out of 9) work preference features, indicating that compared to \textit{Core-Afterhour} contributors, \textit{Core-Workhour} contributors tend to have more complex contribution dynamics, and make more diverse and less balanced contributions across different types of OSS activities. However, the differences are negligible in practice. The null hypothesis \textit{\(H0_{9}\)} is only rejected for 5 work preference features and the effect sizes are negligible. This shows that the work preference differences between \textit{Peripheral-Workhour} and \textit{Peripheral-Afterhour} contributors are generally insignificant.

\paragraph{Technical Importance}
\hfill 

\begin{table}
\caption{Mann-Whitney U test and effect size results for contributor technical importance.}
\label{tab:centrality_mann}
\centering
\begin{adjustbox}{max width=\textwidth}
\begin{tabular}{l|c|c|c}
\hline
\multicolumn{1}{c|}{\multirow{2}{*}{\textbf{Technical Importance}}} & \textbf{Group1: Core} & \textbf{Group1: Core-Afterhour} & \textbf{Group1: Peripheral-Afterhour} \\
\multicolumn{1}{c|}{} & \textbf{Group2:   Peripheral} & \textbf{Group2: Core-Workhour} & \textbf{Group2: Peripheral-Workhour} \\ \hline
\textbf{max\_period\_centrality} & large (0.613) & not significant & negligible (-0.077) \\
\textbf{max\_centrality\_period} & medium (0.447) & negligible (0.069) & not significant \\
\textbf{max\_commit\_centrality} & large (0.595) & not significant & negligible (-0.083) \\
\textbf{max\_centrality\_day} & large (0.523) & negligible (0.07) & not significant \\ \hline
\multicolumn{4}{l}{\begin{tabular}[c]{@{}l@{}}Three columns from left to right correspond to the null hypothesis \textit{\(H0_{10}\)}, \textit{\(H0_{11}\)}, and \textit{\(H0_{12}\)} respectively.\end{tabular}} \\ \hline
\end{tabular}
\end{adjustbox}
\end{table}

\textbf{Core contributors tend to achieve a higher maximum technical importance within a project than peripheral contributors.} As shown in Table~\ref{tab:centrality_mann}, the null hypothesis \textit{\(H0_{10}\)} is rejected for all technical importance metrics. \textit{Core} contributors have significantly higher maximum technical importance than \textit{peripheral} contributors, both at the commit-level (i.e., \textit{max\_commit\_centrality}) and period-level (i.e., \textit{max\_period\_centrality}). This means that when comparing the importance of individual commits, which considers the average influence of files changed within the commit, the most important commit made by \textit{core} contributors tends to have higher importance (i.e., \textit{commit\_centrality}) than the most important commit made by \textit{peripheral} contributors. Similarly, the period of the highest technical importance for \textit{core} contributors, which considers both the importance and intensity of commits made within the period, tends to have higher importance (i.e., \textit{period\_centrality}) than those for \textit{peripheral} contributors. Therefore, \textit{core} contributors tend to reach higher technical importance within a project by making intensive changes to central files, which have a greater impact on the codebase and core functionalities. \textit{Peripheral} contributors tend to stay with modifying peripheral files that have less impact on the system and are often confined to very specific functions, such as updating documentation or fixing a parameter in an API. An example illustrating commit contributions of different levels of technical importance can be found in Figure~\ref{fig:commit_importance} in Appendix~\ref{appendix:B}, where commits with high and low centrality are made to core kernel functions and TensorFlow Lite demo code, respectively.

\textbf{Core contributors tend to require a longer time to progress to their maximum technical importance than peripheral contributors.} As shown in Table~\ref{tab:centrality_mann}, \textit{core} contributors are significantly higher than \textit{peripheral} contributors in terms of both the number of days before making their most important commit (i.e., \textit{max\_centrality\_day}) and the number of periods to reach the period of their highest technical importance (i.e., \textit{max\_centrality\_period}), indicating they require a longer time to reach their maximum technical importance within a project. \textit{Core-Afterhour} and \textit{Core-Workhour} contributors have a median of 173 and 143 days, respectively, to make their most important commit. For both \textit{core} profiles, contributors take a median of 2 periods to reach the period of the highest technical importance. \textbf{This suggests a progression in technical importance for \textit{core} contributors within the first six months of activity, followed by a shift away from highly technical and intensive contributions.} In contrast, \textit{Peripheral-Afterhour} and \textit{Peripheral-Workhour} contributors have a median of reaching the period of the highest technical importance in 1 period and making the most important commit within 5 and 6 days respectively, indicating that \textbf{the technical importance of \textit{peripheral} contributors tends to remain the same as they join the project. The \textit{peripheral} contributors do not make significant progression over time.}

\textbf{The differences in technical importance between the Workhour and Afterhour subgroups are negligible.}  
The null hypothesis \textit{\(H0_{11}\)} is only rejected for the technical importance metrics \textit{max\_centrality\_period} and \textit{max\_centrality\_day} with negligible effect sizes. This means that \textit{Core-Afterhour} and \textit{Core-Workhour} contributors tend to reach the same level of the maximum technical importance. \textit{Core-Afterhour} contributors take a longer time to reach their maximum technical importance, although the difference is negligible in practice. The null hypothesis \textit{\(H0_{12}\)} is only rejected for \textit{max\_commit\_centrality} and \textit{max\_period\_centrality}, also with negligible effect sizes. This indicates that \textit{Peripheral-Workhour} contributors tend to reach higher maximum technical importance than \textit{Peripheral-Afterhour} contributors, with a negligible effect in practice, and both profiles take the same amount of time to reach their maximum technical importance. 


\begin{MySummaryBox}{RQ2 Summary}
The most common workload composition pattern among all contributor profiles is Committer. 
Core contributors demonstrate more complex contribution dynamics and make less balanced contributions compared to peripheral contributors. Core contributors also tend to achieve higher maximum technical importance and exhibit a progression towards increasing technical importance within the first 2 active periods, whereas peripheral contributors typically have the highest technical importance when initially joining the project.

\end{MySummaryBox}

\subsection{RQ3: What are the important factors of contributor OSS engagement for increasing the popularity of a project?} 
\label{sec:rq3}

\subsubsection{\textbf{Motivation}}
\hfill

\noindent
 The diverse distribution of contributors with varying OSS engagements results in varied project dynamics. Analogous to how demographic diversity influences a nation's growth and stability, the distribution of contributors with different workload compositions and work preferences can significantly impact the success and vitality of an OSS project. 
 This becomes particularly valuable during periods of significant turnover, enabling project maintainers to quickly recognize the types of contributions (e.g., issue reporting, issue discussion, committing, pull request discussion, and code review) from departing contributors and assess the potential impact on various project activities in their absence. In this research question, we investigate the impact of the diversity of contributors with different workload compositions and work preferences on the project popularity, in terms of the increase in star ratings and the number of forks.

\subsubsection{\textbf{Approach}}
\hfill

\noindent
 We build four mixed-effect models to explore how contributor workload composition patterns and work preferences are associated with the increase of project stars and forks. Our detailed approach is outlined as follows:

\textbf{Independent variable Preparation:} We prepare two sets of independent variables: workload composition pattern-related variables and work preference-related variables as listed in Table~\ref{tab:rq2_mixed_effect_model}.
\begin{itemize}
    \item \textbf{Workload composition pattern related-variables:} For each period (as defined in Section~\ref{sec:identify_workload_composition_pattern}) within a project, we calculate the ratio of active contributors belonging to each of the five contributor workload composition patterns, and it denotes as \textit{\textless{}Pattern\_name\textgreater{}\_ratio}. For example, \textit{Issue\_Reporter\_ratio} circulates the proportion of contributors belonging to the Issue Reporter pattern among all active contributors in that period. This yields five variables describing the distribution of contributors of different workload composition patterns during that specific period. 

    We also compute the ratio of the five types of OSS contributions (i.e., issue reports, issue comments, commits, pull request comments, and code reviews) made by contributors belonging to each workload composition pattern. We denote it as \textit{\textless{}Pattern\_name\textgreater{}\_\textless{}contribution\_type\textgreater{}\_ratio}. For example, \textit{Issue\_Reporter\_issue\_ratio} denotes the proportion of issue reports raised by contributors belonging to the Issue Reporter pattern among all issue reports raised in that period. Therefore, 25 variables are extracted to capture the diversity of contributions from contributors of five workload composition patterns at that period. 
    
    We add a variable \textit{period} to account for the influence of different developmental stages of software projects on their popularity. In total, 31 workload composition pattern related-variables are extracted. We apply the Spearman Rank Correlation test to measure the variable correlations and remove one from each pair of highly correlated features with a coefficient greater than 0.7~\cite{8613795}. 14 non-correlated variables remain, as shown in Table~\ref{tab:rq2_mixed_effect_model}. 

    \item \textbf{Workload preference related-variables:} For each period in a project, we calculate the median value of each work preference feature (identified in Section~\ref{sec:identify_work_preference}) among the active contributors within that period. A median value represents the collective contributor work preference during that period. As a result, we obtain 9 variables corresponding to the respective median value of each work preference feature. We compute the average contribution made by active contributors to each of the five types of OSS activities within the period, to represent the project-level preference for different types of OSS activities. As previously mentioned, we include a \textit{period} variable to account for the impact of project development stages on the project popularity. In total, we obtain 15 workload preference-related variables. We apply the Spearman Rank Correlation test to identify and remove 7 correlated variables. 8 remaining variables are shown in Table~\ref{tab:rq2_mixed_effect_model}.
\end{itemize}

\textbf{Dependent variable Preparation:} We use the number of stars and forks as project popularity metrics. More specifically, the number of forks reflects the interest in a project from potential contributors, while the number of stars indicates interest from both potential users and contributors. We use Github GraphQL API\footnote{https://docs.github.com/en/graphql} to collect the timestamp of each star received by a project and use Github Forks API\footnote{https://docs.github.com/en/rest/repos/forks} to collect the timestamp when each fork occurs. For each project, we count the number of stars and forks gained within each period. Due to variations in popularity levels among the subject projects, the counts of stars and forks may be at different scales for different projects. Using the counts of stars and forks directly as dependent variables can skew the model results. Therefore, we apply Min-Max Normalization to scale the star and fork counts for different periods within each project to a range of 0 to 1. The formula of Min-Max Normalization is presented in Equation~\ref{eq:min_max}, where features are the subject projects and \(i\) denotes the periods of the projects. The normalized stars and forks serve as our dependent variables.

\textbf{Model Construction:} We construct four mixed-effect models to identify the associations between \textit{(i)} contributor workload composition pattern related-variables and the increase in star ratings, \textit{(ii)} workload composition pattern related-variables and the increase in forks, \textit{(iii)} work preference related-variables and the increase in star ratings, and \textit{(iv)} work preference related-variables and the increase in forks respectively. 

Mixed-effects models are statistical regression models that integrate both fixed and random
effects~\cite{pinheiro2006mixed}. Fixed effects are variables with consistent slopes and intercepts across all data points, while random effects account for variations between experimental groups, such as projects in our case. Mixed-effect models can estimate different slopes or intercepts for different experimental groups. Our mixed-effect models use projects as a grouping factor and assume different intercepts for different projects. We construct the mixed-effect models using Python package \textit{statsmodels} ~\cite{seabold2010statsmodels}. Note that we exclude the first 19 and 13 periods for Pytorch and Theano where no forks were recorded, due to a concern of these two repositories being private during those times. Similarly, the first period for every project is excluded from the model training, because we find that for all projects, the first period has significantly more stars and forks than other periods, which could skew the model results.

\textbf{Model Explanatory Power Evaluation:} We use conditional R\textsuperscript{2} and marginal R\textsuperscript{2} metrics to evaluate the fit of our mixed-effect models. Marginal R\textsuperscript{2} describes the proportion of variance explained by fixed effects. Conditional R\textsuperscript{2} describes the proportion of variance explained by both fixed effects and random effects~\cite{nakagawa2013general}.

\textbf{Model Result Analysis:} To understand the important features in the mixed-effect models, we examine the z-values and their corresponding p-values for each independent variable in the model result. z-value is associated with the t-statistic which tests the significance of each fixed effect coefficient, and it is more reliable than Wald Statistics with limited data points~\cite{bewick2005statistics}. A fixed effect is significant if it has Pr(\textgreater \textbar z\textbar)\, \textless \, 0.05, and such variables are marked with asterisks in Table~\ref{tab:rq2_mixed_effect_model}. We also analyze the coefficient sign of each independent variable. Positive coefficients, denoted by upward arrows ($\nearrow$) in Table~\ref{tab:rq2_mixed_effect_model}, suggest a positive correlation with the dependent variables (i.e., stars or forks), whereas negative coefficients, indicated by downward arrows ($\searrow$), imply a negative correlation. 

\subsubsection{\textbf{Results}}
\hfill

\noindent
\textbf{The mixed-effect models of modeling star increase with work preference features have the highest explanatory power.} Our mixed-effect model for analyzing work preferences and star increase has a conditional R\textsuperscript{2} of 0.73 and marginal R\textsuperscript{2} of 0.5, and the model for work preferences and fork increase has a conditional R\textsuperscript{2} and marginal R\textsuperscript{2} of both 0.56. This result shows that both models have good explanatory power and most of the variances are explained by the fixed effects. The mixed-effect model of workload composition patterns with star increase has both conditional R\textsuperscript{2} and marginal R\textsuperscript{2} of 0.50, and the model of fork increase has both conditional R\textsuperscript{2} and marginal R\textsuperscript{2} of 0.32, indicating that both models have all the variances explained by the fixed effects.

\begin{table}[]
\caption{Mixed-effects models results.}
\label{tab:rq2_mixed_effect_model}

\centering
\begin{adjustbox}{max width=0.8\textwidth}
\begin{tabular}{l|cc|cc}
\hline
 & \multicolumn{2}{c|}{\textbf{Star}} & \multicolumn{2}{c}{\textbf{Fork}} \\
\textbf{Work Preference} & \textbf{Signif.} & \textbf{Rel.} & \textbf{Signif.} & \textbf{Rel.} \\ \hline
period & ** & $\searrow$ &  & $\searrow$ \\
binned\_entropy & . & $\searrow$ & * & $\searrow$ \\
longest\_strike\_below\_mean &  & $\nearrow$ &  & $\nearrow$ \\
balance & *** & $\nearrow$ & *** & $\nearrow$ \\
commit & *** & $\searrow$ & . & $\searrow$ \\
issue &  & $\nearrow$ & . & $\nearrow$ \\
issue\_comment & . & $\nearrow$ & * & $\nearrow$ \\
review & ** & $\nearrow$ & ** & $\nearrow$ \\ \hline
\textbf{Workload   Composition Pattern} & \multicolumn{1}{l}{} & \multicolumn{1}{l|}{} & \multicolumn{1}{l}{} & \multicolumn{1}{l}{} \\ 
Issue\_Reporter\_ratio & *** & $\nearrow$ & * & $\nearrow$ \\
Issue\_Discussant\_ratio & ** & $\nearrow$ &  & $\nearrow$ \\
Committer\_ratio &  & $\nearrow$ &  & $\searrow$ \\
Issue\_Reporter\_issue\_ratio &  & $\searrow$ &  & $\nearrow$ \\
Issue\_Reporter\_review\_ratio &  & $\nearrow$ &  & $\nearrow$ \\
Issue\_Discussant\_issue\_ratio &  & $\nearrow$ &  & $\nearrow$ \\
Issue\_Discussant\_issue\_comment\_ratio & ** & $\searrow$ &  & $\searrow$ \\
Issue\_Discussant\_review\_ratio & *** & $\nearrow$ & * & $\nearrow$ \\
Committer\_issue\_ratio &  & $\nearrow$ &  & $\nearrow$ \\
Committer\_review\_ratio & *** & $\nearrow$ & ** & $\nearrow$ \\
Collaborative\_Committer\_commit\_ratio &  & $\nearrow$ &  & $\nearrow$ \\
Collaborative\_Committer\_issue\_ratio &  & $\nearrow$ &  & $\nearrow$ \\
Collaborative\_Committer\_review\_ratio & . & $\nearrow$ &  & $\searrow$ \\
Code\_Reviewer\_review\_ratio &  & $\nearrow$ & * & $\nearrow$ \\ \hline
\multicolumn{5}{l}{Signif. codes:  Pr(\textgreater \textbar z\textbar) \,=\, \textbf{0 ‘***’ 0.001 ‘**’ 0.01 ‘*’ 0.05} ‘.’ 0.1 ‘ ’ 1} \\
\multicolumn{5}{l}{\begin{tabular}[c]{@{}l@{}}Rel.:  $\nearrow$ stands for a positive association between the independent and \\ dependent variables, and $\searrow$ stands for a negative association.\end{tabular}}
\\ \hline
\end{tabular}
\end{adjustbox}
\end{table}

\paragraph{Important Work Preference Features to Project Popularity}
\hfill 

\textbf{Work preference features such as the balance of contribution and the average contributor code reviews have significant positive associations with project popularity in terms of both star rating and the number of forks.} Table~\ref{tab:rq2_mixed_effect_model} presents the variable importance and coefficients of our mixed effect models. We observe that more balanced contributions across various types of OSS activities (i.e., \textit{balance}) and a higher number of average code reviews conducted by each contributor (i.e., \textit{review}) have a significant positive association with increases in both stars and forks. This association indicates that when contributors distribute their efforts evenly across various OSS activities or engage in more code reviews, the likelihood of attracting new users or contributors tends to rise. The developmental stages of a project (i.e., \textit{Period}) and the average number of commits per contributor (i.e., \textit{commit}) are negatively associated with star increases. This indicates a reduced likelihood of attracting more new users and contributors as project development progresses to later stages and when contributors are heavily taken on in committing responsibilities. 

We observe that the complexity of contributors' contribution dynamics (i.e., \textit{binned\_entropy}) is negatively correlated with fork increases, indicating an increased likelihood of attracting potential contributors when existing contributors have constant and less random contribution frequencies. The average number of issue comments per contributor (i.e., \textit{Issue comments}) is positively associated with fork increases, indicating that newcomers may begin their journey in the project by raising issues or participating in issue discussions. Moreover, the active participation of current contributors in these discussions may foster a welcoming environment and encourage newcomers to contribute to the project.

\paragraph{Important Workload Composition Pattern Features to Project Popularity}
\hfill 

\textbf{Workload composition pattern features, such as the ratio of Issue Reporters among all contributors, and the ratio of code reviews made by Issue Discussants and Committers, have significant positive associations with project popularity in terms of both star rating and the number of forks.} As shown in Table~\ref{tab:rq2_mixed_effect_model}, a higher ratio of Issue Reporters among all contributors in the project (i.e., \textit{Issue\_Reporter\_ratio}) has a significant positive association with the increase in both stars and forks. This suggests that projects are more likely to attract stars and forks when a higher proportion of contributors focus on reporting more issues or requesting more new features. 

A higher ratio of code reviews conducted by Issue Discussants (i.e., \textit{Issue\_Discussant\_review\_ratio}) and a higher ratio of code reviews conducted by Committers (i.e., \textit{Committer\_review\_ratio}) both have a significant positive association with the increase in star ratings and forks. This association indicates that projects tend to gain more star ratings and forks when a larger portion of code reviews are made by Issue Discussants or Committers. A possible explanation is that Issue Discussants are familiar with ongoing project issues and feature requests through their active participation in the issue discussions. Therefore, their code reviews are particularly effective in assisting contributors to resolve their issues or implement features in a way that aligns with the immediate needs of the project. This interaction could be especially beneficial for retaining newcomers, as Issue Discussants may already engage with newcomers during issue discussions before helping them with code reviews. When newcomers begin their journey in the project by discussing their issues with existing contributors and eventually making successful pull requests with their guidance, they are likely to feel welcomed and accepted within the project community. A real-world example is shown in Figure~\ref{fig:issue_discussant_example} in Appendix~\ref{appendix:B}. Meanwhile, Committers only focus on making commits, so they tend to be familiar with the codebase and possess specific technical expertise in their focused areas, which allows them to provide detailed and technical guidance during code reviews, potentially attracting contributors. 

\textbf{A higher ratio of Issue Discussants among existing contributors is significantly positively associated with increased star ratings.} As shown in Table~\ref{tab:rq2_mixed_effect_model}, the ratio of Issue Discussants among all active contributors in the project (i.e., \textit{Issue\_Discussant\_ratio}) is positively associated with the increase in stars. This association implies that projects tend to attract more star ratings when a greater proportion of current contributors actively participate in issue discussions and respond to questions or requests from users or potential contributors. 

\textbf{A higher proportion of code reviews conducted by Code Reviewers is significantly positively associated with the increase in forks.} Table~\ref{tab:rq2_mixed_effect_model} shows that the proportion of code reviews made by Code Reviewers (i.e., \textit{Code\_Reviewer\_review\_ratio}) have a significant positive correlation with fork increases but not with stars. This observation can be explained by the fact that, while both users and potential contributors star projects to express their interest, individuals who fork projects often have the intention of starting to make code contributions. Therefore, the code reviews provided by Code Reviewers hold greater significance for potential contributors than users.

\begin{MySummaryBox}{RQ3 Summary}
Contributors making balanced contributions across various OSS activities and a higher average number of code reviews are significantly associated with the increase in project popularity in terms of star ratings and forks. Moreover, a higher proportion of Issue reporters among existing contributors and a larger portion of code reviews made by Issue Discussants and Committers are significantly associated with greater increases in star ratings and forks. The proportion of Issue Discussants among all contributors is only significantly associated with star ratings. The proportion of reviews made by Code Reviewers is significantly associated with the increase in forks.
\end{MySummaryBox}

\subsection{RQ4: How does contributor OSS engagement evolve?} 

\subsubsection{\textbf{Motivation}}
\hfill

\noindent 
From our observation of contributor engagement within a project in RQ2, we note that their OSS engagements are not static; they shift over time. This motivates us to explore how the workload composition of contributors across different profiles changes as they spend more time on a project, as well as how the demand for contributors with diverse workload compositions changes as project development progresses. Moreover, we are also interested in understanding how contributors' work preferences and technical importance shift as they stay longer in the project and make more contributions. To provide insights into the behavior and trajectory of contributors, we aim to identify if there are significant trends in the evolution of contributors' workload composition, work preferences, and technical importance. Recognizing such trends may assist project maintainers in better comprehending the dynamics of potential long-term contributors, offer a roadmap for motivated new contributors, and suggest potential paths for their progression and advancement within the project.

\subsubsection{\textbf{Approach}}
\hfill

\noindent 
We explore the trends in the evolution of workload compositions, work preferences, and technical importance. Our detailed approach is described as follows:

\paragraph{Evolution of Workload Composition}
\label{par:workload_evolution}
\hfill 

To identify the evolution of a contributor's workload composition, we first extract the contributor's active \textit{periods} (as described in Section~\ref{sec:identify_workload_composition_pattern}), during which the contributor has made OSS contributions to the project. We then extract the workload composition pattern associated with the contributor in each period, thus obtaining a time series of workload composition patterns that reflects the contributor's workload compositions over time. We encode the workload composition patterns into numerical values, with Issue Reporter represented as 1, Issue Discussant as 2, Committer as 3, Collaborative Committer as 4, and Code Reviewer as 5. This numerical representation aligns intuitively with the level of expertise needed to handle the respective workload.

Cox-Stuart trend test~\cite{cox1955some} is a statistical method to detect the presence of a monotonic trend in time series. We employ the Cox-Stuart trend test from the R package \textit{randtests} to identify if a significant increasing or decreasing trend exists in our numerical workload composition time series. For each time series, we use the right-sided Cox-Stuart trend test to detect the increasing trend with null hypothesis\textit{ \(H0_{13}\)} and the left-sided Cox-Stuart trend test to detect the decreasing trend with \textit{\(H0_{14}\)}, with a 95\% confidence level. We test against these two null hypotheses for all the Cox-Stuart trend tests in the rest of the paper.
\begin{itemize}
    \item \textit{\(H0_{13}\): There is no significant increasing trend in the given time series/sequence.}
    \item \textit{\(H0_{14}\): There is no significant decreasing trend in the given time series/sequence.}
\end{itemize}

When applying the Cox-Stuart trend test on workload composition time series, we exclude the contributors who have less than 2 active periods, as the test requires at least two datapoints in the time series.

To pinpoint the changes in contributors' workload composition at different stages of their project involvement, we divide their workload composition time series into three equal-length subsequences, representing their workload compositions in the early, middle, and late stages of the project involvement. We exclude the contributors with fewer than 3 active periods (at least one datapoint in each stage is required) and apply the Cox-Stuart trend test on the jointed sequence of the early and middle stages to examine the changes in workload composition from the early to middle stage, and the same for the middle to late stage.

Additionally, we examine the evolution of both the prevalence and distribution of each workload composition pattern within the subject projects, to understand the sustainability and the demand for contributors with particular workload compositions across different stages of project development. For each project, we count the number of contributors belonging to a workload composition pattern in each period into a time series to analyze the prevalence of the pattern as project development progresses. Similarly, we build another time series for the ratio of contributors belonging to a workload composition pattern among all contributors in each period to analyze how the distribution of the pattern evolves within a project. We apply the Cox-Stuart trend test to examine the trends in the time series of counts and ratios of contributors belonging to each workload composition pattern. To further pinpoint the major changes in contributor workload composition at different project development stages, we divide each time series into three equal-length subsequences, representing the early, middle, and late stages of project development. We apply the Cox-Stuart trend test on the jointed sequence of the early and middle stages to identify the change of contributor workload compositions in a project, from its early to middle stages, and the same for the middle to late stage.

\paragraph{Evolution of Work Preferences}
\hfill 

\begin{figure}
    \centering
    \includegraphics[width=0.65\linewidth]{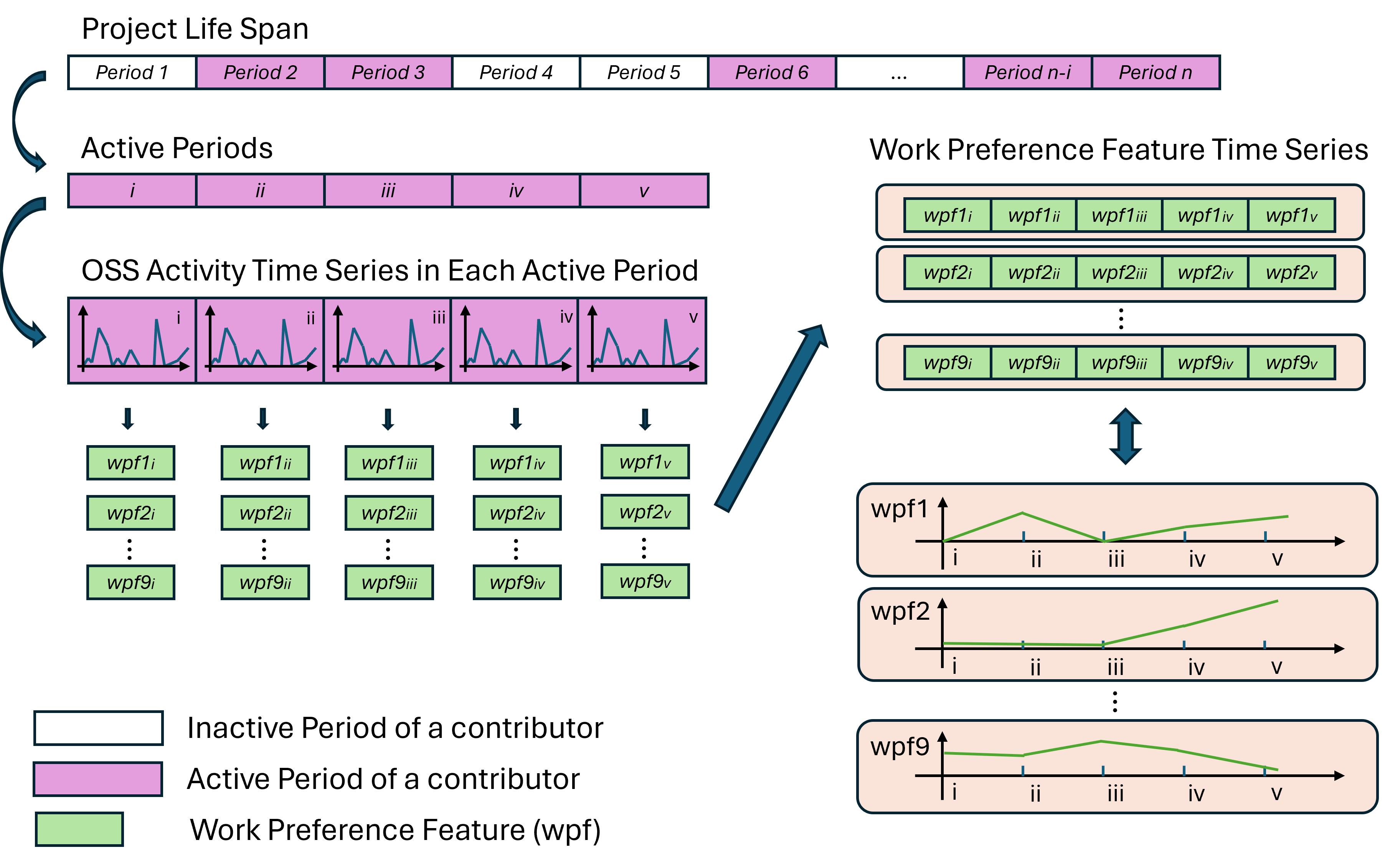}
    \caption{Example of building a contributor's work preference feature time series.}
    \label{fig:work_preference_sequence}
\end{figure}

To identify trends in the evolution of contributors' work preferences, our approach is similar to identifying the evolution of workload composition in Section~\ref{par:workload_evolution}. Figure~\ref{fig:work_preference_sequence} illustrates our detailed approach. For each contributor, we first extract the periods where the contributor is active in the project. We extract nine work preference features (as described in Section~\ref{sec:identify_work_preference}) from the contributor's OSS activity time series within each active period. We build a time series for each work preference feature, representing its evolution across the contributor's active periods. Therefore, we obtain nine work preference feature time series for each contributor to analyze the evolution of their work preferences. We apply the same procedure as in Section~\ref{par:workload_evolution} to select contributors and examine the trends in their work preference feature time series, as well as the trends in the early-to-middle-stage and middle-to-late-stage subsequences.

\paragraph{Evolution of Technical Importance}
\hfill 

We study the evolution of contributors' technical importance at two levels of granularity: per commit and per period. Analyzing contributors' technical importance at the commit level allows us to understand the shift in the importance of their code changes more precisely, such as from making changes to the very specific functions in the peripheral files to modifying the influential files that have a greater impact on the system, or the opposite. Moreover, period-level evolution reflects contributors' changes in technical importance at a higher level, considering both the importance and the intensity of their code contribution over time.

To identify the trend of contributors' commit importance, we extract all commits made by a contributor and sort them into chronological order. We calculate the \textit{commit centrality} (as described in Section~\ref{sec:evaluate_importance}) of each commit, thereby obtaining a sequence of commit centrality that reflects the importance of each commit made by a contributor over time. We apply
the same procedure as in Section~\ref{par:workload_evolution} to select contributors and examine the trends in their commit centrality sequence, as well as the trends in the early-to-middle-stage and middle-to-late-stage subsequences.

Similarly, to identify the trend of contributors' period technical importance, we extract the contributors' active periods, in which a contributor has made OSS contributions to the project. We calculate the \textit{period centrality} (as described in Section~\ref{sec:evaluate_importance}) of each active period and construct a time series of period centrality for each contributor. We examine the trends in their period centrality time series and the trends in the early-to-middle-stage and middle-to-late-stage subsequences.



\subsubsection{\textbf{Results}}
\hfill

\noindent 
Table~\ref{tab:evolution_trend} shows the overall evolution trends of workload composition, work preference, and technical importance for contributors in each profile. Table~\ref{tab:stage_evolution_trend} presents the early-to-middle-stage and middle-to-late-stage trends. In these tables, the upward arrows denote rejection of the null hypothesis \(H0_{13}\),  indicating increasing trends in the given time series, with the percentages indicating the proportion of contributors within the profile exhibiting increasing trends. Likewise, downward arrows represent the rejection of the null hypothesis \(H0_{14}\) and decreasing trends.

\paragraph{Evolution of Workload Composition}
\hfill 

\textbf{A small portion of long-term contributors exhibit significant trends in their workload composition towards handling tasks requiring higher expertise, while the majority of contributors do not exhibit a significant trend in shifting workload composition.} As shown in Table~\ref{tab:evolution_trend}, approximately 2\% \textit{core} contributors and 0.2\% \textit{peripheral} contributors demonstrate significant trends in shifting workload composition. We manually investigate these contributors and observe that they are all long-term contributors with at least 10 active periods (i.e., being active in the subject project for more than 2.5 years). Among these contributors, 84\% have an increasing trend in their workload composition, which means that their workload composition evolves in the direction of Issue Reporters $\rightarrow$ Issue Discussants $\rightarrow$ Committer $\rightarrow$ Collaborative Committer $\rightarrow$ Code Reviewer. This indicates that these long-term contributors tend to shift their focus toward the OSS tasks that are typically considered to require higher expertise. Table~\ref{tab:stage_evolution_trend} shows that the increasing trends of workload composition for long-term contributors primarily occur during their early to middle stage of project involvement. Approximately 1\% \textit{core} contributors exhibit an increasing trend in workload composition during this stage, while only 0.1\% experience a decreasing trend. Conversely, long-term contributors typically maintain their workload composition during their middle to late stages, with 50\% fewer contributors experiencing significant shifts during this stage compared to the early to middle stages.

\textbf{The Committer's and Code Reviewer's contributions are more sustainable compared with other roles as ML project development progresses.} Table~\ref{tab:project_trend} shows the trends in the prevalence (i.e. count) and the proportion (i.e., ratio) of contributors exhibiting each workload composition pattern within the subject projects. In all projects, both in their early stages and overall trends, the number of Committers and their ratio within the project either remains constant or demonstrates increasing trends. This underscores the enduring popularity and community engagement of the selected ML projects. However, in the late stages, MXNet and Theano exhibit a decreasing trend in the number of Committers, possibly due to the delaying popularity of these projects. The number of Code Reviewers and their ratio in the project either remain unchanged or demonstrate increasing overall trends across all projects, indicating the sustained demand for code reviews throughout the ML project development stages. Additionally, all projects exhibit an increasing trend in the ratio of Code Reviewers in their early to middle stages, indicating significant progression of contributors to Code Reviewers during this period, with only a few contributors eligible for code review in the early stage.

\textbf{The Issue Reporter's contributions are less sustainable compared with other roles as ML project development progresses.} As shown in Table ~\ref{tab:project_trend}, 3 (out of 6) projects exhibit a decreasing trend in the prevalence of Issue Reporters, and 5 (out of 6) projects exhibit decreasing ratios of Issue Reporters. This trend likely stems from the emerging phase of the ML library. When a new ML library emerges, there is often a surge of interest due to the growing popularity of ML in recent years. For instance, we observe that TensorFlow, Keras, MXNet, and ONNX reach the peak number of Issue Reporters within 1 to 2 years after they open-sourced. During this time, OSS community members are eager to explore the new ML library and actively provide feedback, report bugs, and request new features. After this initial interest, enthusiastic contributors may shift their focus toward code contributions possibly transitioning to roles like Committers or Collaborative Committers, while others may depart from the project. This leads to a decreasing trend of Issue Reporters from the early to the middle stage of the project development. Additionally, 4 projects (i.e., PyTorch, Keras, MXNet, and ONNX) exhibit no significant trend in the ratio of Issue Reporters in their middle to late stages, while 2 projects (i.e., TensorFlow and Theano) show a decreasing trend. 

\begin{table}
\caption{Project-level trend of workload composition.}
\label{tab:project_trend}
\centering
\begin{adjustbox}{max width=\textwidth}
\begin{tabular}{c|c|cc|cc|cc|cc|cc}
\hline
\multirow{2}{*}{\textbf{Project Stages}} & \multirow{2}{*}{\textbf{Project}} & \multicolumn{2}{c|}{\textbf{Issue Reporter}} & \multicolumn{2}{c|}{\textbf{Issue Dissucant}} & \multicolumn{2}{c|}{\textbf{Committer}} & \multicolumn{2}{c|}{\textbf{Collaborative Committer}} & \multicolumn{2}{c}{\textbf{Code Reviewer}} \\
 &  & \textbf{Count} & \textbf{Ratio} & \textbf{Count} & \textbf{Ratio} & \textbf{Count} & \textbf{Ratio} & \textbf{Count} & \textbf{Ratio} & \textbf{Count} & \textbf{Ratio} \\ \hline
\multirow{6}{*}{\textbf{\begin{tabular}[c]{@{}c@{}}Early - Middle \\ Trend\end{tabular}}} & tensorflow & - & (**)$\searrow$ & - & - & (**)$\nearrow$ & (**)$\nearrow$ & (**)$\nearrow$ & - & (*)$\nearrow$ & (*)$\nearrow$ \\
 & pytorch & - & (**)$\searrow$ & - & - & - & - & (**)$\nearrow$ & (**)$\nearrow$ & (**)$\nearrow$ & (**)$\nearrow$ \\
 & keras & (*)$\searrow$ & (**)$\searrow$ & - & - & - & - & - & (*)$\nearrow$ & - & (*)$\nearrow$ \\
 & mxnet & - & (*)$\searrow$ & (**)$\nearrow$ & (**)$\nearrow$ & - & - & (**)$\nearrow$ & (*)$\nearrow$ & (**)$\nearrow$ & (**)$\nearrow$ \\
 & theano & (**)$\nearrow$ & - & (*)$\nearrow$ & - & (**)$\nearrow$ & - & (*)$\nearrow$ & - & (**)$\nearrow$ & (**)$\nearrow$ \\
 & onnx & - & - & - & - & - & - & - & - & - & (*)$\nearrow$ \\ \hline
\multirow{6}{*}{\textbf{\begin{tabular}[c]{@{}c@{}}Middle - Late \\ Trend\end{tabular}}} & tensorflow & (**)$\searrow$ & (**)$\searrow$ & (*)$\searrow$ & (*)$\searrow$ & - & (**)$\nearrow$ & - & - & - & (*)$\searrow$ \\
 & pytorch & - & - & - & - & - & - & - & - & (**)$\nearrow$ & (**)$\nearrow$ \\
 & keras & - & - & (*)$\searrow$ & - & - & - & (**)$\searrow$ & - & - & - \\
 & mxnet & (*)$\searrow$ & - & (**)$\searrow$ & - & (**)$\searrow$ & (*)$\searrow$ & - & (*)$\nearrow$ & - & (**)$\nearrow$ \\
 & theano & (***)$\searrow$ & (***)$\searrow$ & (***)$\searrow$ & - & (***)$\searrow$ & - & (***)$\searrow$ & - & - & - \\
 & onnx & - & - & - & - & - & - & - & - & (*)$\searrow$ & - \\ \hline
\multirow{6}{*}{\textbf{Overall Trend}} & tensorflow & - & (***)$\searrow$ & - & (***)$\searrow$ & (***)$\nearrow$ & (***)$\nearrow$ & - & - & - & - \\
 & pytorch & - & (***)$\searrow$ & (*)$\nearrow$ & (*)$\nearrow$ & (*)$\nearrow$ & - & (*)$\nearrow$ & - & (***)$\nearrow$ & (***)$\nearrow$ \\
 & keras & (***)$\searrow$ & (***)$\searrow$ & (**)$\searrow$ & - & - & - & (*)$\searrow$ & - & - & - \\
 & mxnet & (**)$\searrow$ & (***)$\searrow$ & - & - & - & - & - & (**)$\nearrow$ & (*)$\nearrow$ & (***)$\nearrow$ \\
 & theano & (*)$\searrow$ & (*)$\searrow$ & (**)$\searrow$ & - & - & - & (*)$\searrow$ & (*)$\searrow$ & (*)$\nearrow$ & (*)$\nearrow$ \\
 & onnx & - & - & (**)$\searrow$ & - & - & - & - & (*)$\nearrow$ & - & - \\ \hline
\multicolumn{12}{l}{\begin{tabular}[c]{@{}l@{}}  $\nearrow$ : The null hypothesis \(H0_{13}\) is rejected and the given time series has an increasing trend. \\ $\searrow$ : The null hypothesis \(H0_{14}\) is rejected and the given time series has a decreasing trend. \\ \,\,-\,\, : Both \(H0_{13}\) and \(H0_{14}\) are accepted. There is no significant trend in the given time series. \end{tabular}} \\
 \multicolumn{12}{l}{Confidence Level: (***) 99.9\%, (**) 99\%, (*) 95\%. } \\
\hline

\end{tabular}
\end{adjustbox}
\end{table}

\paragraph{Evolution of Work Preferences} 
\hfill 

\textbf{A portion of long-term contributors show a tendency towards fewer, constant, and balanced contributions, while the majority of contributors do not demonstrate a clear trend in shifting work preferences}. Similar to our finding on workload compositions, significant trends in work preferences are observed only among long-term contributors (primarily core contributors). For the contributors with significant changes in each work preference feature, we observe that 73\% to 85\% of them have a decreasing trend for the complexity of contribution dynamics (i.e., \textit{binned\_entropy} and \textit{C3 statistics}), the number of peak contributions (i.e., \textit{number\_cwt\_peaks}), and the time with continuous intensive contribution (i.e., \textit{longest\_strike\_above\_mean}). This suggests that these long-term contributors tend to reduce intensive contributions and progress towards making constant contributions. Moreover, among contributors who have significant changes in their contribution balance, we observe 75\% of them have increasing trends. This indicates that long-term contributors tend to progress towards allocating their contributions more evenly across various OSS activities. We also observe a greater proportion of contributors have decreasing trends for the number of commits, issues, issue comments, and pull request comments, while having increasing trends for code reviews. This suggests a shift towards undertaking more review duties and reducing contributions to other OSS tasks. Table~\ref{tab:stage_evolution_trend} shows that significant shifts of work preferences primarily occur in the middle to later stages for the long-term contributors, with few significant trends of work preferences in their early to middle stages.

\paragraph{Evolution of Technical Importance}
\hfill 

\textbf{A portion of long-term contributors progress towards reduced technical importance, while the majority of contributors do not demonstrate a clear trend in shifting technical importance.} We observe that approximately 7\% \textit{core} contributors exhibit significant increasing trends in the evolution of their commit importance (i.e., commit\_centrality), and 9\% show significant decreasing trends. This indicates two different evolution patterns in the commit importance (i.e., commit\_centrality): a gradual increase in the importance of their code contributions or a shift away from highly technical code contributions that modify the influential files in the repository. Period technical importance (i.e., period\_centrality) measures both the volume and the importance of the commits made within a period. Similar to our observations in the evolution of workload compositions, significant trends in period technical importance are observed only among long-term contributors who have been active in the projects for over 2.5 years. Among these contributors with significant changes in their technical importance, we observe a predominant trend toward decreasing technical importance. This finding implies a tendency of long-term contributors to shift away from intensively making highly technical code contributions. Table~\ref{tab:stage_evolution_trend} shows that a greater proportion of long-term contributors have decreasing period technical importance in their middle to later stages, compared to their early to middle stages.

\begin{MySummaryBox}{RQ4 Summary}
The Committer's and Code Reviewer's contributions are more sustainable as project development progresses, while the Issue Reporter's contributions are less sustainable. A portion of long-term contributors show a tendency towards fewer, constant, and balanced contributions, and shift away from highly technical and intensive contributions. Nevertheless, the majority of contributors do not exhibit a significant trend in their evolving OSS engagement. 
\end{MySummaryBox}

\section{Implications}
\label{sec:implications}

In this section, we discuss the implications of our findings for ML contributors, ML project maintainers and managers, and software engineering researchers.

\textbf{Implications for OSS contributors.} In RQ1, our findings reveal that \textit{core} contributors, regardless of their working time, often make large and complex code changes (i.e., have a higher code contribution density). However, they demonstrate a lower pull request approval ratio and pull request approval density than \textit{peripheral} contributors. Prior research by Yu et al.~\cite{yu2016determinants} reports a similar finding that the addition churn of pull requests (i.e., the number of lines of code added) exhibits a significant negative correlation with pull request acceptance and a significant positive correlation with pull request latency. They find that small pull requests tend to be accepted quickly. Our study further indicates that especially for \textit{core} contributors, even though they are working on one pull request, they are recommended to break down their task list and split a large pull request into smaller and more manageable segments, thus avoiding unintended bugs and extensive review efforts.

Furthermore, in RQ1, we identify duration, the number of authored files, and collaborations as the most important factors distinguishing \textit{core} and \textit{peripheral} contributors. Hence, for newcomers who wish to progress to \textit{core} contributors within an ML project, we suggest that they can combine various efforts, including prolonging their involvement in the project, taking ownership at the file level, and diversifying their contributions by collaborating with others.

Additionally, our findings in RQ2 reveal that \textit{core} contributors tend to exhibit a progression toward higher technical importance within the first 6 months of activities in a project, whereas \textit{peripheral} contributors typically maintain the same level of technical importance upon joining the project. Newcomers could start to tackle more challenging tasks after joining an ML project for several months to gradually become a \textit{core} contributor.


\textbf{Implications for project maintainers and managers.} As above mentioned, we observe that \textit{core} contributors make large and complex code contributions to solve difficult tasks, which might require extensive review efforts. We recommend project maintainers to pay extra attention 
to these pull requests and proactively assign code reviewers with higher technical skills (such as Code Reviewers identified in Section~\ref{sec:identify_workload_composition_pattern}) to assist in reviewing code changes and facilitating approval.

Despite peripheral contributors making fewer code contributions, they can play a vital role in other OSS activities, such as raising and resolving issues. Project maintainers should acknowledge the significance of these contributors, as highlighted in RQ3, where projects with a higher ratio of issue-reporting-focused contributors attract greater popularity. \textit{Peripheral} profiles include motivated contributors who begin their contributions by addressing issues they have identified or implementing features they wish to have. However, being part of a \textit{Peripheral} profile, their involvement in the project is typically brief, which may lead to their contributions becoming difficult to maintain. To mitigate this, project maintainers can improve the stickiness of these contributors for better document/feature maintenance. Additionally, as indicated in RQ3, projects gain more popularity with increased code review from Issue Discussants. We suggest project maintainers actively assign Issue Discussants to review code changes from newcomers, especially the newcomers who have previously interacted with in issue discussions. This strategy could foster a welcoming environment for newcoming contributors.

\textbf{Implications for software engineering researchers.} Our research can be leveraged by future research in software engineering that aims to deeply understand ML OSS contributors and delve into their detailed characteristics. We identify four contributor profiles in popular ML libraries and describe their respective contributor characteristics. This categorization approach provides a framework for researchers to extend upon based on their dimensions of interest. For example, they might analyze the quality of code commits from various contributors using code analysis tools, and provide insights into best practices that enhance the code quality from open-source contributors. In RQ3, we find that a higher ratio of contributors focusing on raising issues is associated with the growth of project popularity. This finding can be extended to further investigations, such as the impact of various issue topics towards a project's popularity.
Our finding that increasing code reviews from Issue Discussants and Committers are associated with the growth of a project's popularity can be leveraged to effectively recommend code reviewers not only to examine the quality of code based on code reviewer's expertise and workload, but also to ensure a project's long-term sustainability (e.g., welcoming and retaining newcomers).

\section{Threats To Validity}
\label{sec:threats}

In this section, we discuss possible threats to the validity of our study.

~\textbf{Threats to Internal Validity} concern the effectiveness of our clustering approaches. In this study, we employ various clustering techniques. We select the clustering result from eight clustering algorithms to identify the contributor profiles in Section~\ref{sec:identify_profile}. We also use hierarchical clustering to identify workload composition patterns in Section~\ref{sec:identify_workload_composition_pattern}. To select the optimal number of clusters, we experiment with a range of number of clusters, and use the clustering evaluation metric (i.e., Silhouette score) to find the optimal one. For algorithms not requiring a predetermined cluster number, we conduct a gradient search to find the optimal parameter set that yields the highest Silhouette score. Altering the number of clusters may impact our results. To address this issue, for all the clustering results, beyond the clustering evaluation metric, we also manually investigate the datapoints in the resulting clusters to ensure the clustering yields meaningful grouping and insightful outcomes based on our designated features.

~\textbf{Threats to External Validity} concern the generalizability of our results. Our study has been conducted over 6 ML projects on Github. We also limit our study to analyzing ML library and framework projects, instead of studying all the projects on Github that are implementing ML technology or developing ML libraries. To augment the generalizability of our result, we select the ML projects created for different purposes by different organizations with different sizes to conduct our experiment. Additionally, our selected projects are mainly written in Python and C++. Including more ML frameworks that are developed in other programming languages, such as Java or Matlab, might reveal more meaningful findings regards ML contributors. However, considering that Python is the most preferable programming language for scientific computing and machine learning~\cite{raschka2020machine}, the generalizability of our results to other ML projects is still adequate.

~\textbf{Threats to Construct Validity} concern
whether our analyses measure what we claim to analyze.
In our work, we aim to identify the profiles and understand the characteristics of ML OSS contributors. Although our dataset contains only ML projects, there might still be traditional software developers with little ML knowledge. Unfortunately, we are not able to consult all contributors and ask for their expertise in developing ML software. To minimize this threat, we select the ML projects that are developing ML frameworks and libraries instead of adopting ML techniques for different domains. We believe that there is a greater portion of ML experts in projects developing ML tools than in projects adopting ML tools.

\section{Related Work}
\label{sec:relatedwork}
\subsection{OSS Contributor Characterization}
Historical data in open-source software repositories can provide insights into software developers. Many existing studies in software engineering have categorized the behaviors of OSS contributors and evaluated their expertise in specific areas. 

Researchers begin with classifying OSS contributors according to their amount of code contribution. Mockus et al.~\cite{mockus2000case, mockus2002two} find that the top 20\% active developers are usually the core developers who contribute more than 80\% commits in OSS projects. Based on this finding, Robles et al.~\cite{10.1007/0-387-34226-5_28} further study the turnover of the core developers and propose a quantitative methodology to classify the behaviour of human resources at a project level.
Nakakoji et al.~\cite{nakakoji2002evolution} propose an onion model to classify OSS community members into seven categories: project leaders, core members, active developers, peripheral developers, bug fixers, bug reporters, readers, and passive users. 
Milewicz et al.~\cite{inproceedings} find that open-source scientific software projects are driven by senior members of the team (e.g., professors), and they make 72\% commits on average and mainly tackle architectural problems, while junior members (e.g., graduate students) mainly work on implementing new features. Costa et al.~\cite{da2014unveiling} categorize developers as core, active, and peripheral, and find that core developers make the majority of contributions and are associated with less buggy commits.

The duration of involvement in OSS projects is a key factor in categorizing contributors and has been extensively studied. Pinto et al.~\cite{pinto2016more} find that half of the newcomers in a project only make one contribution and then depart from the project. Steinmacher et al.~\cite{steinmacher2013newcomers} find that more than 80\% newcomers do not become long-term contributors. Zhou et al.~\cite{6880395}, Wang et al.~\cite{wang2018will} and Bao et al.~\cite{fe8a3ca6d43d4253855607d9a2cad816} develop prediction models to identify potential long-term contributors based on their initial behaviors in the project repository.

Technique expertise is an important aspect of differentiating contributors and identifying experts. Dey et al.~\cite{dey2021representation} propose an approach to represent contributors' expertise with their API usage. Dakhel et al.~\cite{Dakhel_2023} represent the domain expertise of contributors based on the projects they have contributed to, issues solved, and their API usage. Ahasanuzzaman et al.~\cite{ahasanuzzaman2024using} propose an approach that uses language-specific functions and APIs to assess one's efficiency in a programming language. Montandon et al.~\cite{DBLP:journals/corr/abs-1903-08113} explore the effectiveness of clustering and machine learning classifiers in assessing contributors' expertise and identifying experts based on GitHub activities. They find that clustering methods yield better results compared to supervised classifiers. Yang et al.~\cite{YANG2020106336} develop a tool to visualize developers' portraits based on their Github profile pages and coding details.

There are studies exploring the impact of personal aspects on OSS contributors. Cataldo et al.~\cite{geo} investigate the correlation between the quality of software projects and the geographical distributions of developers and find that projects with imbalanced distributions of the development team tend to have more defects compared to projects with balanced distributions. Claes et al.~\cite{claes2018programmers} conduct a large-scale study on the working hours of developers and find that approximately two-thirds of developers follow normal office hours, and no correlation between project maturation and the reduction in abnormal working hours. Yue et al.~\cite{yue2022off} find that the preference of contribution dynamics in the early career affects the contributor's technical success in the project. In contrast, we study the work preferences of the contributor's entire career in ML projects as well as the impact of contributors' work preferences on the project's popularity. Elbaum et al.~\cite{munson1998code} identify a strong association between code churn and defects. Shin et al.~\cite{shin2010evaluating} report code churn as one of the software vulnerability indicators, indicating that the style of contributors' contributions, such as the size of their commits, can significantly impact the quality and security of the codebase. 

In contrast to previous research that primarily examines traditional software projects, our study concentrates on contributors within popular machine learning projects. Unlike existing studies that often investigate one specific aspect of contributors, we aim to construct a comprehensive profile of open-source machine learning contributors by considering multiple perspectives and providing a holistic understanding of the characteristics and behaviors of members of the ML OSS community.

\subsection{Research for Machine Learning Contributors}

Research efforts to study ML software developers have been arising within the past few years. Cai et al.~\cite{mldeveloper} deploy a survey to investigate the motivations, challenges, and demands of software developers when they begin learning ML, and find that the major challenge of ML beginners is the lack of understanding of ML concepts and mathematics, and they demand ML frameworks to provide such support. Hill et al.~\cite{MLdeveloperexp} interview 11 ML professionals from a company and find that their main challenge is creating a repeatable workflow for ML system development process. Ishikawa et al.~\cite{8836142} conduct a questionnaire survey on 278 ML practitioners and report the main difficulties perceived by ML contributors including low accuracy, absence of an oracle, and uncertainty regarding system behavior. Gangash et al.~\cite{8816808} conduct an empirical study analyzing ML-related posts on Stack Overflow and explore the topics frequently discussed by ML developers. Islam et al.~\cite{islam2019developers} map the posts related to 10 popular ML libraries on Stack Overflow to ML system development stages. They identify model construction and data preparation as the most challenging stage of ML system development perceived by ML developers. Han et al.~\cite{10398589} study the onboard process for deep learning newcomers based on their activities before their first accepted commit in deep learning projects. In contrast, we study the complete duration of ML contributors in OSS ML projects and their evolution.

Previous research on ML contributors has primarily explored their motivations for learning ML and the challenges encountered in ML software development. These studies often use methods like surveys, interviews, and analyzing posts from Q\&A systems to delve into the contributors' mentalities, focusing on their thoughts and philosophies. In contrast, our approach is centered on understanding ML contributors based on their contributions and activities within software repositories. Our work emphasizes the contributors' actions rather than their mindset. We aim to establish a categorization of ML contributors based on their OSS activities, so we can pinpoint their characteristics and provide suggestions accordingly.

\section{Conclusion}
\label{sec:conclusion}
This paper conducts an empirical study aimed at advancing the understanding of machine learning contributors within the Open Source Software community. By studying 7,640 contributors from 6 popular ML libraries, we identify four contributor profiles and reveal significant features associated with the profiles, such as project experience, authorship diversity, collaborations, and geological location. We study the OSS engagement of contributors from three aspects: workload composition, work preferences, and technical importance, and reveal the variations in the OSS engagement across contributors in different profiles. Moreover, we explore the impact of contributor OSS engagement on project popularity and identify the important factors of workload composition and work preference associated with the increase in project popularity. The study of the evolution of contributor OSS engagement highlights that long-term contributors evolve towards fewer, constant, and balanced contribution dynamics, moving away from highly technical and intensive contributions. These findings contribute to a deeper understanding of the ML OSS ecosystem and offer practical implications for project managers, maintainers, and researchers aiming to foster a welcoming and collaborative OSS environment. Additionally, our findings offer insights into potential pathways for newcomers to progress within ML projects. Moving forward, we aim to expand our empirical studies to include contributors from more ML frameworks and applications.

\section{Replication Package}
\label{sec:replication}

\mbox{To facilitate the replication of our study, we provide our } \mbox{data and code publicly available at:} 

\noindent\url{https://github.com/seal-tosem/replication}.




\bibliographystyle{ACM-Reference-Format}
\bibliography{ref}
\appendix

\afterpage{\clearpage}
\section{RQ1 Supplementary Result}
\label{appendix:rq1}

\begin{table}[H]
\caption{Result of Mann-Whitney U test and effect size results for contributor features.}
\label{tab:rq1_mann}
\centering
\begin{adjustbox}{max width=0.9\textwidth}
\begin{tabular}{l|c|c|c}
\hline
\multirow{2}{*}{\textbf{Contributor Features}} & \textbf{Group1: Core} & \textbf{Group1:   Core-Afterhour} & \textbf{Group1: Peripheral-Afterhour} \\
 & \textbf{Group2: Peripheral} & \textbf{Group2: Core-Workhour} & \textbf{Group2: Peripheral-Workhour} \\ \hline
\textbf{Duration} & large (0.55) & not significant & not significant \\
\textbf{Timezone} & negligible (0.057) & medium (0.439) & small (0.268) \\
\textbf{Authored files} & large (0.94) & negligible (0.076) & negligible (-0.035) \\
\textbf{Commit Rate} & not significant & not significant & not significant \\
\textbf{Code Commits} & large (0.841) & negligible (0.067) & not significant \\
\textbf{Code Commit Rate} & small (0.25) & not significant & not significant \\
\textbf{Other Commits} & medium (0.381) & negligible (0.089) & not significant \\
\textbf{Other Commit Rate} & small (0.199) & negligible (0.084) & not significant \\
\textbf{Code Contribution} & large (0.988) & negligible (0.051) & negligible (-0.038) \\
\textbf{Code Contribution Rate} & large (0.661) & not significant & not significant \\
\textbf{Total Issues} & small (0.175) & negligible (-0.124) & not significant \\
\textbf{Issue participated} & medium (0.395) & negligible (-0.054) & not significant \\
\textbf{Issue Solved} & small (0.186) & negligible (-0.067) & negligible (-0.022) \\
\textbf{Issue Contribution} & small (0.246) & negligible (-0.103) & not significant \\
\textbf{Issue Contribution Rate} & negligible (0.118) & negligible (-0.108) & not significant \\
\textbf{Issue Solving Ratio} & negligible (0.141) & negligible (-0.062) & negligible (-0.022) \\
\textbf{Issue Solving Density} & negligible (0.14) & negligible (-0.067) & negligible (-0.022) \\
\textbf{Total Pull Requests} & medium (0.431) & small (-0.227) & not significant \\
\textbf{PR Merged} & medium (0.389) & small (-0.177) & negligible (0.038) \\
\textbf{PR Reviewed} & medium (0.394) & negligible (-0.081) & negligible (-0.024) \\
\textbf{PR Participated} & large (0.553) & small (-0.162) & negligible (-0.038) \\
\textbf{PR Contribution} & large (0.519) & small (-0.164) & negligible (-0.03) \\
\textbf{PR Contribution Rate} & negligible (-0.121) & small (-0.19) & not significant \\
\textbf{PR Approval Ratio} & small (-0.28) & not significant & negligible (0.059) \\
\textbf{PR Approval Density} & medium (-0.423) & not significant & negligible (0.077) \\
\textbf{Followers} & negligible (0.03) & not significant & not significant \\
\textbf{Collaborations} & medium (0.352) & small (-0.206) & negligible (-0.051) \\ \hline
\multicolumn{4}{l}{\begin{tabular}[c]{@{}l@{}}Groups with no significant difference (i.e., null hypothesis is accepted) are indicated with 'not significant'. Groups \\that are significantly different examined by Mann-Whitney U test are indicated with Cliff's delta value and its \\ interpretation (i.e., negligible, small, medium, or large). A positive value means group1 is greater than group2, and \\vice versa. Three columns from left to right correspond to the null hypothesis \textit{\(H0_{1}\)}, \textit{\(H0_{2}\)}, and \textit{\(H0_{3}\)} respectively. \end{tabular}} \\ \hline
\end{tabular}
\end{adjustbox}
\end{table}

\clearpage

\afterpage{\clearpage}
\section{RQ2 and RQ3 Supplementary Result}
\label{appendix:B}

Figure~\ref{fig:issue_discussant_example} shows an example of a newcomer being encouraged by Issue Discussants to become a contributor in  TensorFlow. Issue Discussant A and Issue Discussant B are two existing contributors in TensorFlow with the workload composition pattern of Issue Discussant during the period from Oct 27th, 2017, to January 25th, 2018. In this period, a newcomer raises one's first two issue reports (i.e., Issue 1 and Issue 2) on TensorFlow. Issue Discussant A actively engages in discussions for Issue 1 and Issue Discussant B participates in the discussions for both issues. In both issue reports, Issue Discussant B encourages the newcomer to make contributions. The newcomer submits one's first pull request to address the observed issue (i.e., Pull Request 1), which is then reviewed and approved by Issue Discussant B. This newcomer eventually becomes a long-contributor to Tensorflow and remains active in the project for more than 4 years.

\begin{figure}[]
    \centering
    \begin{subfigure}[b]{0.85\textwidth}
        \includegraphics[width=\textwidth]{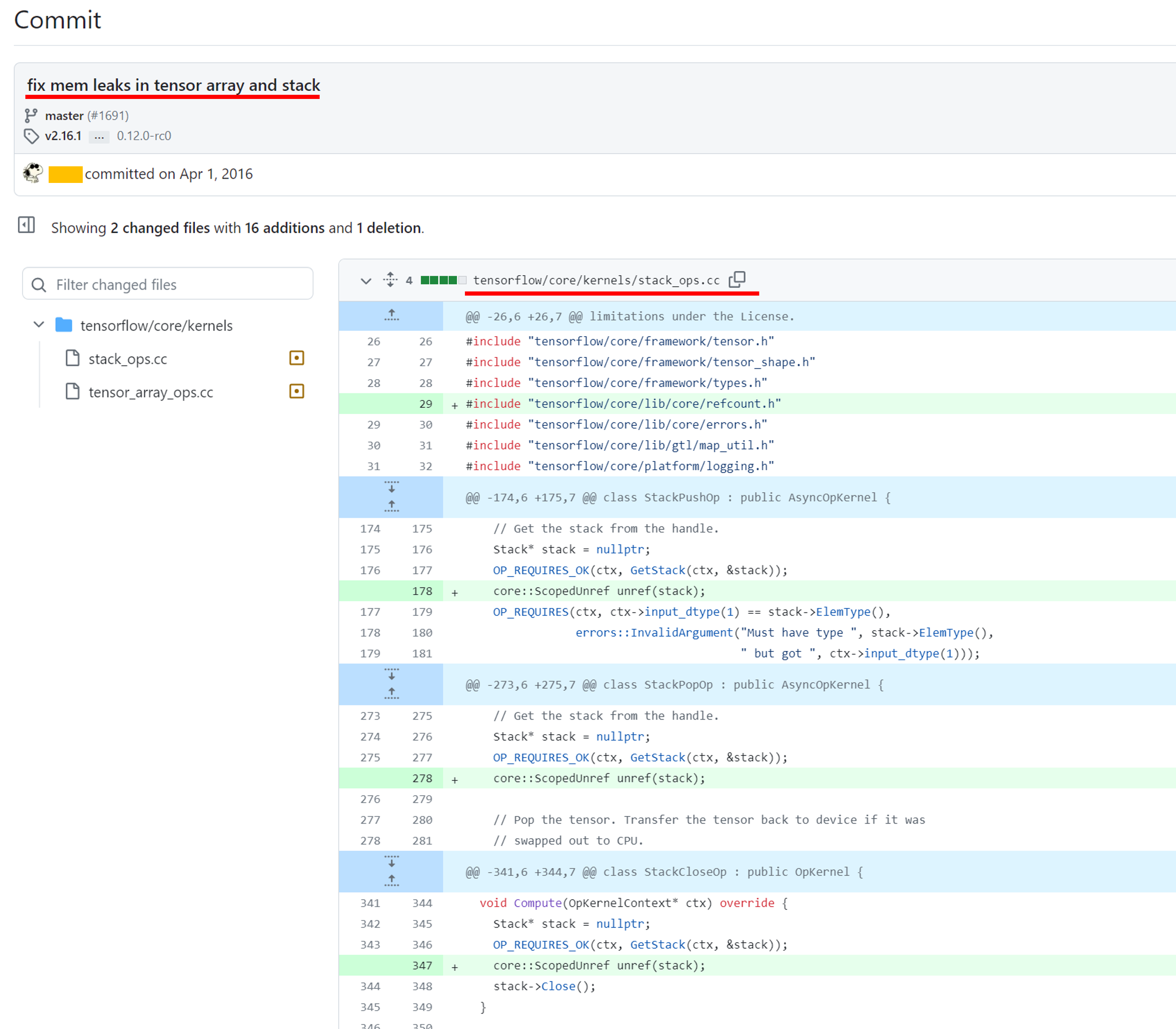}
        \caption{High commit centrality.}
        \label{fig:high_commit_importance}
    \end{subfigure}
    \begin{subfigure}[b]{0.85\textwidth}
        \includegraphics[width=\textwidth]{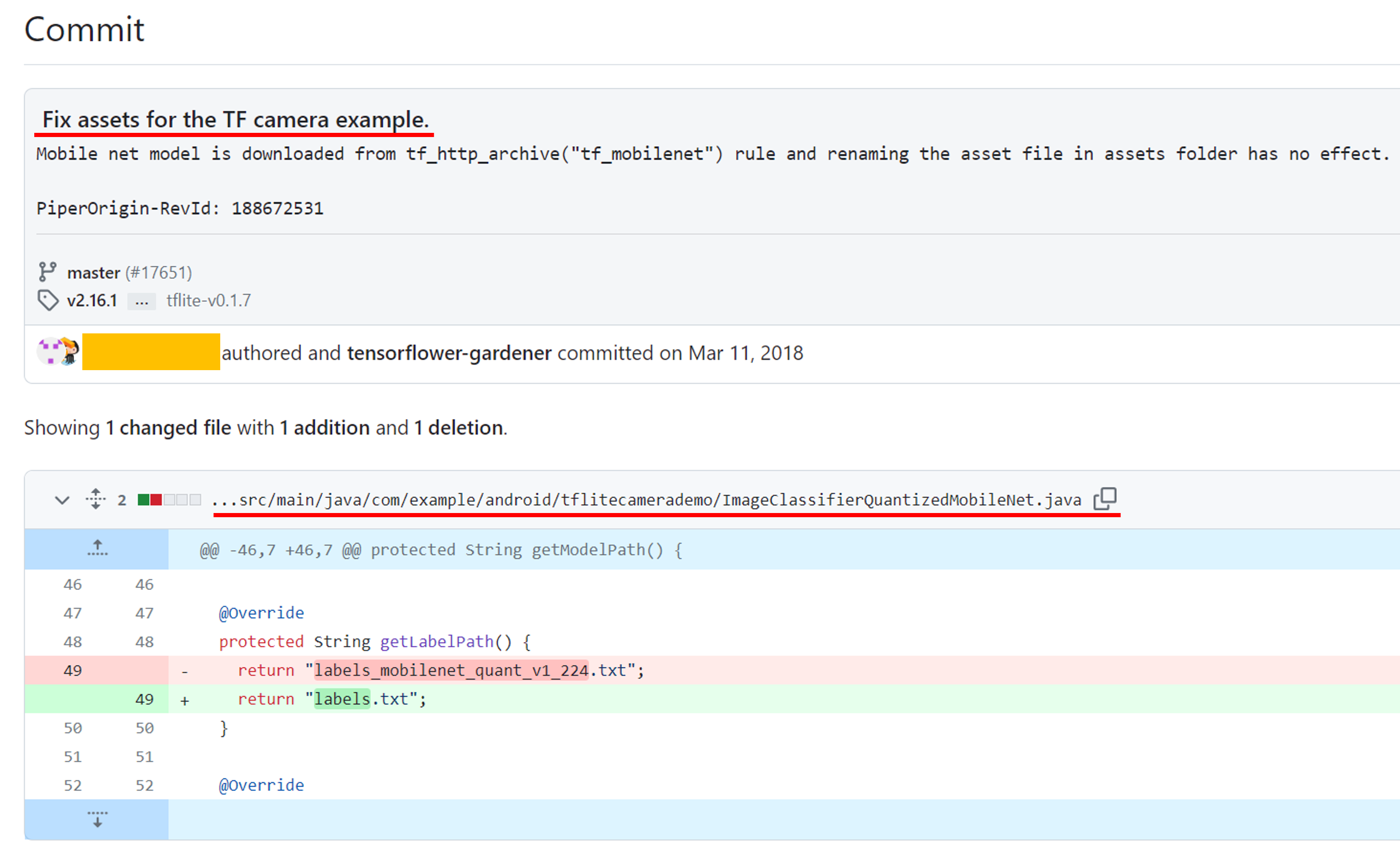}
        \caption{Low commit centrality.}
        \label{fig:low_commit_importance}
    \end{subfigure}
    \caption{An example of commits with different levels of commit importance (commit centrality).}
    \label{fig:commit_importance}
\end{figure}

Figure~\ref{fig:high_commit_importance} shows a commit with high commit importance (i.e., \textit{commit centrality} = 0.018). This commit fixes the memory leak in the tensors, which is the fundamental functionality of the deep learning frameworks. The two files being modified in this commit contain the core and kernel functions, and changing these files can affect all computations happening within the framework. Figure~\ref{fig:low_commit_importance} shows a commit with relatively low commit importance (i.e., \textit{commit centrality} = 1.22e-24). This commit makes changes to the demo code for a specific implementation of TensorFlow Lite on an Android device's camera. The impact of such modifications is often limited to the specific functionalities and unlikely to cause large-scale disruptions.

Note that the figures in this appendix have been cropped to omit unimportant details. The original content can be found on TensorFlow GitHub repository \footnote{https://github.com/tensorflow/tensorflow/issues/15046}\footnote{https://github.com/tensorflow/tensorflow/issues/15374}\footnote{https://github.com/tensorflow/tensorflow/pull/15533}\footnote{https://github.com/tensorflow/tensorflow/commit/2e8726c84a70ff962e11fac26472603aa92cc94e}\footnote{https://github.com/tensorflow/tensorflow/commit/4b7511f4ecd6d0bd491ec557fe05fdfe731ecdae}.

\begin{figure}
    \centering
    \begin{subfigure}{\textwidth}
        \centering
        \includegraphics[width=0.67\textwidth]{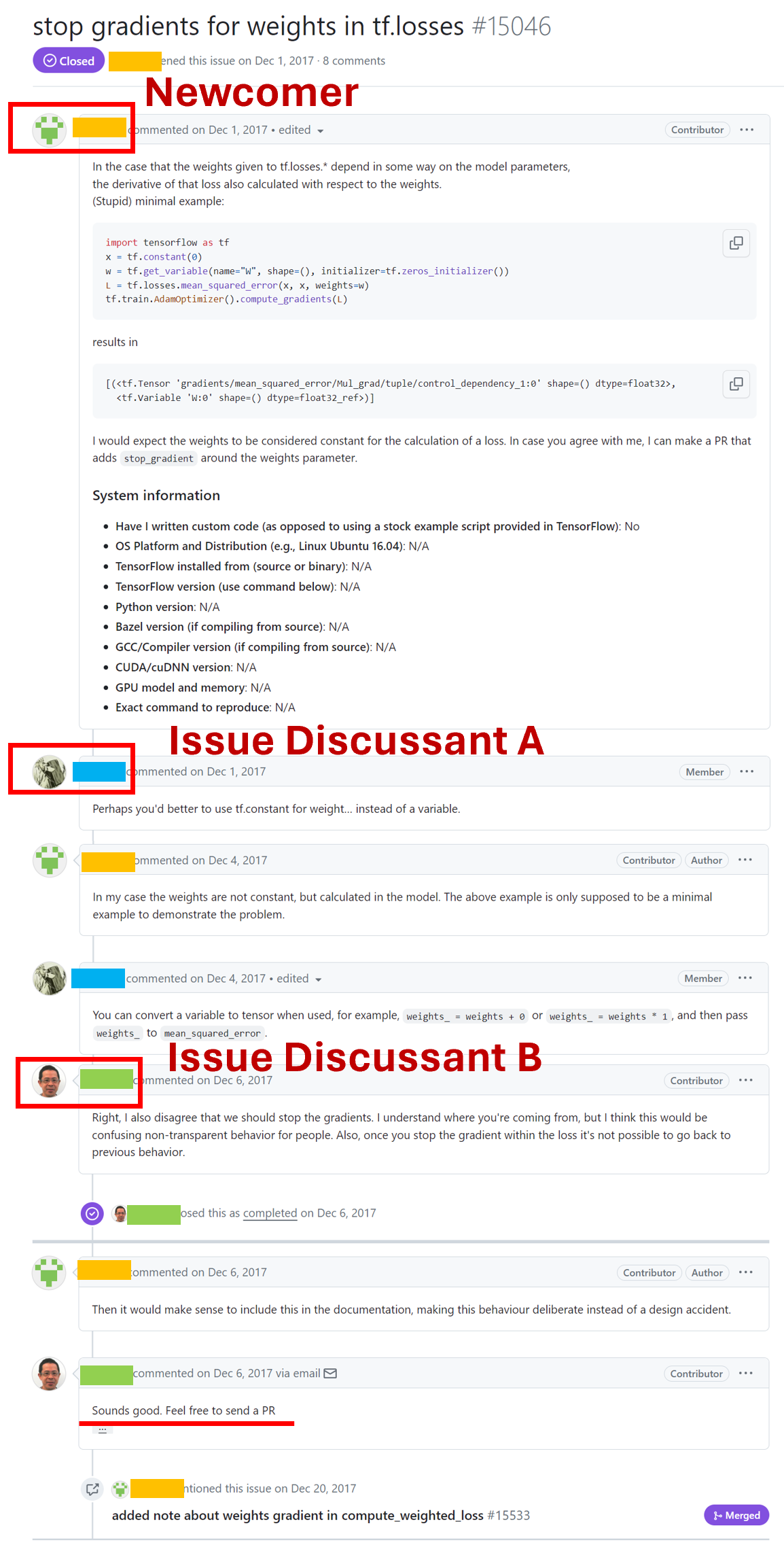}
        \caption{Issue 1}
    \end{subfigure}
\end{figure}
\begin{figure}\ContinuedFloat
    \centering
    \begin{subfigure}{0.68\textwidth}
        \centering
        \includegraphics[width=\textwidth]{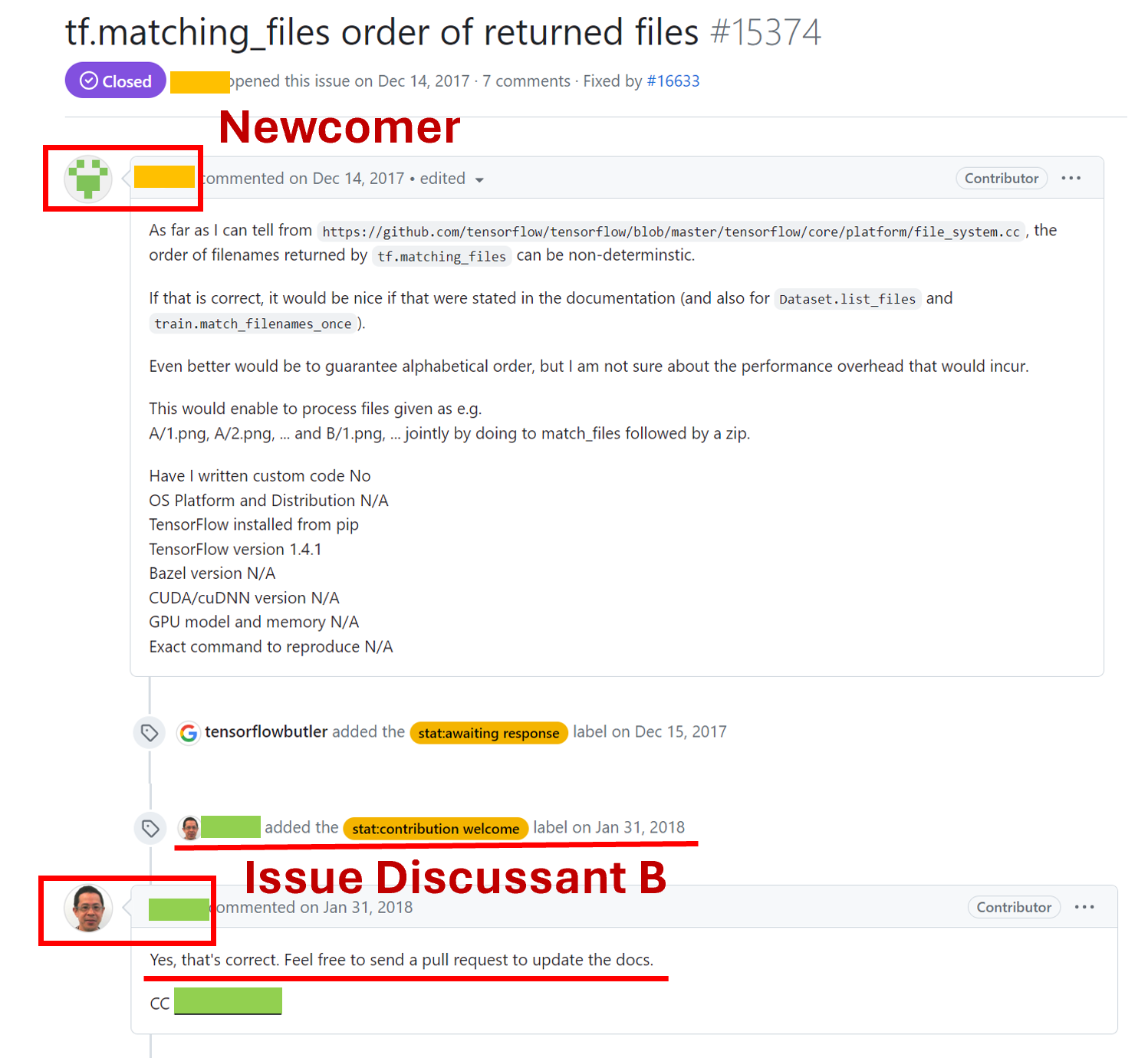}
        \caption{Issue 2}
    \end{subfigure}
    \begin{subfigure}{0.68\textwidth}
        \centering
        \includegraphics[width=\textwidth]{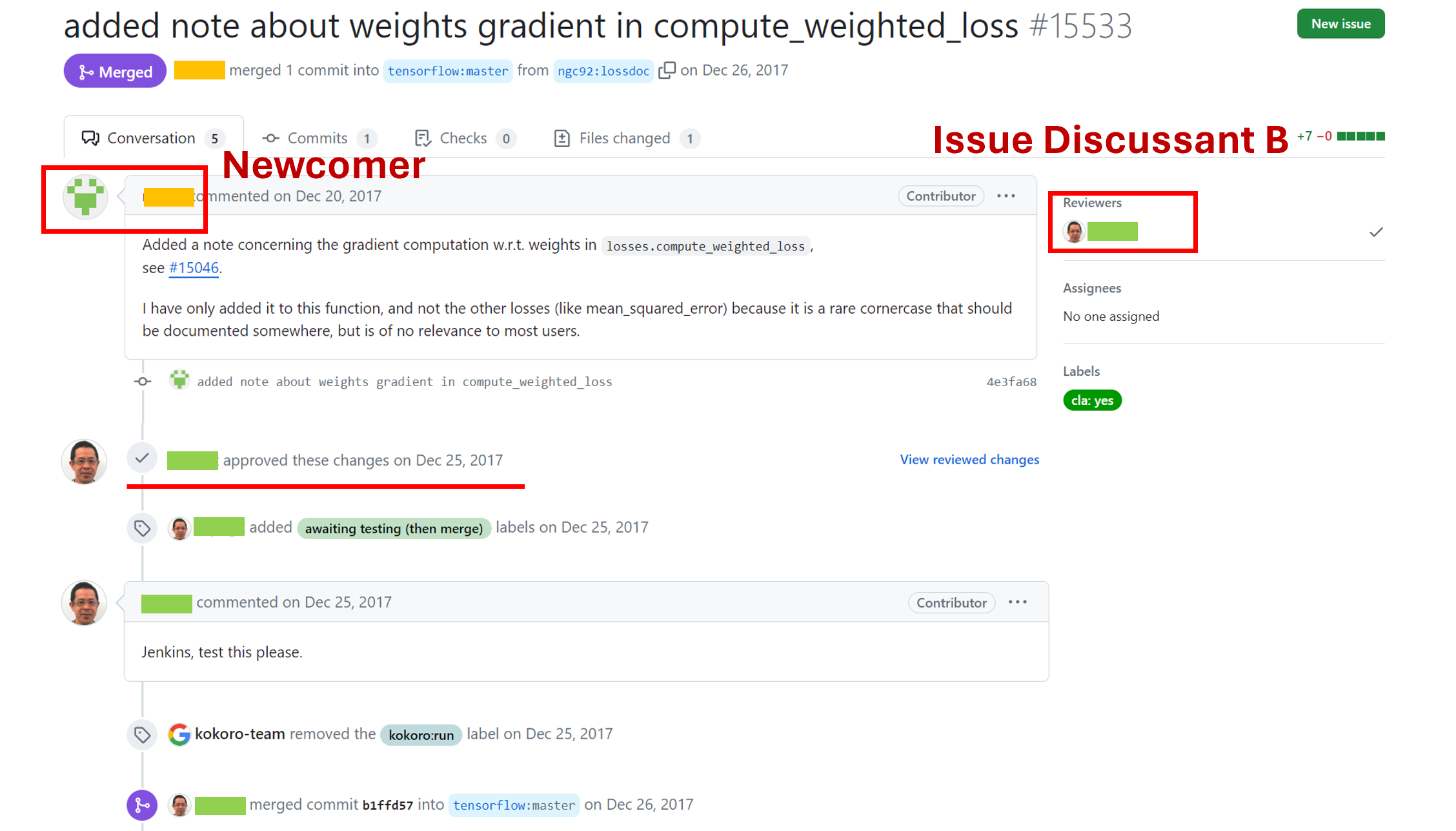}
        \caption{Pull Request 1}
    \end{subfigure}
\caption{A real-world example of newcomers encouraged by Issue Discussants to make contributions and become long-term contributors.}
\label{fig:issue_discussant_example}
\end{figure}

\afterpage{\clearpage}

\afterpage{\clearpage}

\newpage
\afterpage{\clearpage}
\section{RQ4 Supplementary Result}
\label{appendix:A}

\begin{table}[H]
\caption{Overall trend for Workload Composition, Work preference, and technical importance.} 
\label{tab:evolution_trend}
\centering
\begin{adjustbox}{max width=0.9\textwidth}
\begin{tabular}{l|c|c|c|c}
\hline
\textbf{Profile} & \textbf{Core-Afterhour} & \textbf{Core-Workhour} & \textbf{Peripheral-Afterhour} & \textbf{Peripheral-Workhour} \\ \hline
\textbf{workload composition} & $\nearrow$(1.8\%) $\searrow$(0.4\%) & $\nearrow$(1.6\%) $\searrow$(0.4\%) & $\nearrow$(0.2\%) $\searrow$(0.0\%) & $\nearrow$(0.2\%) $\searrow$(0.0\%) \\ \hline
\textbf{binned\_entropy} & $\nearrow$(1.0\%) $\searrow$(4.2\%) & $\nearrow$(0.8\%) $\searrow$(2.7\%) & $\nearrow$(0.0\%) $\searrow$(0.2\%) & $\nearrow$(0.1\%) $\searrow$(0.3\%) \\
\textbf{c3(1)} & $\nearrow$(0.5\%) $\searrow$(3.9\%) & $\nearrow$(1.3\%) $\searrow$(2.3\%) & $\nearrow$(0.0\%) $\searrow$(0.1\%) & $\nearrow$(0.0\%) $\searrow$(0.2\%) \\
\textbf{c3(2)} & $\nearrow$(0.6\%) $\searrow$(3.2\%) & $\nearrow$(0.8\%) $\searrow$(2.0\%) & $\nearrow$(0.1\%) $\searrow$(0.1\%) & $\nearrow$(0.0\%) $\searrow$(0.1\%) \\
\textbf{c3(3)} & $\nearrow$(0.8\%) $\searrow$(2.7\%) & $\nearrow$(1.0\%) $\searrow$(2.0\%) & $\nearrow$(0.0\%) $\searrow$(0.1\%) & $\nearrow$(0.0\%) $\searrow$(0.1\%) \\
\textbf{number\_cwt\_peaks} & $\nearrow$(0.6\%) $\searrow$(3.9\%) & $\nearrow$(0.7\%) $\searrow$(3.0\%) & $\nearrow$(0.0\%) $\searrow$(0.1\%) & $\nearrow$(0.1\%) $\searrow$(0.4\%) \\
\textbf{longest\_strike\_above\_mean} & $\nearrow$(0.6\%) $\searrow$(3.5\%) & $\nearrow$(1.0\%) $\searrow$(2.4\%) & $\nearrow$(0.0\%) $\searrow$(0.1\%) & $\nearrow$(0.0\%) $\searrow$(0.4\%) \\
\textbf{longest\_strike\_below\_mean} & $\nearrow$(2.5\%) $\searrow$(1.0\%) & $\nearrow$(2.3\%) $\searrow$(1.1\%) & $\nearrow$(0.2\%) $\searrow$(0.1\%) & $\nearrow$(0.2\%) $\searrow$(0.2\%) \\
\textbf{diverse} & $\nearrow$(0.4\%) $\searrow$(2.3\%) & $\nearrow$(0.9\%) $\searrow$(2.4\%) & $\nearrow$(0.1\%) $\searrow$(0.2\%) & $\nearrow$(0.0\%) $\searrow$(0.2\%) \\
\textbf{balance} & $\nearrow$(4.1\%) $\searrow$(1.1\%) & $\nearrow$(2.7\%) $\searrow$(1.2\%) & $\nearrow$(0.1\%) $\searrow$(0.1\%) & $\nearrow$(0.2\%) $\searrow$(0.0\%) \\ \hline
\textbf{commit} & $\nearrow$(1.0\%) $\searrow$(4.6\%) & $\nearrow$(1.2\%) $\searrow$(4.1\%) & $\nearrow$(0.0\%) $\searrow$(0.0\%) & $\nearrow$(0.0\%) $\searrow$(0.0\%) \\
\textbf{issue} & $\nearrow$(0.2\%) $\searrow$(1.4\%) & $\nearrow$(0.4\%) $\searrow$(2.0\%) & $\nearrow$(0.0\%) $\searrow$(0.1\%) & $\nearrow$(0.0\%) $\searrow$(0.3\%) \\
\textbf{issue comment} & $\nearrow$(0.9\%) $\searrow$(4.5\%) & $\nearrow$(0.9\%) $\searrow$(2.8\%) & $\nearrow$(0.0\%) $\searrow$(0.1\%) & $\nearrow$(0.1\%) $\searrow$(0.2\%) \\
\textbf{pr comment} & $\nearrow$(0.9\%) $\searrow$(2.8\%) & $\nearrow$(1.5\%) $\searrow$(2.4\%) & $\nearrow$(0.1\%) $\searrow$(0.2\%) & $\nearrow$(0.1\%) $\searrow$(0.2\%) \\
\textbf{review} & $\nearrow$(1.5\%) $\searrow$(1.1\%) & $\nearrow$(2.7\%) $\searrow$(0.9\%) & $\nearrow$(0.2\%) $\searrow$(0.1\%) & $\nearrow$(0.0\%) $\searrow$(0.1\%) \\ \hline
\textbf{commit\_centrality} & $\nearrow$(8.1\%) $\searrow$(10.6\%) & $\nearrow$(6.1\%) $\searrow$(9.6\%) & $\nearrow$(0.0\%) $\searrow$(0.1\%) & $\nearrow$(0.1\%) $\searrow$(0.4\%) \\
\textbf{period\_centrality} & $\nearrow$(1.5\%) $\searrow$(4.7\%) & $\nearrow$(0.8\%) $\searrow$(4.8\%) & $\nearrow$(0.0\%) $\searrow$(0.0\%) & $\nearrow$(0.0\%) $\searrow$(0.1\%) \\ \hline

\multicolumn{5}{l}{\begin{tabular}[c]{@{}l@{}}  (\%)$\nearrow$ : The percentage of contributors reject the null hypothesis \(H0_{13}\) and have an increasing trend in the given feature. \\ (\%)$\searrow$ : The percentage of contributors reject the null hypothesis \(H0_{14}\) and have a decreasing trend in the given feature. \end{tabular}} \\
\hline
\end{tabular}
\end{adjustbox}
\end{table}

\afterpage{\clearpage}

\setlength\rotFPtop{0pt plus 1fil}
\begin{sidewaystable}
\footnotesize
\caption{Early-middle and middle-late stage trend for workload composition, work preference, and technical importance.} 
\label{tab:stage_evolution_trend}
\centering
\begin{adjustbox}{max width=\textwidth}
 \begin{tabular}{l|cc|cc|cc|cc}
\hline
\multirow{2}{*}{\textbf{Profile}} & \multicolumn{2}{c|}{\textbf{Core-Afterhour}} & \multicolumn{2}{c|}{\textbf{Core-Workhour}} & \multicolumn{2}{c|}{\textbf{Peripheral-Afterhour}} & \multicolumn{2}{c}{\textbf{Peripheral-Workhour}} \\
 & \textbf{Early-Middle} & \textbf{Middle-Late} & \textbf{Early-Middle} & \textbf{Middle-Late} & \textbf{Early-Middle} & \textbf{Middle-Late} & \textbf{Early-Middle} & \textbf{Middle-Late} \\ \hline
\textbf{workload   composition} & $\nearrow$(0.9\%) $\searrow$(0.1\%) & $\nearrow$(0.2\%) $\searrow$(0.0\%) & $\nearrow$(0.7\%) $\searrow$(0.1\%) & $\nearrow$(0.3\%) $\searrow$(0.2\%) & $\nearrow$(0.0\%) $\searrow$(0.0\%) & $\nearrow$(0.0\%) $\searrow$(0.0\%) & $\nearrow$(0.0\%) $\searrow$(0.0\%) & $\nearrow$(0.1\%) $\searrow$(0.0\%) \\ \hline
\textbf{binned\_entropy} & $\nearrow$(1.6\%) $\searrow$(1.0\%) & $\nearrow$(0.1\%) $\searrow$(4.2\%) & $\nearrow$(0.9\%) $\searrow$(0.5\%) & $\nearrow$(0.2\%) $\searrow$(3.3\%) & $\nearrow$(0.2\%) $\searrow$(0.0\%) & $\nearrow$(0.0\%) $\searrow$(0.0\%) & $\nearrow$(0.1\%) $\searrow$(0.0\%) & $\nearrow$(0.0\%) $\searrow$(0.1\%) \\
\textbf{c3(1)} & $\nearrow$(0.9\%) $\searrow$(0.9\%) & $\nearrow$(0.6\%) $\searrow$(3.5\%) & $\nearrow$(0.8\%) $\searrow$(0.6\%) & $\nearrow$(0.3\%) $\searrow$(3.3\%) & $\nearrow$(0.2\%) $\searrow$(0.0\%) & $\nearrow$(0.0\%) $\searrow$(0.2\%) & $\nearrow$(0.0\%) $\searrow$(0.0\%) & $\nearrow$(0.0\%) $\searrow$(0.0\%) \\
\textbf{c3(2)} & $\nearrow$(0.9\%) $\searrow$(0.7\%) & $\nearrow$(0.3\%) $\searrow$(2.9\%) & $\nearrow$(0.8\%) $\searrow$(0.6\%) & $\nearrow$(0.2\%) $\searrow$(2.6\%) & $\nearrow$(0.0\%) $\searrow$(0.0\%) & $\nearrow$(0.0\%) $\searrow$(0.2\%) & $\nearrow$(0.0\%) $\searrow$(0.0\%) & $\nearrow$(0.0\%) $\searrow$(0.0\%) \\
\textbf{c3(3)} & $\nearrow$(1.0\%) $\searrow$(0.4\%) & $\nearrow$(0.2\%) $\searrow$(3.0\%) & $\nearrow$(1.0\%) $\searrow$(0.3\%) & $\nearrow$(0.2\%) $\searrow$(2.6\%) & $\nearrow$(0.0\%) $\searrow$(0.0\%) & $\nearrow$(0.0\%) $\searrow$(0.2\%) & $\nearrow$(0.0\%) $\searrow$(0.0\%) & $\nearrow$(0.0\%) $\searrow$(0.0\%) \\
\textbf{number\_cwt\_peaks} & $\nearrow$(0.7\%) $\searrow$(0.7\%) & $\nearrow$(0.1\%) $\searrow$(3.3\%) & $\nearrow$(0.9\%) $\searrow$(0.5\%) & $\nearrow$(0.1\%) $\searrow$(2.6\%) & $\nearrow$(0.0\%) $\searrow$(0.0\%) & $\nearrow$(0.0\%) $\searrow$(0.0\%) & $\nearrow$(0.1\%) $\searrow$(0.0\%) & $\nearrow$(0.0\%) $\searrow$(0.0\%) \\
\textbf{longest\_strike\_above\_mean} & $\nearrow$(0.7\%) $\searrow$(0.3\%) & $\nearrow$(0.4\%) $\searrow$(2.1\%) & $\nearrow$(0.5\%) $\searrow$(0.6\%) & $\nearrow$(0.2\%) $\searrow$(1.8\%) & $\nearrow$(0.0\%) $\searrow$(0.0\%) & $\nearrow$(0.0\%) $\searrow$(0.0\%) & $\nearrow$(0.1\%) $\searrow$(0.0\%) & $\nearrow$(0.0\%) $\searrow$(0.1\%) \\
\textbf{longest\_strike\_below\_mean} & $\nearrow$(0.6\%) $\searrow$(1.8\%) & $\nearrow$(2.4\%) $\searrow$(0.1\%) & $\nearrow$(0.1\%) $\searrow$(0.7\%) & $\nearrow$(2.1\%) $\searrow$(0.0\%) & $\nearrow$(0.0\%) $\searrow$(0.0\%) & $\nearrow$(0.5\%) $\searrow$(0.0\%) & $\nearrow$(0.0\%) $\searrow$(0.1\%) & $\nearrow$(0.4\%) $\searrow$(0.0\%) \\
\textbf{diverse} & $\nearrow$(0.6\%) $\searrow$(0.4\%) & $\nearrow$(0.0\%) $\searrow$(2.8\%) & $\nearrow$(0.8\%) $\searrow$(0.1\%) & $\nearrow$(0.5\%) $\searrow$(1.5\%) & $\nearrow$(0.0\%) $\searrow$(0.0\%) & $\nearrow$(0.0\%) $\searrow$(0.2\%) & $\nearrow$(0.1\%) $\searrow$(0.0\%) & $\nearrow$(0.0\%) $\searrow$(0.0\%) \\
\textbf{balance} & $\nearrow$(1.1\%) $\searrow$(1.0\%) & $\nearrow$(4.3\%) $\searrow$(0.2\%) & $\nearrow$(0.7\%) $\searrow$(1.1\%) & $\nearrow$(2.7\%) $\searrow$(1.0\%) & $\nearrow$(0.0\%) $\searrow$(0.0\%) & $\nearrow$(0.2\%) $\searrow$(0.2\%) & $\nearrow$(0.0\%) $\searrow$(0.0\%) & $\nearrow$(0.3\%) $\searrow$(0.0\%) \\ \hline
\textbf{commit} & $\nearrow$(1.0\%) $\searrow$(1.1\%) & $\nearrow$(0.2\%) $\searrow$(3.5\%) & $\nearrow$(0.3\%) $\searrow$(1.3\%) & $\nearrow$(0.2\%) $\searrow$(2.5\%) & $\nearrow$(0.0\%) $\searrow$(0.0\%) & $\nearrow$(0.0\%) $\searrow$(0.0\%) & $\nearrow$(0.0\%) $\searrow$(0.0\%) & $\nearrow$(0.0\%) $\searrow$(0.0\%) \\
\textbf{issue} & $\nearrow$(0.2\%) $\searrow$(0.3\%) & $\nearrow$(0.0\%) $\searrow$(0.6\%) & $\nearrow$(0.6\%) $\searrow$(0.3\%) & $\nearrow$(0.1\%) $\searrow$(0.5\%) & $\nearrow$(0.0\%) $\searrow$(0.0\%) & $\nearrow$(0.0\%) $\searrow$(0.0\%) & $\nearrow$(0.0\%) $\searrow$(0.0\%) & $\nearrow$(0.0\%) $\searrow$(0.1\%) \\
\textbf{issue comment} & $\nearrow$(1.3\%) $\searrow$(1.6\%) & $\nearrow$(0.2\%) $\searrow$(4.5\%) & $\nearrow$(0.9\%) $\searrow$(0.3\%) & $\nearrow$(0.1\%) $\searrow$(2.6\%) & $\nearrow$(0.2\%) $\searrow$(0.0\%) & $\nearrow$(0.0\%) $\searrow$(0.2\%) & $\nearrow$(0.1\%) $\searrow$(0.0\%) & $\nearrow$(0.0\%) $\searrow$(0.0\%) \\
\textbf{pr comment} & $\nearrow$(0.9\%) $\searrow$(0.6\%) & $\nearrow$(0.3\%) $\searrow$(3.9\%) & $\nearrow$(1.3\%) $\searrow$(0.3\%) & $\nearrow$(0.5\%) $\searrow$(3.0\%) & $\nearrow$(0.2\%) $\searrow$(0.0\%) & $\nearrow$(0.0\%) $\searrow$(0.0\%) & $\nearrow$(0.1\%) $\searrow$(0.0\%) & $\nearrow$(0.0\%) $\searrow$(0.0\%) \\
\textbf{review} & $\nearrow$(1.8\%) $\searrow$(0.0\%) & $\nearrow$(0.4\%) $\searrow$(3.0\%) & $\nearrow$(2.1\%) $\searrow$(0.0\%) & $\nearrow$(0.6\%) $\searrow$(2.5\%) & $\nearrow$(0.2\%) $\searrow$(0.0\%) & $\nearrow$(0.0\%) $\searrow$(0.0\%) & $\nearrow$(0.1\%) $\searrow$(0.0\%) & $\nearrow$(0.0\%) $\searrow$(0.0\%) \\ \hline
\textbf{commit\_centrality} & $\nearrow$(6.7\%) $\searrow$(8.9\%) & $\nearrow$(5.8\%) $\searrow$(7.6\%) & $\nearrow$(3.5\%) $\searrow$(8.1\%) & $\nearrow$(5.4\%) $\searrow$(7.6\%) & $\nearrow$(0.0\%) $\searrow$(0.0\%) & $\nearrow$(0.2\%) $\searrow$(0.0\%) & $\nearrow$(0.0\%) $\searrow$(0.0\%) & $\nearrow$(0.0\%) $\searrow$(0.2\%) \\
\textbf{period\_centrality} & $\nearrow$(0.8\%) $\searrow$(1.0\%) & $\nearrow$(0.4\%) $\searrow$(2.1\%) & $\nearrow$(0.1\%) $\searrow$(0.9\%) & $\nearrow$(0.1\%) $\searrow$(2.1\%) & $\nearrow$(0.0\%) $\searrow$(0.0\%) & $\nearrow$(0.0\%) $\searrow$(0.0\%) & $\nearrow$(0.0\%) $\searrow$(0.0\%) & $\nearrow$(0.0\%) $\searrow$(0.0\%) \\ \hline
\multicolumn{9}{l}{\begin{tabular}[c]{@{}l@{}}  (\%)$\nearrow$ : The percentage of contributors reject the null hypothesis \(H0_{13}\) and have an increasing trend in the given feature. \\ (\%)$\searrow$ : The percentage of contributors reject the null hypothesis \(H0_{14}\) and have a decreasing trend in the given feature.\end{tabular}} \\
\hline
\end{tabular}
\end{adjustbox}
\end{sidewaystable}
\clearpage

\end{document}